%% file: artmonr2-vfinal2.tex
\documentclass[structabstract]{aa}  
\usepackage{graphicx}
\usepackage{txfonts}
\begin{document}
   \title{Spectral line survey of the ultracompact {\sc Hii} region Mon R2}
   \subtitle{}
\author{D.~Ginard\inst{1}
\and
M.~Gonz\'alez-Garc\'{\i}a\inst{2}
\and
A.~Fuente\inst{1}
\and
J.~Cernicharo\inst{3}
\and
T.~Alonso-Albi\inst{1}
\and
P.~Pilleri\inst{1,3}
\and
M. Gerin\inst{4}
\and
S. Garc\'{\i}a-Burillo\inst{1}
\and
V.~Ossenkopf\inst{5,6}
\and
J.R.~Rizzo\inst{3}
\and
C.~Kramer\inst{2}
\and
J.~R.~Goicoechea\inst{3}
\and
J.~Pety\inst{7}
\and
O.~Bern\'e\inst{8}
\and 
C.~Joblin\inst{9,10}
}

\institute{ Observatorio Astron\'omico Nacional (OAN). Apdo. 112.		% 1
             E-28800 Alcal\'a de Henares. Madrid. Spain
\and
       Instituto de Radio Astronom\'ia Milim\'etrica (IRAM), Avenida Divina Pastora 7, Local 20, 18012 Granada, Spain			% 2
\and
Centro de Astrobiolog\'ia, CSIC-INTA, Crta M-108, km.~4, E-28850 Torrej\'on de Ardoz, Spain% 3
         \and
  LERMA, Observatoire de Paris, 61 Av. de l'Observatoire, 75014 Paris, France % 4
\and
I. Physikalisches Institut der Universit\"at 
zu K\"oln, Z\"ulpicher Stra\ss{}e 77, 50937 K\"oln, Germany% 5
\and
SRON Netherlands Institute for Space Research, P.O. Box 800, 9700 AV % 6
Groningen, Netherlands
\and
Institut de Radioastronomie Millim\'etrique, 300 Rue de la Piscine, 38406 Saint Martin d'H\'eres, France % 7
\and
Leiden Observatory, Leiden University, The Netherlands	% 8
\and
Universit\'e de Toulouse; UPS-OMP; IRAP;  Toulouse, France
 \and
%8
CNRS; IRAP; 9 Av. colonel Roche, BP 44346, F-31028 Toulouse cedex 4, France 
              }
             
 %\abstract{}{aims}{methods}{results}{} 
% 5 {} token are mandatory
 \abstract
 % context heading (optional)
{Ultracompact (UC) \ion{H}{II} regions constitute one of the earliest phases in the formation of a massive
star and are characterized by extreme physical conditions ($G_{\mathrm 0}$$>$10$^5$ Habing field
and $n$$>$10$^6$~cm$^{-3}$). The UC \ion{H}{II} Mon~R2 is the closest one and therefore an excellent target to study the
chemistry in these complex regions.} 
  % aims heading (mandatory)
  {Our goal is to investigate the chemistry of the molecular gas around the UC \ion{H}{II} and the possible variations due
  to the different local physical conditions.}
  % methods heading (mandatory)
 {We carried out a 3mm and 1mm spectral survey using the IRAM 30-m telescope towards three positions
  that represent different physical environments in Mon~R2: (i) the ionization front (IF) at (0$\arcsec$,0$\arcsec$);  two peaks in the molecular cloud (ii) MP1 at the 
  offset (+15$\arcsec$,-15$\arcsec$) and (iii) MP2 at the farther offset (0$\arcsec$,40$\arcsec$). In addition, we carried out extensive modeling 
 to explain the chemical differences between the three observed regions.}
  % results heading (mandatory)
  {We detected more than thirty different species  (including isotopologues and deuterated compounds). 
  In particular, we detected SO$^+$ and C$_4$H confirming that UV radiation plays an important role 
  in the molecular chemistry of this region. In agreement with this interpretation we detected the typical PDR molecules CN, HCN, HCO, C$_2$H, 
  and c-C$_3$H$_2$. There are chemical differences between the observed positions. While the IF and the MP1 
  have a chemistry similar to that found in high UV
  field and dense PDRs like the Orion Bar, the MP2 is more similar to lower UV/density PDRs  like the Horsehead nebula. Our chemical
  modeling supports this interpretation.
 
  In addition to the PDR-like species, we also detected complex molecules such as CH$_3$CN, H$_2$CO, HC$_3$N, CH$_3$OH or CH$_3$C$_2$H that are not usually 
  found in PDRs. A wealth of sulfur compounds, CS, HCS$^+$, C$_2$S, H$_2$CS, SO and SO$_2$ and the deuterated species
  DCN and C$_2$D were also identified.  The origin of the complex species requires further study. The observed deuteration fractionations,
  [DCN]/[HCN]$\sim$0.03 and [C$_2$D]/[C$_2$H]$\sim$0.05, are among the highest in warm regions.
  %We discuss the possible influence of time-dependent effects on the abundance of complex molecules.
  }
  % conclusions heading (optional). leave it empty if necessary 
   {Our results show that the high UV/dense PDRs present a different chemistry from that of the low UV case. Some 
   abundance ratios like [CO$^+$]/[HCO$^+$] or [HCO]/[HCO$^+$] are good diagnostics to differentiate between them. In Mon R2 we have
   the two classes of PDRs, a high UV PDR towards the IF and the adjacent molecular bar and a low-UV PDR which extends towards the north-west
   following the border of the cloud. }
\keywords{Surveys - Stars: formation - ISM: molecules - Line: identification - Astrochemistry - ISM:individual objects:Mon R2}
\maketitle
%
%________________________________________________________________

\section{Introduction}
 
During the earliest stages of star formation, the UV radiation from the newly-born star 
ionizes the most exposed layers of the host molecular cloud, creating a layer of ionized gas (constituted mainly by {\sc Hii})
and originating a so-called photo-dissociation region (PDR).

Ultracompact HII regions (UC {\sc Hii}) are defined as regions of ionized gas with
diameters smaller than $\sim$0.1pc (see Churchwell 2002 for a review). UC {\sc Hii} regions are expected to expand at velocities of the order of the sound
speed (10~km~s$^{-1}$) until reaching equilibrium at dimensions of a few pc. In regions with density of $n\sim$10$^5$~cm$^{-3}$,
{\sc Hii} regions should remain ultracompact for $\sim$3000~yr and only a few dozen should exist in the Galaxy. However,
observations suggest that many more UC {\sc Hii}s exist and that lifetimes should be one to two orders of magnitude
larger. Several models have been proposed to explain this $``$lifetime paradox$"$ but all have shortcomings. The
paradox could be resolved if the molecular gas in which an O star forms is denser and warmer than previously
believed, resulting in an initial Stromgren sphere much smaller than originally estimated. This suggestion is
supported by observations which show that in dense molecular cloud cores densities of $\sim$10$^7$~cm$^{-3}$ and temperatures
of $\sim$100~K are not atypical (Rizzo et al. 2005). The study of the physics and chemistry of UC {\sc Hii} and of the surrounding PDRs
is the key to understand this evolutionary stage of a massive star.
 
UC {\sc Hii} are characterized by extreme UV radiation field ($G_{\mathrm 0}$$>$~10$^5$ in units of the Habing field) and gas
densities  often higher than $>$~10$^6$~cm$^{-3}$. UC {\sc Hii} may represent the best example of highly UV-irradiated PDRs 
and can be used as a template for other more complex systems. Highly UV-irradiated PDRs are found in astrophysical environments of high
interest such as the surface of circumstellar disks and in the nuclei of starburst galaxies.
Unfortunately, our knowledge of the chemistry and physics 
of UC {\sc Hii} is still far from complete.

Monoceros R2 (Mon~R2) hosts the closest (d=830~pc; Herbst \& Racine 1976) and brightest galactic UC {\sc Hii}, and the only one that can be 
spatially resolved in the mm domain with single-dish telescopes. It is therefore the best case-study to determine the physical and chemical evolution of massive star forming regions.
The central, spherical UC {\sc Hii} was created by the interaction of the central infrared source IRS1 (Wood \& Churchwell 1989) with its host molecular cloud. 
The formation of the B0V star associated to IRS1 created
a huge bipolar outflow ($\sim$15$\arcmin$ = 3.6 pc long, Massi, Felli \& Simon 1985; Henning, Chini \& Pfau 1992, Tafalla et al. 1994). 
Later CO 3$\rightarrow$2 mapping of the region showed that this very extended outflow is now inactive. The highest velocity gas is detected towards
IRS~3 suggesting that this young star is associated with a compact ($<$14$"$) bipolar outflow (Giannakopoulou et al. 1997).

Further studies at mm-wavelengths (Giannakopoulou et al. 1997; Tafalla et al. 1997; Choi et al. 2000; Rizzo et al. 2003, 2005)
showed that the UC {\sc Hii} region is located inside a cavity bound by a dense molecular ridge at the south-east.
The peak of this molecular ridge is located at the offset (+10$\arcsec$,$-$10$\arcsec$) relative to the peak of
the ionized gas, and presents a molecular hydrogen column density of 2$-$6$\times$10$^{22}$~cm$^{-2}$. 
The detection and subsequent analysis of the millimeter lines of the cyclic hydrocarbon species c-C$_3$H$_2$ with the IRAM 30m telescope revealed 
gas densities of few $10^6$\,cm$^{-3}$ (Rizzo et al. 2005) in the PDR surrounding the {\sc Hii} region. Recently,
Spitzer observations of the H$_2$ rotational lines and the polycyclic aromatic hydrocarbons  (PAHs) mid-infrared bands 
have revealed a thin layer ($n$=4$\times$10$^5$ cm$^{-3}$, $N(H_{\mathrm 2})$=1$\times$10$^{21}$~cm$^{-2}$) of warm gas ($T_{\mathrm k}$=574($\pm$20)~K) 
between the ionized gas and the dense molecular gas traced by previous millimeter observations (Bern\'e et al. 2009). 
%Figure 1 displays a summary of the morphology of Mon~R2 at both mm ($^{13}$CO \textit{2$\rightarrow$1}) and mid-IR wavelengths. 
They interpreted the differences between the spatial distribution of the emission
of the H$_2$ rotational lines and the emission from PAHs around the {\sc Hii} region as the consequence of 
variations of the local UV field and density in this molecular gas layer.

In this paper, we present a mm molecular survey towards the UC {\sc Hii} region Mon R2. Our goal is to
investigate the molecular chemistry in the PDR that surrounds the UC {\sc Hii} region
and try to construct a pattern to interpret future observations.

\begin{figure*}
\centering
\includegraphics[width=0.9\textwidth]{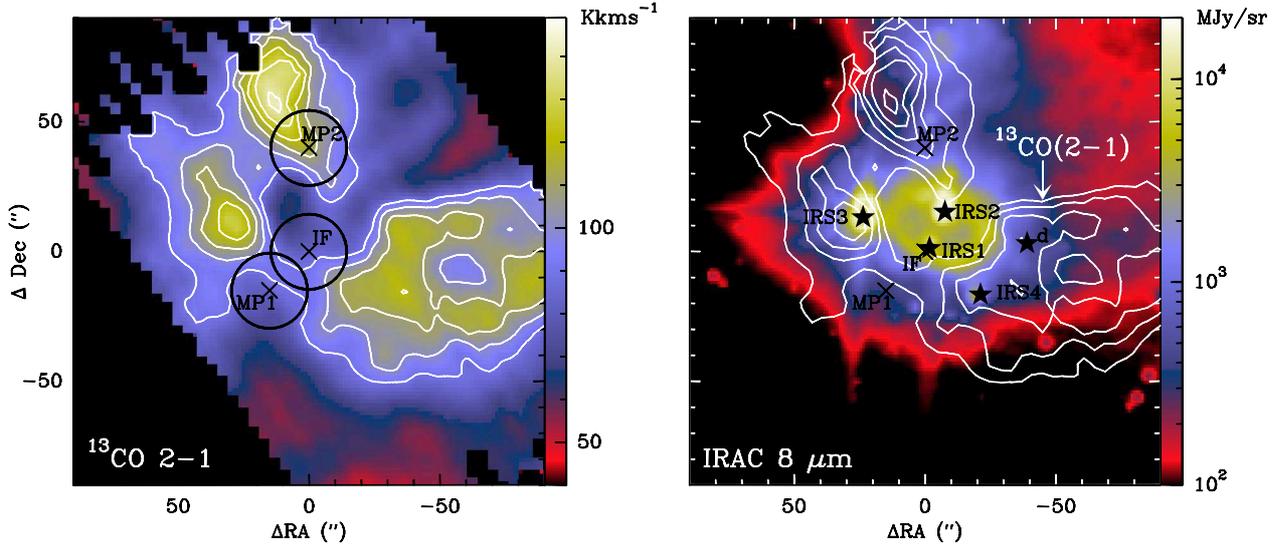}
     \caption{{\it Left:} Map of the integrated intensity between -5 and 15~km~s$^{-1}$ of 
the $^{13}$CO \textit{2$\rightarrow$1} line observed 
at the IRAM 30m telescope (lower contour at 90\,K\,km\,s$^{-1}$ and steps of 10\,K\,km\,s$^{-1}$, from Pilleri et al. 2012a). The three observed positions, IF, MP1 and MP2, are indicated. 
The beam of the 30m at 3mm towards the 3 observed positions is also drawn.
{\it Right:} IRAC 8~$\mu$m map towards Mon2. The extended emission is produced by the PAH bands at 7.7~$\mu$m. This figure shows 
the existence of an extended PDR in the 
cloud.}
\end{figure*}

\begin{table}
\caption{Summary of 30m observations.}
%{\scriptsize
\begin{tabular}{llccccc} \hline 
\multicolumn{1}{l}{Offset} &  \multicolumn{1}{c}{Freq} & \multicolumn{1}{c}{HPBW} &
\multicolumn{1}{c}{$\eta_{MB}$} & \multicolumn{1}{c}{rms$^1$} &  \multicolumn{1}{c}{t} \\ 
\multicolumn{1}{l}{} & \multicolumn{1}{c}{(GHz)} & \multicolumn{1}{c}{($\arcsec$)} &
\multicolumn{1}{c}{} &  \multicolumn{1}{c}{(mK)} &  \multicolumn{1}{c}{(min)}\\ 
\hline
IF    &  [83.746, 87.461] &  29 & 0.81  & 6 & 55 \\
(0$\arcsec$,0$\arcsec$)        &  [87.517, 91.232] &  29 & 0.80  & 4 & 46 \\
        &  [91.227, 94.942] &  29 & 0.80  & 4 & 46 \\
        &  [103.197, 106.913] & 24 & 0.80  & 3 & 46 \\
        &  [106.907, 110.622] & 24 & 0.80  & 4 & 46 \\
        &  [204.583, 208.294] & 12 &0.63 & 20 & 41 \\
        &  [207.184, 210.900] & 12 &0.63 & 13 & 74  \\
        &  [216.344.220.059] & 11 & 0.63  & 15 & 28  \\
%($-$15$\arcsec$.$-$15$\arcsec$)  &  [87.517.91.232] &  29$\arcsec$ & 0.80  & 0.003 \\
%        &  [91.227.94.942] &  29$\arcsec$ & 0.80  & 0.003 \\
%        &  [103.197.106.913] & 24$\arcsec$ & 0.80  & 0.003 \\
%        &  [106.907.110.622] & 24$\arcsec$ & 0.80  & 0.004 \\
MP1  & [83.746, 87.461] &  29 & 0.81  & 3 & 55 \\
(15$\arcsec$,$-$15$\arcsec$)        &  [87.517, 91.232] &  29 & 0.80  & 3 & 55 \\
        &  [91.227, 94.942] &  29 & 0.80  & 4 & 55 \\
        &  [103.197, 106.913] & 24 & 0.80  & 3 & 55 \\
        &  [106.907, 110.622] & 24 & 0.80  & 3 & 55 \\
        &  [204.583, 208.294] & 12 &0.63  & 20 & 41 \\
        &  [207.184, 210.900] & 12 &0.63  & 13 & 64 \\
        &  [216.344, 220.059] & 11 & 0.63  & 10 & 28  \\
MP2 & [83.746, 87.461] &  29 & 0.81  & 4 & 111 \\
(0$\arcsec$,$+$40$\arcsec$)        &  [87.517, 91.232] &  29 & 0.80  & 3 & 46 \\
        &  [91.227, 94.942] &  29 & 0.80   &  3 & 46 \\
        &  [103.197, 106.913] & 24 & 0.80  &  3 & 46  \\
        &  [106.907, 110.622] & 24 & 0.80  &  4 & 46 \\
        &  [204.583, 208.294] & 12 &0.63   &  20 & 27 \\
        &  [207.184, 210.900] & 12 &0.63   &  20 & 24 \\
        &  [216.344.220.059] & 11 & 0.63   &  10 & 55 \\
\hline
\end{tabular}

\noindent
$^1$ rms in units of T$_{MB}$ with WILMA (spectral resolution of 2~MHz).
%}
\end{table}
 
\section{Observations}

Figure 1 displays a summary of the morphology of Mon~R2 at both mm ($^{13}$CO \textit{2$\rightarrow$1}) and mid-IR wavelengths. As expected, the maximum of 
the emission at 8~$\mu$m is found towards the {\sc Hii} region and the infrared star IRS~3. Extended emission is detected around the
{\sc Hii} region that is associated with the PAH band at 7.7~$\mu$m. This emission does not distribute uniformly around
the {\sc Hii} region. It extends more than one arcminute away from IRS~1 towards the north-west whereas it declines very rapidly towards the south-east.
 
We performed a 3mm and 1mm spectral survey towards three selected positions that are expected to represent different
physical and chemical conditions. Our center position (RA=06h07m46.2s, DEC=-06$^\circ$23$\arcmin$08.3$\arcsec$ (J2000)) corresponds to the position 
of the ionization front (IF) and that of the ionizing star IRS~1.
This offset together with the ($+$15$\arcsec$, $-$15$\arcsec$) form a strip across the molecular bar with a sampling of $\approx$1 beam (see Fig. 1). 
For simplicity, we will refer to this offset as MP1 hereafter. The farther offset (0$\arcsec$,$+$40$\arcsec$) (hereafter MP2) is within the extended PAH 
emission towards the north-west.

The observations were carried out with the IRAM 30m telescope at Pico Veleta (Spain) during July 2009. To optimize the observing time, we
used the dual sideband offered by EMIR. This way we can simultaneously observe 8~GHz in the lower sideband and 8~GHz in the upper sideband, 
separated by a gap of 8~GHz. The rejection from one band to the other one has been measured by the IRAM
staff to be $\approx$13~dB. This observing mode allows to cover a band of
16~GHz at 3mm with only one tuning. Initially we observed 16~GHz single polarization (from 87.517 to 94.942~GHz and from 103.197 to 110.622 GHz)
using the dual sideband of the 3mm receiver.
Afterwards we completed our survey by observing simultaneously the 3mm and 1mm bands (4~GHz at 3mm and 4~GHz at 1mm) and integrating
the two polarizations. The 3mm and 1mm EMIR receivers were centered at 85.339~GHz and 217.937~GHz respectively. In the 1mm receiver, we need to
drop one of the polarizations due to the poorer quality of the baselines.  
Two additional settings were used to observe the SO$^+$~\textit{9/2$\rightarrow$7/2} (207.800~GHz) and  SO~\textit{5$_4$$\rightarrow$3$_2$} (206.176~GHz) lines. 
Simultaneously with the observations of these lines at 1mm, we used the 3mm receiver to obtain the high spectral resolution spectra of 
the HCO$^+$ \textit{1$\rightarrow$0}, HCN \textit{1$\rightarrow$0}, C$_2$H \textit{1$\rightarrow$0} and C$_4$H \textit{11$\rightarrow$10} lines shown in Fig.~2.

The observed spectral ranges, telescope beam efficiencies, half power beam width (HPBW) and rms of our observations are
listed in Table~1. The observing procedure was position switching with
the reference located at an offset ($+$400$\arcsec$, $-$400$\arcsec$). To cover these spectral
ranges we used the WILMA autocorrelator which provides a spectral
resolution of 2~MHz (5.4$-$6.8~km~s$^{-1}$ in the 3mm band). This spectral resolution does not allow to resolve the lines at
3mm and therefore it gives information only on the integrated line intensity. The WILMA spectra and the line integrated intensities are shown in Appendix A.
Selected lines were observed simultaneously with the VESPA autocorrelator,
providing a spectral resolution of around 40~kHz. These high spectral resolution
spectra are shown in Fig.~2 and Gaussian fit parameters to SO, SO$_2$, SO$^+$ and C$_4$H lines are given in Table~3. All the intensities are 
in unit of main beam brightness temperature (T$_{MB}$).

\begin{figure*}
\begin{center}
\includegraphics{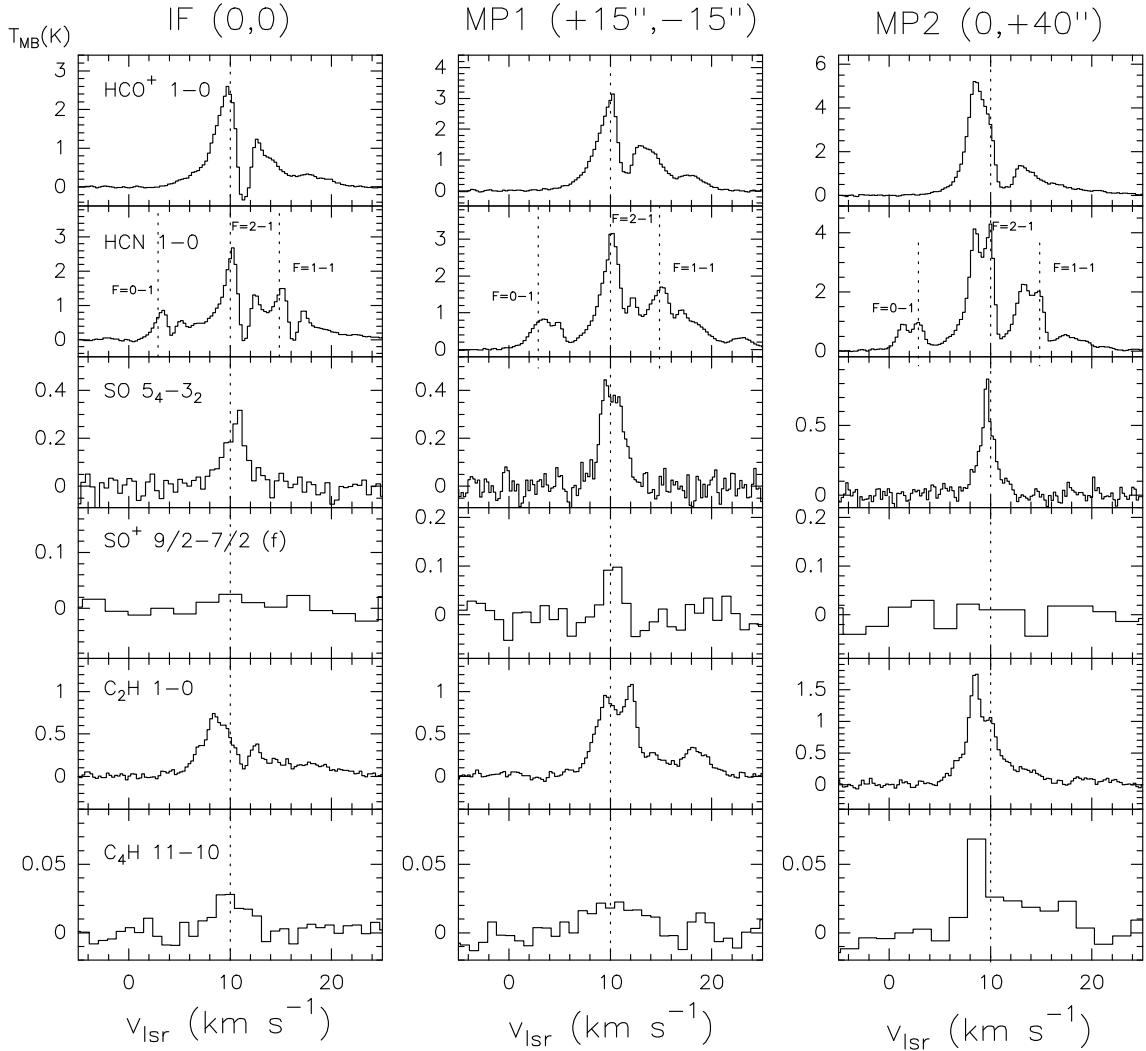}
      \caption{High resolution spectra observed with the 30m telescope towards the positions (0$\arcsec$,0$\arcsec$), (+15$\arcsec$, -15$\arcsec$) and (0$\arcsec$,+40$\arcsec$).
All the lines were observed with a spectral resolution of $\sim$40~kHz. The spectra of the SO$^+$ \textit{9/2$\rightarrow$7/2} and
C$_4$H \textit{11$\rightarrow$10} lines have been rebinned to obtain a higher signal to noise ratio.}
         \end{center}
 \end{figure*}

\section{Results}
\subsection{Line identification}
To identify the lines, we have used three different catalogues, the JPL (http://spec.jpl.nasa.gov), the CDMS (http://www.astro.uni-koeln.de/cdms/catalog) and  that
from J. Cernicharo  (private communication)
The observations were performed with 2SB receivers with an image band rejection of $\sim$13dB. Therefore, we have some bright lines from the image band in our spectra (see Fig. A1, A2, A3). 
%{\bf In the 3mm setting, we have the observations of the signal and image band separately, which allows to avoid fortuitous identifications.}
In Appendix A, we show the spectra of our complete mm survey, and give a list of the detected
lines ($\geq$3-$\sigma$), their identification and line integrated intensities (Table~ A1). 
We also list the $\geq$4-$\sigma$ unidentifed lines and the corresponding frequencies in the signal and image bands.
For clarity, upper limits are not included in Table~ A1. Since the molecular lines are unresolved, the upper limit to the integrated intensity is simply
$3 \times rms \times channel~ velocity~ width$. The Gaussian fit parameters to the recombination lines are given in
Table A2.

We have identified 87 lines, out of a total of 105 lines, towards the (0$\arcsec$,0$\arcsec$) position. The large number of recombination 
lines shows the presence of ionized gas.
In addition, we have also a wealth of lines of complex molecules, more typical of warm and dense molecular
clouds. Towards the MP1, we have detected 101 lines, and identified all but 5 lines. At this position the recombination lines
are very weak. Towards the MP2 we have detected a total of 99 lines with only 8 unidentified lines.
We have tentatively detected the H$_2$CN \textit{3$_{0,2}$$\rightarrow$2$_{0,2}$} line towards the MP1 and MP2. H$_2$CN was firstly detected in TMC1 by Ohishi et al. (1994). If confirmed,
this region would be the second one in which this radical has been detected and would corroborate that  Mon~R2 is a source with
an extremely rich carbon chemistry.

In Table~2, we show a summary of the molecular species detected in our survey towards the three positions. 
We distinguish between certain and tentative detections. Tentative detections
correspond to species with only one line detected and at a level of 3$-$4-$\sigma$. In these cases, we checked that there were not
any other intense (and undetected) line of the same species lying in the spectral ranges covered by our survey. 
In Table~2  we added  the reactive ions CO$^+$ and HOC$^+$ previously
detected by Rizzo et al. (2003) because of their importance to interpret the PDR chemistry. Rizzo et al. (2003) did not detect these ions 
towards the MP1 and they did not search for them 
in the MP2. Like CO$^+$ and HOC$^+$, the reactive ion SO$^+$ is expected to present very low fractional abundance in the shielded
part of the cloud. In this work, SO$^+$ has been searched in the three positions but only detected towards the MP. Many molecules have
been detected in the three positions.
Some are well-known tracers of PDRs like HCN, HCO, C$_2$H, c-C$_3$H$_2$ and C$_4$H. Others, like CH$_3$CN and CH$_3$OH, are
complex molecules usually not found in PDRs. At the end of Table~2,
we show those species with certain detections only towards the MP1 and the MP2 peaks, such as C$_4$H, HCO, SiO, C$_2$D and SO$_2$.
The sulphuratted carbon-chain C$_2$S has been detected in the MP2 and tentatively in the IF. HNCO has only been detected in the MP.

The detection of recombination lines, PDR tracers and complex molecules is consistent with the interpretation of the 
PDR as an expanding envelope around the UC {\sc Hii} region firstly proposed by Rizzo et al. (2005) and recently confirmed by Fuente et al. (2010)
and  Pilleri et al. (2012a). Even in this situation, 
we expect to see some chemical differences because of the different incident UV fields and mixing ratios of the different
PDR layers.

\begin{table}
\caption{Detected molecules.}
\begin{center}
\begin{tabular}{p{1cm}p{0.7cm}|p{1cm}p{1cm}|p{1cm}p{1cm}} \hline 
\multicolumn{2}{c|}{IF (0$\arcsec$,0$\arcsec$)} &  \multicolumn{2}{c|}{MP1 ($+$15$\arcsec$,$-$15$\arcsec$)} & 
\multicolumn{2}{c}{MP2 (0$\arcsec$,$+$40$\arcsec$)} \\ 
\multicolumn{1}{c}{Detected} & \multicolumn{1}{c|}{Tentative} &
\multicolumn{1}{c}{Det.} & \multicolumn{1}{c|}{Tent.} &
\multicolumn{1}{c}{Det.} & \multicolumn{1}{c}{Tent.} \\
\hline
CO$^+$   & & & & & \\
HOC$^+$                                                    &                &                                               &              &                    \\
                                                    &                 &  SO$^+$   &                                             &               &                \\
$^{13}$CO       &                                         & $^{13}$CO   &                     &  $^{13}$CO      & \\  
C$^{18}$O      &                                            & C$^{18}$O   &                 &  C$^{18}$O        &\\
SO                    &                                             & SO                  &                                         &  SO &          \\
      &                                             & $^{13}$CN     &                                         &  $^{13}$CN   &       \\
                                    &                                   &  C$^{15}$N       &                                      &  C$^{15}$N    &     \\
$^{13}$CS     &                                    & $^{13}$CS                   &                           &  $^{13}$CS          &  \\
N$_2$H$^+$    &                                           & N$_2$H$^+$      &                                   &  N$_2$H$^+$ &          \\
%    &          &             &    NH$_2$D     &      &       \\
HCS$^+$     &                                                 & HCS$^+$              &                                   &  HCS$^+$        &    \\
HNC       &                                                        & HNC                        &                                 &  HNC              &  \\
HCN      &                                                         & HCN                          &                               &  HCN               & \\
H$^{13}$CN  &                                             & H$^{13}$CN              &                             &  H$^{13}$CN    &    \\
HN$^{13}$C  &                                               & HN$^{13}$C                &                           &  HN$^{13}$C      &   \\
HC$^{15}$N   &                                              & HC$^{15}$N                  &                         &  HC$^{15}$N       &  \\
DCN       &                                                       &       DCN               &                                       &    DCN  &                         \\
HCO$^+$      &                                               & HCO$^+$              &                                  &  HCO$^+$  &           \\
H$^{13}$CO$^+$  &                                       & H$^{13}$CO$^+$  &                                &  H$^{13}$CO$^+$  &        \\
C$_2$H       &                                                  &  C$_2$H                     &                          &  C$_2$H           &   \\
c-C$_3$H$_2$     &                                       & c-C$_3$H$_2$           &                           &  c-C$_3$H$_2$ &         \\
         
H$_2$CO     &                                           & H$_2$CO                   &                         &  H$_2$CO
        &     \\ 
HC$_3$N       &                                             & HC$_3$N                        &                        &  HC$_3$N             &  \\
H$_2$CS       &                                             & H$_2$CS                          &                       &  H$_2$CS             &  \\
CH$_3$CN       &                                          & CH$_3$CN                        &                     &  CH$_3$CN            &  \\
CH$_3$OH      &                                         &   CH$_3$OH    &                      &    CH$_3$OH                &               \\
CH$_3$C$_2$H    &                                   & CH$_3$C$_2$H                   &                 &    CH$_3$C$_2$H    &               \\
                 &  C$_4$H                          & C$_4$H                                      &          &  C$_4$H        & \\                     
                &  C$_2$S                           & HCO                            &                 &     HCO            &                \\
               &                                    &  SiO                                            &            &  SiO                  &  \\ 
               &                                    &   C$_2$D                                         &      &      C$_2$D        &          \\
               &                                    &  SO$_2$                                          &  & SO$_2$ &                    \\
               &                                    &  HNCO                                            &   &  C$_2$S                \\
               &                                    &          &H$^{15}$NC                             &                          & H$_2$CN        \\

                & &  & H$_2$CN                                 &       &                                 \\
    &          &             &    NH$_2$D     &      &       \\
 \hline
\end{tabular}
\end{center}
%}
\end{table}

\subsection{ High resolution observations: Detection of SO$^+$ and C$_4$H }
In Fig. 2 we show the spectra of the HCO$^+$ \textit{1$\rightarrow$0}, HCN \textit{1$\rightarrow$0},
SO \textit{5$_4$$\rightarrow$3$_2$}, SO$^+$ \textit{9/2$\rightarrow$7/2}, C$_2$H \textit{1$\rightarrow$0}, and C$_4$H \textit{11$\rightarrow$10} lines.
This is the first detection of SO$^+$ and C$_4$H towards Mon~R2.
These lines were observed with the $\sim$40 kHz spectral resolution provided by the VESPA autorrelator.
Gaussian fits to the SO, SO$^+$ and C$_4$H lines
are shown in Table~3.

\begin{table*}
\caption{SO, SO$_2$, SO$^+$,C$_4$H observations.}
\begin{tabular}{p{2.2cm} p{2.2cm} p{1.5cm} p{1.8cm} p{2cm} p{2cm} p{1.8cm} p{1.0cm} p{0cm}} \hline 
\multicolumn{2}{l}{Line} &  \multicolumn{1}{c}{HPBW($\arcsec$)} & \multicolumn{1}{c}{Freq(GHz)} &
\multicolumn{1}{c}{I(K km~s$^{-1}$)} & \multicolumn{1}{c}{v$_{lsr}$(km~s$^{-1}$)} & 
\multicolumn{1}{c}{$\Delta$v(km~s$^{-1}$)} & \multicolumn{1}{c}{T$_{MB}$(K)} \\ 
\hline
\multicolumn{8}{l}{Offset (0$\arcsec$,0$\arcsec$)} \\
SO        &  \textit{5$_4$$\rightarrow$4$_3$}  &   \centering{12}   &  \centering{206.176} & \centering{0.70(0.05)} & \centering{10.56(0.08)} &  \centering{2.55(0.27)} & \centering{0.26} & \\
SO$^+$    &  \textit{9/2$\rightarrow$7/2} (f)  &  \centering{12} &  \centering{208.965}  &  \centering{0.11(0.05)$^t$} & \centering{10.09(1.00)} &  \centering{3.87(1.72)} & \centering{0.026} & \\
%$<$ 0.10 K km/s 2xsigma detection
SO$^+$    &  \textit{9/2$\rightarrow$7/2} (e)         &  \centering{12} &  \centering{208.590} & \centering{$<$ 0.16} & & & & \\
SO$_2$    &  \textit{3$_{2,2}$$\rightarrow$2$_{1,1}$} &  \centering{12} &  \centering{208.700} & \centering{$<$ 0.16} & & & & \\
C$_4$H    &  \textit{11$\rightarrow$10}  &  \centering{24} &  \centering{104.666} &  \centering{0.09(0.02)} &  \centering{9.77(0.46)}  &  \centering{3.56(1.11)}  & \centering{0.025}  & \\
C$_4$H    &  \textit{11$\rightarrow$10}  &  \centering{24} &  \centering{104.705} &  \centering{0.11(0.02)} &  \centering{10.17(0.39)} &  \centering{3.77(0.80)} & \centering{0.028} & \\
\multicolumn{8}{l}{Offset ($+$15$\arcsec$,$-$15$\arcsec$)} & \\
SO        &  \textit{5$_4$$\rightarrow$4$_3$}  & \centering{12} &  \centering{206.176} &  \centering{1.16(0.04)}  & \centering{10.13(0.04)}  &  \centering{2.59(0.09)} & \centering{0.42} & \\
SO$^+$    &  \textit{9/2$\rightarrow$7/2} (f)  & \centering{12} &  \centering{208.965} &  \centering{0.26(0.03)}  & \centering{10.17(0.12)}  &  \centering{2.06(0.21)} & \centering{0.12} & \\
SO$^+$    &  \textit{9/2$\rightarrow$7/2} (e)  & \centering{12} &  \centering{208.590} &  \centering{0.13(0.04)}  & \centering{10.20(0.21)}  &  \centering{1.47(0.53)} & \centering{0.08} & \\
SO$_2$    &  \textit{3$_{2,2}$$\rightarrow$2$_{1,1}$} &  \centering{12} &  \centering{208.700} &  \centering{$<$ 0.07} & & & & \\
C$_4$H    &  \textit{11$\rightarrow$10}  &  \centering{24} &  \centering{104.666} &  \centering{0.09(0.03)} &  \centering{8.70(0.81)}  &  \centering{5.14(1.64)} & \centering{0.017}  & \\
C$_4$H    &  \textit{11$\rightarrow$10}  &  \centering{24} &  \centering{104.705} &  \centering{0.14(0.03)} &  \centering{9.78(0.63)}  &  \centering{5.90(1.17)} & \centering{0.022} & \\
\multicolumn{8}{l}{Offset (0$\arcsec$,$+$40$\arcsec$)} & \\
SO        &  \textit{5$_4$$\rightarrow$4$_3$}  & \centering{12} &  \centering{206.176} & \centering{1.19(0.03)}  &  \centering{9.65(0.02)} & \centering{1.70(0.06)} & \centering{0.66}  & \\
SO$^+$    &  \textit{9/2$\rightarrow$7/2} (f)  & \centering{12} &  \centering{208.965} &  \centering{$<$ 0.17} & & & & \\
SO$^+$    &  \textit{9/2$\rightarrow$7/2} (e)  & \centering{12} &  \centering{208.590} &  \centering{$<$ 0.17} & & & & \\
SO$_2$    &  \textit{3$_{2,2}$$\rightarrow$2$_{1,1}$} &  \centering{12} &  \centering{208.700} & \centering{$<$ 0.22} & & & & \\
C$_4$H    &  \textit{11$\rightarrow$10}  &  \centering{24} &  \centering{104.666} &  \centering{0.14(0.02)} & \centering{6.53(0.07)}  &  \centering{1.11(0.16)} & \centering{0.12}  & \\
C$_4$H    &  \textit{11$\rightarrow$10}  &  \centering{24} &  \centering{104.705} &  \centering{0.09(0.02)} & \centering{7.99(0.09)}  &  \centering{0.90(0.18)} & \centering{0.10} & \\
%C$_4$H    &  12$\rightarrow$11  &  24$\arcsec$ &  114.181d &  0.18(0.02) & 7.12(0.10)  &  1.34(0.28) & 0.12 & \\
%C$_4$H    &  12$\rightarrow$11  &  24$\arcsec$ &  114.221 &  0.14(0.02) & 8.182(0.05) &  0.92(0.12) & 0.15 & \\
\hline
\end{tabular}

\noindent
$^t$ Tentative, $<$3-$\sigma$
\end{table*}

There are significant differences among the profiles of the lines shown in Fig.~2 which testify the existence of
several gas components.
A clear absorption is seen in the HCO$^+$ \textit{1$\rightarrow$0}, HCN \textit{1$\rightarrow$0} and C$_2$H \textit{1$\rightarrow$0}
lines towards the IF position. This self-absorbed feature lies at a velocity of $\sim$10.8~km~s$^{-1}$ which is the central
velocity of the SO \textit{5$_4$$\rightarrow$4$_3$} line emission. This suggests that the SO \textit{5$_4$$\rightarrow$4$_3$} emission is 
coming from the cloud around the {\sc Hii} region.
Towards the MP1 the profiles of the SO \textit{5$_4$$\rightarrow$4$_3$} and SO$^+$ \textit{9/2$\rightarrow$7/2} lines 
are wider and centered at $\sim$10.0~km~s$^{-1}$.
The line profiles towards the MP2 are even more complicated (see Fig. 2). Similarly to what happens
in the other positions, there is a self-absorbed feature at $\sim$10.8~km~s$^{-1}$ which is the velocity of the external cloud, but the HCN 1$\rightarrow$0 line also
presents self-absorption at $\sim$9~km~s$^{-1}$.
The C$_2$H \textit{1$\rightarrow$0} and C$_4$H \textit{11$\rightarrow$10} lines show wide profiles, with a red-shifted wing up to
velocities of  $\sim$18~km~s$^{-1}$. The similarity between the profiles of the C$_2$H  \textit{1$\rightarrow$0} and C$_4$H \textit{11$\rightarrow$10} lines 
supports the interpretation of these two
species share a similar chemistry and come from the same region. Since C$_4$H is found to be abundant in low UV PDRs and has not been detected in the high velocity 
wings of bipolar outflows (see e.g. Bachiller \& P\'erez-Guti\'errez 1997), 
these high velocity wings are very likely tracing the PDR
around the {\sc Hii} region (Fuente et al. 2010, Pilleri et al. 2012a).
The existence of several bipolar outflows in this regions makes difficult, however, to disentangle the origin of this high velocity gas.
The profile of the SO \textit{5$_4$$\rightarrow$4$_3$} line is, however, Gaussian and centered at the systemic velocity 
suggesting that its emission originates in the dense cloud.

\begin{table}
\caption{Collisional coefficients.} % title of Table
\begin{tabular}{ll} 
\hline \hline
C$^{18}$O  & Yang et al. 2010 \\
SO  &  Lique et al. 2006 \\
CS &  Lique et al. 2007 \\
SiO  & Dayou \& Balanca, 2006\\
HCN, HNC  & Sarrasin et al. 2010 \\
HCO$^+$ &  Flower, D.R. 1999 \\
HCS$^+$ & Monteiro 1984 \\
N$_2$H$^+$ & Daniel et al. 2005 \\
SO$_2$ &  Cernicharo et al. 2011 \\
H$_2$CO & Green et al. 1991 \\
c-C$_3$H$_2$ & Avery \& Green 1989 \\
HC$_3$N & Wernli et al. 2007 \\
 \hline \hline
\end{tabular}
\end{table}

\section{Gas Physical Conditions and column densities}
To derive accurate column densities it is important to have a good estimate of the
gas physical conditions. For the species C$^{18}$O, SO, $^{13}$CN, SiO, C$_2$S, SO$_2$, H$_2$CO, H$_2$CS, c-C$_3$H$_2$, HC$_3$N, 
C$_4$H, CH$_3$OH, CH$_3$CN and CH$_3$C$_2$H
we detected several transitions and a multitransitional study was possible.
We used the rotational diagram technique to derive beam averaged rotation temperatures and column densities.
This technique is based on the assumption of optically thin emission for all the lines of a given species. For
optically thin emission, the integrated intensity of each line is proportional to
the population of the upper level of the corresponding transition. In this case, 
a rotational diagram provides good estimates of the rotation temperature and total column density
(see e.g. Schloerb et al. 1983). In the case of optically thick
lines, the integrated intensity of the line is not proportional to the population of the upper level anymore and the rotation temperature cannot be
derived with this technique. 

The lines of the most abundant isotopologues are optically thick and in many cases the lines are self-absorbed (see Fig. 2).
For this reason, we only made rotational diagrams for the rarer isotopologues C$^{18}$O, H$^{13}$CO$^+$ and H$^{13}$CN. 
The column densities of CO, HCO$^+$ and HCN were derived assuming the isotopic
ratios $^{12}$C/$^{13}$C=50 (Savage et al. 2002 and references therein) and $^{16}$O/$^{18}$O=500.
The information of our survey was completed with the H$^{13}$CO$^+$ \textit{3$\rightarrow$2}, H$^{13}$CN \textit{3$\rightarrow$2}
and C$_2$H \textit{3$\rightarrow$2} lines obtained from our unpublished HERA maps (Pilleri et al. 2012b, in preparation)
after degrading their angular resolution to that of the observed 3mm transitions. The resulting integrated line intensities
are shown at the end of Tables A1. In these cases, the rotational temperatures and column densites derived from the rotational diagrams are not 
affected by the unknown beam filling factor.  For the other cases,
we assumed a beam filling factor of unity for the 3mm and 1mm lines. 
This is reasonable taking into account 
that the molecular emission extends over a  region of about 2$\arcmin$$\times$2$\arcmin$ (see Fig. 1), which is larger than the beam of the 30m telescope 
at both 3mm and 1mm wavelengths but it would overestimate the rotational temperature if the size of the emission is smaller than 29$"$. Comparing the rotational 
temperatures thus derived with those from  H$^{13}$CO$^+$, H$^{13}$CN 
and C$_2$H lines  we estimate that the uncertainty in the column density estimates because of the unknown filling factor is of a factor of 3.  
%The spatial distribution of some species could have a size smaller than  $\sim$29$\arcsec$, the 3mm beam. In these cases, we would 
%overestimate the rotational temperature and the column density estimate would be an averaged values in 
%the 3mm beam rather than the real column density.

%We did not find significant differences in the rotational temperatures among the observed
%positions. However, the rotational temperature of the different species varies towards the same position. 
%The molecules, C$^{18}$O, H$_2$CO, HC$_3$N, SO, SO$_2$, CH$_3$CN  and CH$_3$C$_2$H
%have systematically higher rotation temperatures ($\sim$30$-$50~K) than the other species ($\sim$10~K). 
%{\bf This is easily understood for C$^{18}$O since the observed lines thermalize at densities $<$10$^5$~cm$^{-3}$
%and their rotation temperature only depends on the gas kinetic temperature. 
%CH$_3$CN is a symmetric top and hence transitions of differing K ladder
%for the same rotational transition have relative populations
%which are close to the Boltzmann value.} However, the
%H$_2$CO, HC$_3$N, SO, and SO$_2$ lines require higher densities to be thermalized and the high rotation temperatures
%prove that they arise from dense and warm gas. 

Where possible we estimated the molecular column densitites and the molecular hydrogen density using the large velocity gradient radiative transfer code by 
Cernicharo et al. (2006). The collisional coefficients adopted in our calculations are shown in Table~4. For $^{13}$CS, H$^{13}$CO$^+$, H$^{13}$CN, HC$^{15}$N, HN$^{13}$C and DCN we used
the collisional coefficients of the main isotopologue.
We fixed the gas kinetic temperature and varied the density and column density to fit the line intensities.
Based on the rotation temperatures of C$^{18}$O and CH$_3$CN data, we assumed T$_k$=50~K for the IF and MP1 positions and T$_k$=70~K for the 
MP2. Since the kinetic temperature is fixed in our calculations, the ratio between the intensities of the two
lines observed is mainly dependent on the density while the intensity of each line is more dependent on the total column density. 
The obtained densities vary between $\sim$ a few 10$^5$~cm$^{-3}$ to $\sim$10$^6$~cm$^{-3}$.
Our results are in agreement with Tafalla et al. (1997) who derived densities of a few 10$^5$~cm$^{-3}$
from the CS lines and its isotopes. For some species we have detected only one line.
In these cases we adopted a density n(H$_2$)=5$\times$10$^5$~cm$^{-3}$ in our LVG calculations for the three positions.

In Table~5 we show the column densities derived towards the three observed positions for all the detected species. The corresponding rotational 
diagrams are shown in appendix B (see Fig. B1, B2, B3). The column densities obtained using the 
LVG approach are in agreement with the rotational diagrams calculations which confirms our assumption that
the emission is optically thin.
For those species for which the collisional coefficients were not known and only one transition was observed, 
we estimated the column densities assuming optically thin emission and Local Thermodynamic Equilibrium with the level$'$s population
described by a given rotation temperature. The assumed rotation temperatures are based on the results of the rotational diagrams of other molecules with
similar excitation conditions. For SO$^+$, $^{13}$CN, C$_2$D, C$_2$S, HCO, HNCO, and C$_4$H, we assumed T$_{rot}$=10~K, similar to those obtained using
rotational diagrams for C$_2$H and H$^{13}$CO$^+$. For CH$_3$CN and CH$_3$C$_2$H, we adopted T$_{rot}$=30~K.
%Taking into account that in most cases we observed the ground state transitions and the emission is optically thin, we consider that the uncertainty in the 
%estimated column densities is of a factor of 3. 
Some cases are marked with the label $``$bad estimate" in Table~5 because the uncertainty  in the column density estimates could be as high as a factor of 10. This is the case 
of CH$_3$CN and CH$_3$OH. The main source of uncertainty in these molecules is  that our spectral resolution does not allow to resolve all the K-components of the same rotational 
line. The lack of detection of the CH$_3$CN 5$_0$$\rightarrow$4$_0$ line towards the IF suggests that the emission of this transition could be self-absorbed. 

Molecular abundances were calculated assuming a canonical C$^{18}$O wrt H$_2$ abundance 
of 1.7$\times$10$^{-7}$. The abundance of each species $X$ wrt H$_2$ is then given 
by the expression $f_{X}=N_X/N_{H_2}= N_X/N_{C^{18}O}\times1.7~10^{-7}$.

\begin{table*}
\caption{Physical parameters.} % title of Table
%\centering % used for centering table
\begin{tabular}{p{1.5cm} p{1.0cm} p{2.0cm} p{1.5cm} p{1.0cm} p{1.5cm}  p{1.5cm} p{1.5cm}  p{2.5cm}} % centered columns (4 columns)
\hline \hline
\multicolumn{1}{c}{}         & \multicolumn{2}{c|}{RD$^a$}    & \multicolumn{2}{c|}{LVG$^b$} &  \multicolumn{2}{c}{LTE$^c$} &  f$_X$$^d$ &   \\
\multicolumn{1}{c}{Molecule} & \multicolumn{1}{c}{T$_{rot}$}  & \multicolumn{1}{c|}{N}       &  \multicolumn{1}{c}{n}   &  \multicolumn{1}{c|}{N}  &  
\multicolumn{1}{c}{T$_{rot}$$^e$}  &   \multicolumn{1}{c}{N}                   &            &            \\ % table heading
         & \multicolumn{1}{c}{(K)} &  \multicolumn{1}{c|}{(cm$^{-2}$)}  &\multicolumn{1}{c}{(cm$^{-3}$)}  &   \multicolumn{1}{c|}{(cm$^{-2}$)}   &  
\multicolumn{1}{c}{(K)}       &  \multicolumn{1}{c}{(cm$^{-2}$)}  &            &            \\ % table heading
\hline % inserts single horizontal line
\hline
\\
\multicolumn{9}{c}{{\large IF (0$\arcsec$,0$\arcsec$)}} \\
\\
%$^{13}$CO  & 15(1) & 4.8(0.1)$\times$10$^{16}$ & ...  \\
C$^{18}$O  & 40(5)    & 7.3(1.0)$\times$10$^{15}$ & $>$3$\times$10$^4$  &  9.0$\times$10$^{15}$           &                 &                        &      1.7$\times$10$^{-7}$$^e$ &  \\
SO         &  100(600)  & 2.8(10)$\times$10$^{13}$ &     &            &                 &                        &      6.5$\times$10$^{-10}$    &
  Bad estimate \\
SO$^+$     &          &                           &                     &             &    \multicolumn{1}{c}{10}           &  $<$1.0$\times$10$^{12}$  &      $<$2.3$\times$10$^{-11}$    &  \\  
$^{13}$CN  &       &                              &                     &             &    \multicolumn{1}{c}{10}           & $<$1.2$\times$10$^{13}$ &     $<$2.8$\times$10$^{-10}$  &  \\
$^{13}$CS  &       &                              & 5$\times$10$^5$$^f$ & 1.7$\times$10$^{12}$    &                    &                             &      3.9$\times$10$^{-11}$ &  Col CS \\
SiO        &       &                              & 5$\times$10$^5$$^f$ & $<$3.0$\times$10$^{11}$ &       &                       &    $<$7.0$\times$10$^{-12}$ & \\
HNC        &       &                              & 5$\times$10$^5$$^f$ &  4.2$\times$10$^{12}$    &       &                       &      9.8$\times$10$^{-11}$ & \\
H$^{13}$CN & 9(1)  & 3.0(1.3)$\times$10$^{12}$  & 1$\times$10$^6$ &   3.0$\times$10$^{12}$ &       &                       &      7.0$\times$10$^{-11}$ &  Col HCN \\
HN$^{13}$C &       &                              & 5$\times$10$^5$$^f$ & 4.0$\times$10$^{11}$    &       &        &      9.3$\times$10$^{-12}$ & Col  HNC \\ 
HC$^{15}$N &       &                              & 5$\times$10$^5$$^f$ & 4.0$\times$10$^{11}$    &       &        &      9.3$\times$10$^{-12}$ & Col  HCN\\ 
DCN        &       &                              &  5$\times$10$^5$$^f$         &  3.0$\times$10$^{12}$         &              &   &     7.0$\times$10$^{-11}$ & Col  HCN \\
C$_2$H     & 16(1) & 3.7(0.1)$\times$10$^{14}$   &                     &                         &       &                       &      8.6$\times$10$^{-9}$ & self-absorption\\
C$_2$D     &       &                              &                     &                         &  \multicolumn{1}{c}{10}   & $<$7.5$\times$10$^{12}$ &  $<$1.7$\times$10$^{-10}$ & \\
H$^{13}$CO$^+$ & 22(5)     & 8.7(4.7)$\times$10$^{11}$ & 1$\times$10$^6$ & 1$\times$10$^{12}$ &       &                       &     2.0$\times$10$^{-11}$ & \\
HCS$^+$        &           &                           & 5$\times$10$^5$$^f$           &     8.0$\times$10$^{12}$      &    &   &     1.9$\times$10$^{-10}$ & \\
N$_2$H$^+$     &           &                        & 5$\times$10$^5$$^f$          &     2.1$\times$10$^{12}$ &     &                       &     4.9$\times$10$^{-11}$ & \\
C$_2$S         &           &                        &                              &                                & \multicolumn{1}{c}{10}  & 2.0$\times$10$^{12}$  &     4.6$\times$10$^{-11}$ & 
Tentative\\ 
HCO            &           &                           &                     &                      & \multicolumn{1}{c}{10}  & $<$3.0$\times$10$^{12}$ &   $<$7.0$\times$10$^{-11}$ & \\
SO$_2$         &           &                           & 5$\times$10$^5$$^f$   &  $<$4.3$\times$10$^{12}$     &    &  &  $<$1.0$\times$10$^{-10}$ & \\
H$_2$CO        & 65(13)    & 6.7(4.3)$\times$10$^{13}$ & 5$\times$10$^5$ &   5.6$\times$10$^{13}$     &       &                       &     1.5$\times$10$^{-9}$ &  ortho/para=3 \\
H$_2$CS        & 33(11)    & 5.6(7.5)$\times$10$^{12}$ &                       &                  &       &                       &     1.3$\times$10$^{-10}$  & ortho/para=3\\
HNCO           &           &                           &                       &                  & \multicolumn{1}{c}{10}    & $<$8.0$\times$10$^{11}$ &   $<$1.8$\times$10$^{-11}$ & \\
c-C$_3$H$_2$   &  11(2)    & 7.1(7.0)$\times$10$^{12}$ &  7$\times$10$^5$      &  8.5$\times$10$^{12}$             &       &             &   2.0$\times$10$^{-10}$ & 
ortho/para=3,self-absorption\\
HC$_3$N        &  33(2)    & 3.4(0.5)$\times$10$^{12}$ &  7$\times$10$^5$      &  3.4$\times$10$^{12}$             &       &                         &   7.9$\times$10$^{-11}$ & \\
C$_4$H         &           &                           &                       &                  & \multicolumn{1}{c}{10}    & 2.7$\times$10$^{13}$    &   6.3$\times$10$^{-10}$ & \\
CH$_3$OH       &  9(3)     & 6.4(10)$\times$10$^{14}$  &                 &                  &       &                               &   1.5$\times$10$^{-8}$ & Bad estimate\\
CH$_3$CN       &           &                           &                       &                  & \multicolumn{1}{c}{30}    & 1.6$\times$10$^{12}$    &   3.7$\times$10$^{-11}$ & Bad estimate  \\
CH$_3$C$_2$H   &  27(6)    & 8.6(8.6)$\times$10$^{13}$ &                 &                  &       &                               &   2.0$\times$10$^{-9}$ &  Bad estimate \\  
\\
\multicolumn{9}{c}{{\large  MP1 ($+$15$\arcsec$,$-$15$\arcsec$)}} \\
\\
%$^{13}$CO &  12(1)   & 4.5(0.3)$\times$10$^{16}$ & ...  \\ % inserting body of the table
C$^{18}$O &  38(2)   & 9.7(0.5)$\times$10$^{15}$ & $>$3$\times$10$^4$  &  9.7$\times$10$^{15}$    &       &                        &   1.7$\times$10$^{-7}$$^d$ & \\
SO        &  54(15)   & 3.0(1.2)$\times$10$^{13}$ & 2$\times$10$^6$     & 3.2$\times$10$^{13}$                       &       &                        &   5.2$\times$10$^{-10}$   & \\
SO$^+$    &          &                           &                     &                          & \multicolumn{1}{c}{10}    & 3.0$\times$10$^{12}$   &   5.2$\times$10$^{-11}$   & \\
$^{13}$CN &  3(2)    & 1.4(3.5)$\times$10$^{15}$ &                     &                          &  \multicolumn{1}{c}{10}   &  9.0$\times$10$^{12}$  &   1.6$\times$10$^{-10}$   & Bad estimate\\
$^{13}$CS &          &                           & 5$\times$10$^5$$^f$ & 2.5$\times$10$^{12}$     &       &                        &   4.4$\times$10$^{-11}$   & Col CS\\
SiO       &  11(2)   & 6.0(6.0)$\times$10$^{11}$ & 8$\times$10$^5$     & 6.0$\times$10$^{11}$     &       &                        &   1.0$\times$10$^{-11}$   & \\
HNC        &       &                             & 5$\times$10$^5$$^f$ & 1.5$\times$10$^{13}$     &       &                        &   2.6$\times$10$^{-10}$  & \\
H$^{13}$CN &  8(1)   & 3.7(1.6)$\times$10$^{12}$ & 7$\times$10$^5$ &   3.0$\times$10$^{12}$       &       &                        &   6.5$\times$10$^{-11}$ & Col HCN \\
HN$^{13}$C &       &                             & 5$\times$10$^5$$^f$ & 8.0$\times$10$^{11}$  &     &    &   1.4$\times$10$^{-11}$ & Col HNC\\
HC$^{15}$N &       &                             & 5$\times$10$^5$$^f$ & 8.0$\times$10$^{11}$  &     &    &   1.4$\times$10$^{-11}$  & Col HCN\\ 
H$^{15}$NC &       &                             & 5$\times$10$^5$$^f$ & 1.8$\times$10$^{11}$  &     &    &   3.1$\times$10$^{-12}$  & Col HNC\\ 
DCN        &       &                             & 5$\times$10$^5$$^f$ & 6.0$\times$10$^{12}$ &     &     &   1.0$\times$10$^{-10}$ &  Col HCN\\
C$_2$H     & 13(1) & 4.6(0.5)$\times$10$^{14}$   &                     &                            &       &                        &   8.1$\times$10$^{-9}$ & \\
C$_2$D     &       &                               &                     &                         &  \multicolumn{1}{c}{10}   & 1.3$\times$10$^{13}$  &  2.3$\times$10$^{-10}$ & \\
H$^{13}$CO$^+$ & 11(1) & 1.5(0.1)$\times$10$^{12}$ & 3$\times$10$^5$     & 1.4$\times$10$^{12}$  &       &                        &  2.6$\times$10$^{-11}$  & \\
HCS$^+$        &       &                           & 5$\times$10$^5$$^f$ & 1.6$\times$10$^{12}$  &   &                        &  2.8$\times$10$^{-11}$ & \\
N$_2$H$^+$     &       &                           & 5$\times$10$^5$$^f$ & 1.2$\times$10$^{13}$    &       &                        &  2.1$\times$10$^{-10}$ & \\
C$_2$S         &       &                           &                     &                         & \multicolumn{1}{c}{10}    & $<$1.1$\times$10$^{12}$ & 1.9$\times$10$^{-11}$ & \\
HCO            &       &                           &                     &                         &  \multicolumn{1}{c}{10}   & 5.0$\times$10$^{12}$ &   8.8$\times$10$^{-11}$ & \\
SO$_2$         &       &                           &  5$\times$10$^5$$^f$      &     4.5$\times$10$^{12}$     &     &  &   7.9$\times$10$^{-11}$ & \\
H$_2$CO        & 53(1) & 1.1(0.1)$\times$10$^{14}$ & 5$\times$10$^5$ &  8.9$\times$10$^{13}$    &       &                       &   1.9$\times$10$^{-9}$ & ortho/para=3\\
H$_2$CS        & 22(6) & 9.0(5.9)$\times$10$^{12}$  &                     &                         &       &                       &   1.6$\times$10$^{-10}$ &  ortho/para=3 \\ 
HNCO           &       &                           &                     &                         & \multicolumn{1}{c}{10}    &  1.5$\times$10$^{12}$ &   2.6$\times$10$^{-11}$ & \\
c-C$_3$H$_2$   & 11(2) & 9.3(3.6)$\times$10$^{12}$ & 3$\times$10$^5$     &  8.0$\times$10$^{12}$     &       &                       &   1.6$\times$10$^{-10}$ & ortho/para=3\\
HC$_3$N        & 26(1) & 9.3(1.3)$\times$10$^{12}$ & 5$\times$10$^5$     &  9.0$\times$10$^{12}$       &       &                       &   1.6$\times$10$^{-10}$ & \\
C$_4$H         &       &                           &                     &                         & \multicolumn{1}{c}{10}    & 4.0$\times$10$^{13}$  &   7.0$\times$10$^{-10}$ & \\
CH$_3$OH       & 16(8) & 1.1(4.8)$\times$10$^{13}$ &               &                         &       &                             &   1.9$\times$10$^{-10}$ & Bad estimate\\
CH$_3$CN       & 30(6) & 4.2(3.6)$\times$10$^{12}$ &               &                         &       &                             &   7.3$\times$10$^{-11}$ & Bad estimate \\
CH$_3$C$_2$H   & 39(16)  & 7.6(10)$\times$10$^{13}$ &              &                         &       &                             &   1.3$\times$10$^{-9}$ & \\  \hline
\hline %inserts single lin
\end{tabular}

\noindent
$^a$ Rotational diagrams. Numbers in parenthesis are the mathematical error of the fitting.
$^b$LVG calculations assuming T$_k$=50~K, $\Delta$v=5~km~s$^{-1}$ for the IF and MP1 and T$_k$=70~K for the MP2 positions. When only one transition is 
observed, a molecular hydrogen density of 5$\times$10$^5$~cm$^{-3}$.$^c$ Local Thermodynamic Equilibrium calculations. $^d$ Fractional abundance derived assuming 
N(C$^{18}$O)/N(H$_2$)=1.7$\times$10$^{-7}$. $^e$ Assumed C$^{18}$O wrt H$_2$ abundance. $^f$Assumed molecular hydrogen density. \\
\end{table*}

\addtocounter{table}{-1}

\begin{table*}
\caption{Physical parameters (continuation).} % title of Table
%\centering % used for centering table
\begin{tabular}{p{1.5cm} p{1.0cm} p{2.0cm} p{1.5cm} p{1.0cm} p{1.5cm}  p{1.5cm} p{1.5cm}  p{2.5cm}} % centered columns (4 columns)
\hline \hline
\multicolumn{1}{c}{}         & \multicolumn{2}{c|}{RD$^a$}    & \multicolumn{2}{c|}{LVG$^b$} &  \multicolumn{2}{c}{LTE$^c$} &  f$_X$$^d$ &   \\
\multicolumn{1}{c}{Molecule} & \multicolumn{1}{c}{T$_{rot}$}  & \multicolumn{1}{c|}{N}       &  \multicolumn{1}{c}{n}   &  \multicolumn{1}{c|}{N}  &  
\multicolumn{1}{c}{T$_{rot}$$^e$}  &   \multicolumn{1}{c}{N}                   &            &            \\ % table heading
         & \multicolumn{1}{c}{(K)} &  \multicolumn{1}{c|}{(cm$^{-2}$)}  &\multicolumn{1}{c}{(cm$^{-3}$)}  &   \multicolumn{1}{c|}{(cm$^{-2}$)}   &  
\multicolumn{1}{c}{(K)}       &  \multicolumn{1}{c}{(cm$^{-2}$)}  &            &            \\ % table heading
\hline % inserts single horizontal line
\hline
\\
\multicolumn{9}{c}{{\large MP2 (0$\arcsec$,$+$40$\arcsec$)}} \\
\\
%$^{13}$CO  & 22(1) & 5.3(0.1)$\times$10$^{16}$ & ... \\
C$^{18}$O     & 65(8)  & 1.3(0.2)$\times$10$^{16}$ &  $>$3$\times$10$^4$ &   1.3$\times$10$^{16}$   &       &                       &    1.7$\times$10$^{-7}$$^d$ & \\
SO            & 21(8)  & 2.9(2.3)$\times$10$^{13}$ & 2$\times$10$^5$     &  5.0$\times$10$^{13}$    &       &                       &    3.8$\times$10$^{-10}$    & \\
SO$^+$        &        &                           &                     &                         & \multicolumn{1}{c}{10}    & $<$1.8$\times$10$^{12}$ &  $<$2.3$\times$10$^{-11}$ & \\
$^{13}$CN     & 4(2)   & 3.5(1.5)$\times$10$^{13}$ &       &                         &       &       &  4.6$\times$10$^{-10}$  & \\
$^{13}$CS     &        &                           & 5$\times$10$^5$$^f$ & 2.5$\times$10$^{12}$   &       &                      &   3.3$\times$10$^{-11}$  & Col CS\\
SiO           & 11(1)  & 6.5(1.2)$\times$10$^{11}$ & 5$\times$10$^5$     &  1.0$\times$10$^{12}$  &       &                        &   8.5$\times$10$^{-12}$  & \\
HNC           &       &                            & 5$\times$10$^5$$^f$ & 1.2$\times$10$^{13}$    &       &                       &   1.6$\times$10$^{-10}$ & \\ 
H$^{13}$CN    & 8(1)  & 2.9(0.5)$\times$10$^{12}$  & 6$\times$10$^5$     & 2.6$\times$10$^{12}$    &       &                       &   3.8$\times$10$^{-11}$ & Col HCN \\
HN$^{13}$C    &       &                            & 5$\times$10$^5$$^f$ & 5.0$\times$10$^{11}$    &       &                       &   6.5$\times$10$^{-12}$ & Col HNC\\
HC$^{15}$N    &       &                            & 5$\times$10$^5$$^f$ & 5.0$\times$10$^{11}$    &       &                        &   6.5$\times$10$^{-12}$ & Col HCN\\
DCN           &       &                            & 5$\times$10$^5$$^f$ &  3.2$\times$10$^{12}$   &       &                        &   4.2$\times$10$^{-11}$ & Col HCN\\
C$_2$H        & 12(1) & 4.1(0.3)$\times$10$^{14}$  &                     &                         &       &                       &   5.4$\times$10$^{-9}$  & \\
C$_2$D        &       &                            &                     &                         &  \multicolumn{1}{c}{10}   & 2.0$\times$10$^{13}$  &   2.6$\times$10$^{-10}$ & \\
H$^{13}$CO$^+$ & 10(1) & 1.1(0.3)$\times$10$^{12}$ &  2$\times$10$^5$    & 1.3$\times$10$^{12}$    &       &                       &   1.4$\times$10$^{-11}$ & \\
HCS$^+$        &       &                           & 5$\times$10$^5$$^f$ & 5.0$\times$10$^{12}$   &  &  & 6.5$\times$10$^{-11}$ & \\
N$_2$H$^+$     &       &                           & 5$\times$10$^5$$^f$ &  4.0$\times$10$^{12}$    &       &                        &  5.2$\times$10$^{-11}$ & \\
C$_2$S         & 11(7) & 1.9(5.3)$\times$10$^{12}$ &               &                         &       &                        &  2.5$\times$10$^{-11}$ & \\
HCO            &       &                           &                     &                         &  \multicolumn{1}{c}{10}   & 1.3$\times$10$^{13}$   &  1.7$\times$10$^{-10}$ & \\
SO$_2$         & 28(15) & 1.0(1.4)$\times$10$^{13}$ & 5$\times$10$^6$ &    1.0$\times$10$^{13}$       &       &                        &  1.3$\times$10$^{-10}$ & only 3mm lines fitted \\
H$_2$CO        & 52(4)  & 9.3(2.0)$\times$10$^{13}$ & 1$\times$10$^7$ &    8.0$\times$10$^{13}$     &       &                        &  1.2$\times$10$^{-9}$  & ortho/para=3 \\
H$_2$CS        & 60(38) & 1.3(2.1)$\times$10$^{13}$ &                 &                         &       &                        &  1.7$\times$10$^{-10}$ & ortho/para=3\\
HNCO           &        &                                 &                 &                         & \multicolumn{1}{c}{10}    &  $<$9.0$\times$10$^{11}$ &  $<$1.2$\times$10$^{-11}$ & \\
c-C$_3$H$_2$   & 12(1) & 1.3(0.1)$\times$10$^{13}$  & 4$\times$10$^5$   &   1.3$\times$10$^{13}$       &       &                         &   1.7$\times$10$^{-10}$ & ortho/para=3 \\
HC$_3$N        & 35(3) & 6.1(1.2)$\times$10$^{12}$  & 5$\times$10$^5$   &   6.0$\times$10$^{12}$        &       &                         & 8.0$\times$10$^{-11}$  & \\
C$_4$H         & 9(4)  & 1.2(1.6)$\times$10$^{14}$  &                   &                       &       &        &   1.6$\times$10$^{-9}$ & \\
CH$_3$OH       & 41(74) & 2.1(7.0)$\times$10$^{14}$  &                  &                       &       &        & 2.7$\times$10$^{-9}$ & Bad estimate \\
CH$_3$CN         & 59(20)    & 5.6(2.2)$\times$10$^{12}$ &              &                       &       &                       &   7.3$\times$10$^{-11}$ & Bad estimate \\
CH$_3$C$_2$H     &           &    &              &                       &  \multicolumn{1}{c}{30}     &  9.0$\times$10$^{13}$  &   1.2$\times$10$^{-10}$ & Bad estimate \\ 
\hline \hline %inserts single line
\end{tabular}
\end{table*}

\begin{table*}[!htb]
\caption{Comparison with prototypical PDRs.} % title of Table
\begin{center} % used for centering table
\begin{tabular}{lcccc|cc|cc|cc} % centered columns (4 columns) 
\hline \hline
 &\multicolumn{3}{c}{Mon~R2}  &  Ref &   \multicolumn{1}{c}{Orion Bar} &  Ref  &  \multicolumn{1}{c}{NGC 7023}   & Ref &  \multicolumn{1}{c}{Horsehead}  & Ref \\
 & (0$\arcsec$,0$\arcsec$)   &   ($+$15$\arcsec$,$-$15$\arcsec$)  &  (0$\arcsec$,$+$40$\arcsec$) &	      & IF &     &  ($-$25$\arcsec$,$+$40$\arcsec$)   &       &  PDR peak	&   \\
%&	East 	&  \\ % table heading
\hline % inserts single horizontal line
$G_{\mathrm 0}$ (Habing)  & \multicolumn{3}{c}{5$\times$10$^5$}         & {\it 1,2} &  \multicolumn{1}{c}{2$\times$10$^4$} & {\it 3,4} &\multicolumn{1}{c}{2$\times$10$^3$} & 
{\it 5} & \multicolumn{1}{c}{100}   & {\it 6,7,8}  \\
$n$ (cm$^{-3}$)$^a$              & \multicolumn{3}{c}{4$\times$10$^6$}   &  & \multicolumn{1}{c}{$\sim$5$\times$10$^5$} & &  \multicolumn{1}{c}{$\sim$2$\times$10$^5$}& 
& \multicolumn{1}{c}{$\sim$10$^5$} & \\ \hline
%&  \multicolumn{1}{c}{M82} & Ref \\
H$^{13}$CN /HN$^{13}$C  &  7.5 &  5 &  4 & {\it 1} & $>$ 2  & {\it 3}  &        &            &        &        \\
HCN /HNC                &     &    &    &         &    3   &  {\it 3} &    2   &  {\it 5}   &        &       \\
\hline
$^{13}$CN/H$^{13}$CN   	& $<$4    &     2     &   5$-$12  & {\it 1}   &      &	       &	      &                 &      &       \\ %&         &      \\
CN/HCN			&	&	   &	    &	        &  3.0 &  {\it 3}      &   4.5        &  {\it 5}	&      &       \\ % &  6.0    &  8,9	\\
\hline
H$^{13}$CO$^+$/H$^{13}$CN &  0.29 &  0.40  &   0.37 & {\it 1}   &      &   	       &	      &	         &      &       \\ %&	  &             \\
HCO$^+$/HCN               &       &        &        &           &  0.9 &  {\it 3}      &  3.3	      &  {\it 5} &      &       \\ %&  1	  &             \\
\hline
CO$^+$/H$^{13}$CO$^+$	 &  0.53  &	   &        & {\it 1,2} & 0.52 & {\it 3,4}     &  0.6	        &  {\it 4,5}  & $<$0.03    &   {\it 6}    \\ % &  2$^a$     &	18,19	\\
H$^{13}$CO$^+$/HOC$^+$  &   5.8   &        &        & {\it 1,2} & 3.8  & {\it 3,4}     &  5--14$^b$     &  {\it 4,5}  & $\sim$4   &   {\it 6}    \\% &  0.9$^a$   &   18,20     \\
\hline
c-C$_3$H$_2$/C$_2$H	& 0.02	&  0.02   &  0.03  & {\it 1}   & 0.016    & {\it 3,4}     &  0.03  &  {\it 4,5}  &    0.06    &   {\it 7}     \\ % &  0.04      & 18 \\
C$_4$H/C$_2$H           & 0.07   &  0.09    &  0.3   & {\it 1}   &          &               &                      &            &  0.1 &   {\it 7}     \\
%C$_2$H/H$^{13}$CO$^+$	& 273  	&  267     &  354   & 1   & 644$^{iso}$  & {\it 3,5}     &  111         &  {\it 7}	  & 205-295    &   {\it 8,9}     \\%&  19	     & 18,19  \\
HCO/H$^{13}$CO$^+$	& $<$3 &  $<$3 &  12    & {\it 1}   &  	     &               &          &            & 50$^c$	& {\it 8,9}     \\ %&  3.6        &	21	\\
\hline
SO$^+$/SO	        & 	         &   0.1           & $<$0.06         & {\it 1}       &         &             &	      &	       &	  &          \\ % &     &
CS/HCS$^+$              &  11$^{d}$     &    78$^{d}$     & 25$^{d}$      & {\it 1}       &         &             &        &        &   0.008  &  {\it 10}  \\
%DCN/HCN                 &  $<$0.005$^{e}$  &  0.06$^{e}$    &   0.03$^{e}$   & {\it 1}       &  0.008     &  {\it 10}    &   &       &      &         \\ 
%C$_2$D/C$_2$H           &  $<$0.03         &  0.04          &   0.05         & {\it 1}       &  $<$0.04   &  {\it 3,10}  &   &       &      &       \\
\hline \hline %inserts single line
\end{tabular}
\end{center}

\noindent
$^a$ The density is not uniform in these regions. We indicate a representative density of the clumps/filaments.

\noindent
$^b$ Depending on the assumed excitation conditions.

\noindent
$^c$ It strongly depends on the position, varying between $\sim$1 in the shielded core to $\sim$50 in the HCO peak. The value in the table corresponds to 
the HCO peak derived from $\sim$14$\arcsec$ angular resolution observations. 

\noindent
$^{d}$ Obtained from our $^{13}$CS data assuming $^{12}$C/$^{13}$C=50.

\noindent
References:
(1) This work;
(2) Rizzo et al. 2003;
(3) Fuente et al. 1996;
(4) Fuente et al. 2003;
(5) Fuente et al. 1993;
(6) Goicoechea et al. 2009;
(7) Teyssier et al. 2004;
(8) Gerin et al. 2009a;
(9) Gerin et al. 2009b;
(10) Goicoechea et al. 2006.

\end{table*}

\section{Comparison with protypical PDRs}

%Note that $^{13}$CO, HCN and HCO$^+$ do not appear. The lines of these species are optically thick and we
%estimated their abundances from the rarer isotopologues C$^{18}$O, H$^{13}$CN and H$^{13}$CO$^+$ assuming the isotopic
%ratios $^{12}$C/$^{13}$C=50 (Savage et al. 2002 and references therein) and $^{16}$O/$^{18}$O=500. 

The 8~$\mu$m IRAC image undoubtedly shows the existence of an extended PDR in this region.
One of the goals of this paper is to establish the role of the UV photons, relative to other
phenomena linked to the star formation process like bipolar outflows, in the chemistry of this molecular cloud.
For this aim, we follow a two-fold strategy. First of all we compare the chemical composition in the three observed positions.
Since the average kinetic temperature and density in the three positions are quite similar, differences in the chemistry
can be interpreted as due to the different incident UV field.
Second, we compare Mon~R2 with that in other prototypical PDRs like the Orion Bar, NGC~7023 and the Horsehead.

The IF and MP1 define a cut along the dense PDR in Mon~R2, the MP1 being the more shielded position. Comparing the molecular abundances between 
these two positions, we find that only one molecule, HCS$^+$, has a larger (by a factor of 6)
abundance towards the IF than towards the MP1. This suggests that HCS$^+$ comes from the inner PDR like 
the reactive molecular ions CO$^+$ and HOC$^+$ (Rizzo et al. 2003, 2005). 
The reactive ion SO$^+$ is, however, more abundant towards the MP1. Fuente et al. (2003) already suggested that
this molecular ion comes from a more shielded layer than the ions CO$^+$ and HOC$^+$. 
There are some species that present significantly lower abundance towards the IF. 
These species are N$_2$H$^+$ and HC$_3$N proving
that are easily destroyed by UV photons and are only found in the more shielded gas. 
A peculiar chemistry is detected towards the MP2. This position is very rich in carbon chains such
as C$_2$H, C$_4$H, C$_2$S, and c-C$_3$H$_2$. As we discuss below this different chemistry could be due to MP2 
being related with a low UV PDR.
%As discussed bellow, this would be consistent with the results of
%Bern\'e et al. (2009) who found a lower UV field towards this position.
%compared with the IF. 

In Fig. 3,  we compare the molecular abundances relative to C$^{18}$O measured in Mon~R2  
with those in other prototypical PDRs. This kind of comparison is not straightforward since data
with different spatial resolutions and calibrations are used. Despite this, we can extract some qualitative
conclusions. Mon~R2 seems richer in the carbon chains HCN and CN 
than other prototypical PDRs in star forming regions such as the Orion Bar, NGC~7023 and the Horsehead (see Fig. 3).
One could think that this high HCN and CN abundance could be related with the bipolar outflows in the region.
This interpretation will be discussed in Sect. 5.2.
The reactive ion CO$^+$ is only detected in highly UV irradiated positions such as the IF in Mon~R2,
the Orion Bar and NGC~7023. The chemistry of the MP2 position in Mon~R2 resembles that of the Horsehead, with a 
high C$_4$H and HCO abundance. 

Abundance ratios of chemically related  molecules and with similar
excitations conditions are more accurate than absolute molecular abundances. On one hand, the spatial extent of the emission
is expected the same, and the abundance ratio is not sensitive to the unknown beam filling factor. Moreover, as long as the excitation is similar 
for the two
species, the abundance ratio is not sensitive to uncertainties in the assumed physical conditions. For these reasons, abundance ratios
instead of fractional abundances are often used to compare with models. In Table~6 we show some interesting abundance ratios in Mon~R2 and compared them
with the values in the other prototypical PDRs.

\subsection{HCN/HNC}
In  this source, the HCN and HNC lines are optically thick. 
Moreover, the clear self-absorption features of these lines prevent from calculating reliable molecular abundances. 
Instead, we use the rare isotopes H$^{13}$CN and HN$^{13}$C. Using a $^{12}$C/$^{13}$C
ratio of 50
we derived X(HCN)$\sim$5$\times$10$^{-9}$ and [HCN]/[HNC]$\approx$5$-$8
in the three observed positions. The [HCN]/[HNC] ratio is a factor $\sim$2 larger than in NGC 7023 and the Orion Bar (Fuente et al. 1996, 2003).
In addition to H$^{13}$CN, we have detected HC$^{15}$N towards the three
targeted positions. The  [H$^{13}$CN]/[HC$^{15}$N] is about 10 towards the three offsets, which would imply $^{14}$N/$^{15}$N=500 with 
our assumed  $^{12}$C/$^{13}$C isotopic ratio. This ratio is similar to that measured by Gerin et al. (2009c) in dark clouds and low mass star forming regions, and consistent
with the protosolar value of $\sim$424. Assuming a higher  $^{12}$C/$^{13}$C isotopic ratio, $\sim$89, would imply 
 $^{14}$N/$^{15}$N=890. In MP1, we have a tentative detection of the H$^{15}$NC \textit{1$\rightarrow$0}
line. Using this 3-$\sigma$ detection we derive [HC$^{15}$N]/[H$^{15}$NC] = 4, 
similar to the values found in NGC 7023 and the Orion Bar. 
A complete multitransitional study of HCN and its isotopologues and detailed source modelling is required to have a more accurate estimate of the 
[HCN]/[HNC] ratio and how it changes across the envelope in Mon~R2

\subsection{CN/HCN}
Fuente et al. (1993) detected a [CN]/[HCN] abundance ratio larger than unity
in the north-west PDR of NGC 7023 and interpreted
it as the consequence of the photodissociation of HCN into CN. Since then, the [CN]/[HCN] ratio has been successfully used as a PDR tracer in different environments,
PDRs associated with star forming regions such as the Orion Bar and NGC~7023, planetary nebulae, protoplanetary disks and even external galaxies.
In Mon~R2 we use the rarer isotopes $^{13}$CN and H$^{13}$CN to derive the [CN]/[HCN] abundance ratio. We derive a  [CN]/[HCN] abundance ratio $\sim$2$-$12 towards the three positions. 
This high [CN]/[HCN] abundance ratio is consistent with that found
in other environments such as NGC~7023, the Orion Bar, the planetary nebula NGC~7027 and the starburst galaxy M~82. 
The [CN]/[HCN] ratio in bipolar outflow is $<$1 (see Bachiller \& P\' erez-Guti\'errez 1997). This support the interpretation of CN and HCN coming from the PDR instead of
from the bipolar outflows.
%This suggests that the [CN]/[HCN] ratio is a robust PDR indicator
%but its value is not very sensitive to the specific impinging UV field $G_{\mathrm 0}$, or to the density for the
%range of values typically found in reflection nebulae and {\sc Hii} regions ($G_{\mathrm 0}$$>$100).

\subsection{CS/HCS$^+$}
We have detected HCS$^+$ towards the ionization front with an abundance of $\sim$1.9$\times$10$^{-10}$. This is one of the few molecules that shows enhanced
abundance at this position and argues in favor of HCS$^+$ arising from the surface of the PDR. The HCS$^+$ abundance measured towards the IF 
is  a factor of$\sim$ 5 larger than that observed in the Horsehead nebula (Goicoechea et al. 2006) but comparable with that measured by Lucas \& Liszt (2002) towards
diffuse clouds. The HCS$^+$ abundance towards the other two positions, MP1 and the MP2 peaks, is lower by a factor of $>$4 and more similar to
that measured in the Horsehead nebula. The derived [CS]/[HCS$^+$] ratio is uncertain because the CS lines are optically thick. From our $^{13}$CS 
observations and assuming $^{12}$CS/$^{13}$CS=50, we obtain that the  [CS]/[HCS$^+$]$\sim$11 in IF and $\sim$25 in MP2.  
Tafalla et al. (1997) derived column densities of
N(CS)=5.6$\times$10$^{13}$~cm$^{-2}$ towards the IF from an LVG analysis using the CS J=\textit{2$\rightarrow$1},
J=\textit{3$\rightarrow$2} and \textit{5$\rightarrow$4} lines which would imply [CS]/[HCS$^+$]$\sim$7. These values 
are closer to the value [CS]/[HCS$^+$]$\sim$10 measured in diffuse clouds than to the value 
$\sim$175 measured towards the Horsehead nebula.
% and suggests
%a very low sulphur depletion in this PDR.
% Goicoechea et al. (2006) discussed that the [CS]/[HCS$^+$] ratio is also
%sensitive to degree of sulphur depletion.

\subsection{SO$^+$,SO, SO$_2$, H$_2$CS}
SO$^+$ has been detected in warm and cold clouds with abundances ranging from $\sim$10$^{-9}$--10$^{-11}$. These fractional
abundances can be explained by gas phase ion-molecule chemistry (Turner 1994). In PDRs, SO$^+$ has been detected in NGC~7023 and
the Orion Bar. The SO$^+$ abundance in these regions is 5$\times$10$^{-11}$ and 1$\times$10$^{-10}$ respectively. In Mon~R2, we
have detected SO$^+$ in the MP1, with an abundance of $\sim$5$\times$10$^{-11}$. 
We have not detected SO$^+$ towards the MP2 peak. In PDRs, SO$^+$ is primarily formed via S$^+$ + OH$\rightarrow$SO$^+$ + H and removed 
by dissociative recombination in S + O. The lack of SO$^+$ in the MP2 peak is very likely related to a lower abundance of OH at
this position. The radical OH is formed by endothermic reactions in the outer layers of the PDR and is very sensitive to the gas kinetic
temperature.

In addition to CS, HCS$^+$, SO and SO$^+$, we have also detected the more complex sulphuretted species C$_2$S, H$_2$CS  and 
SO$_2$. SO has similar fractional abundance towards the three positions, but there are some differences in the abundances of the other compounds. 
The carbon chains like C$_2$S and H$_2$CS are more abundant towards the MP2 than towards the IF and MP1.  SO$_2$ is not detected towards the IF, corroborating previous interpretation of
this molecule being easily destroyed by UV radiation (see e.g. Fuente et al. 2003).

\subsection{SiO}
SiO was detected in Mon~R2 by Rizzo et al. (2005). They derived a SiO column density of 1.1$\times$10$^{11}$~cm$^{-2}$ towards the IF which is consistent
with our upper limit of $<$3.0$\times$10$^{11}$~cm$^{-2}$.
We have detected SiO towards MP1 and MP2 with an abundance of $\sim$10$^{-11}$ which is 
lower than that determined by Schilke et al. (2001) towards the Orion Bar ($\sim$7$\times$10$^{-11}$ in the IF) and consistent with that expected in PDRs. 
There is no hint of enhanced SiO abundance produced by the shocks associated with the bipolar outflows or the expected
shock front associated to the expansion of the UC {\sc Hii}. This is not surprising since an expansion velocity of $\lesssim$1.0 km/s (Fuente et al. 2010; Pilleri et al. 2012a) is 
not high enough to destroy the core of the silicates grains. The low SiO abundance argues against a dominant role of shocks in the chemistry of this region.

\subsection{C$_2$H, c-C$_3$H$_2$, C$_4$H}
Intense C$_2$H and c-C$_3$H$_2$ emission is detected in all the observed positions with the maximum column density towards MP2. The maximum fractional abundances 
are X(C$_2$H)$\sim$5$\times$10$^{-9}$ and X(c-C$_3$H$_2$)$\sim$1.7$\times$10$^{-10}$. These values are a factor of $\sim$3 lower than those measured towards the Horsehead and
the Orion Bar but similar to those measured in NGC~7023. Taking into account the uncertainties in the calculation of the fractional abundances, these differences 
are not significant. Note that for the Horsehead we have taken the single-dish column densities published by Teyssier et al. (2004) to compare with our data because 
interferometric observations filter out the extended emission tracing only the small spatial scales.
The [c-C$_3$H$_2$]/[C$_2$H] ratio is quite uniform with a value of $\sim$0.03. The large chain C$_4$H had only been detected in low-UV PDRs (IC~63, $\rho$-Oph and the Horsehead) 
thus far (Teyssier et al. 2004; Pety et al. 2005).
The [C$_4$H]/[C$_2$H] ratio is maximum at the MP2 confirming that the PDR at this position is particularly rich in carbon chains.

 \begin{figure*}
% \centering
\includegraphics[width=0.33\textwidth]{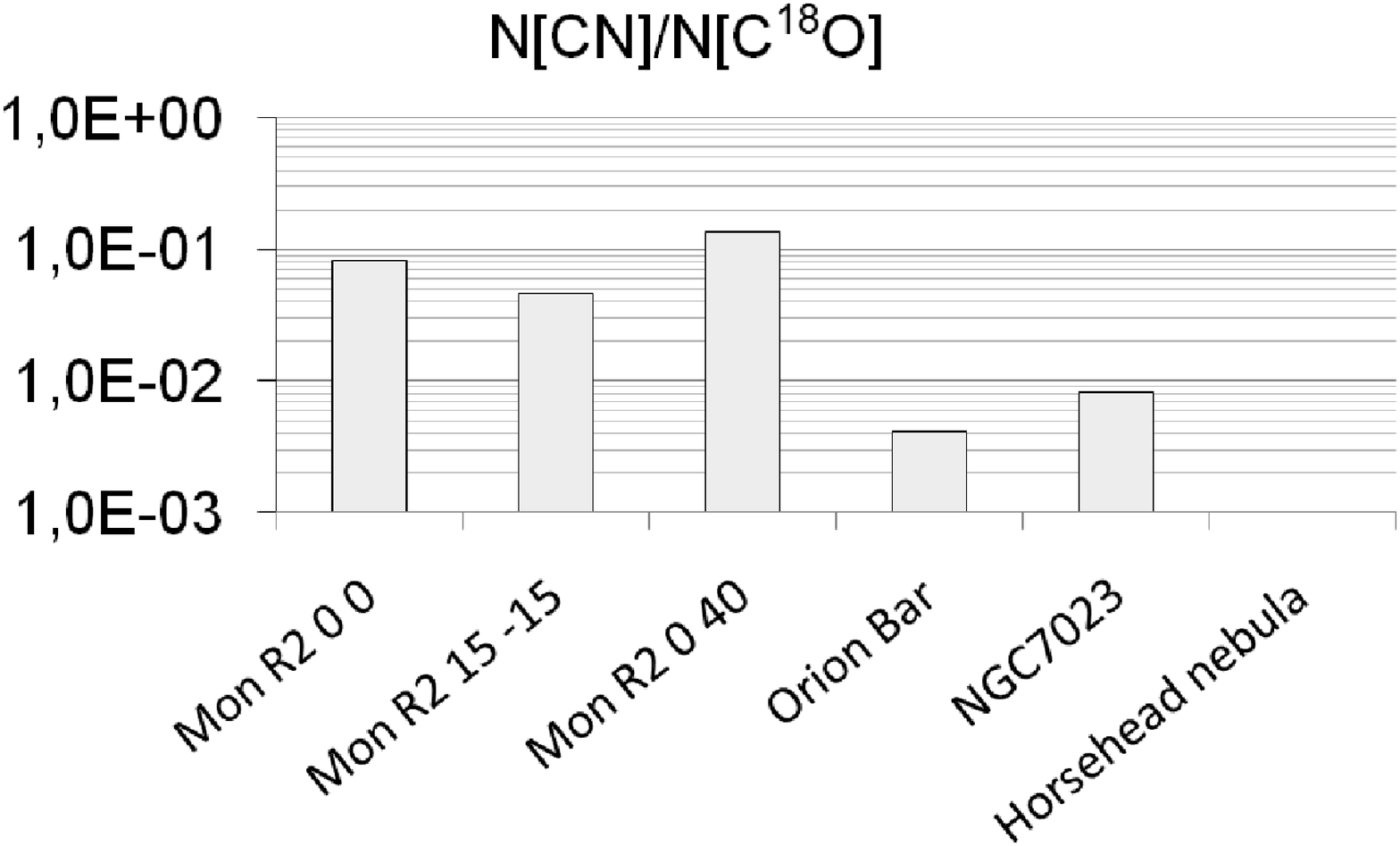}
\includegraphics[width=0.33\textwidth]{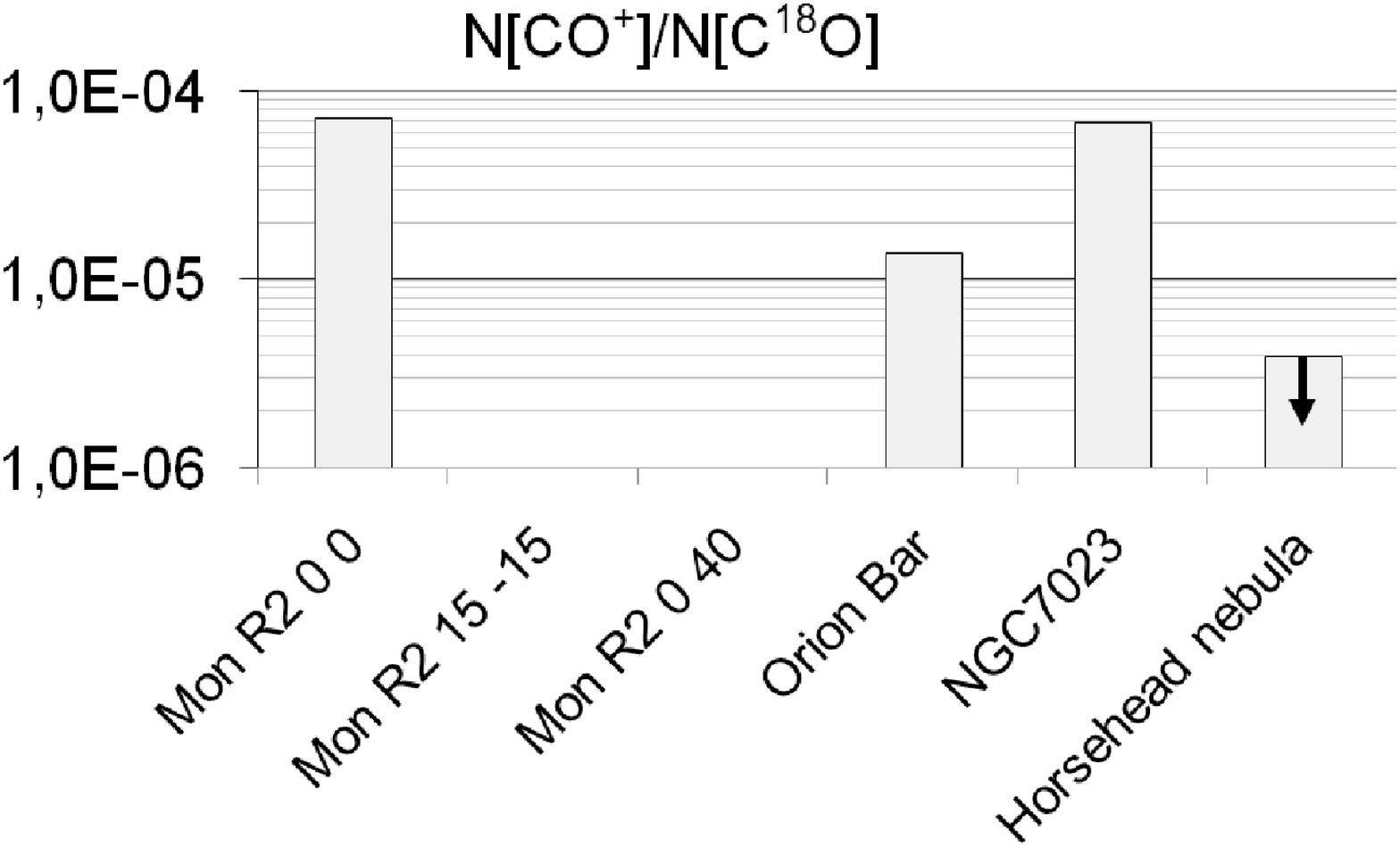}
\includegraphics[width=0.33\textwidth]{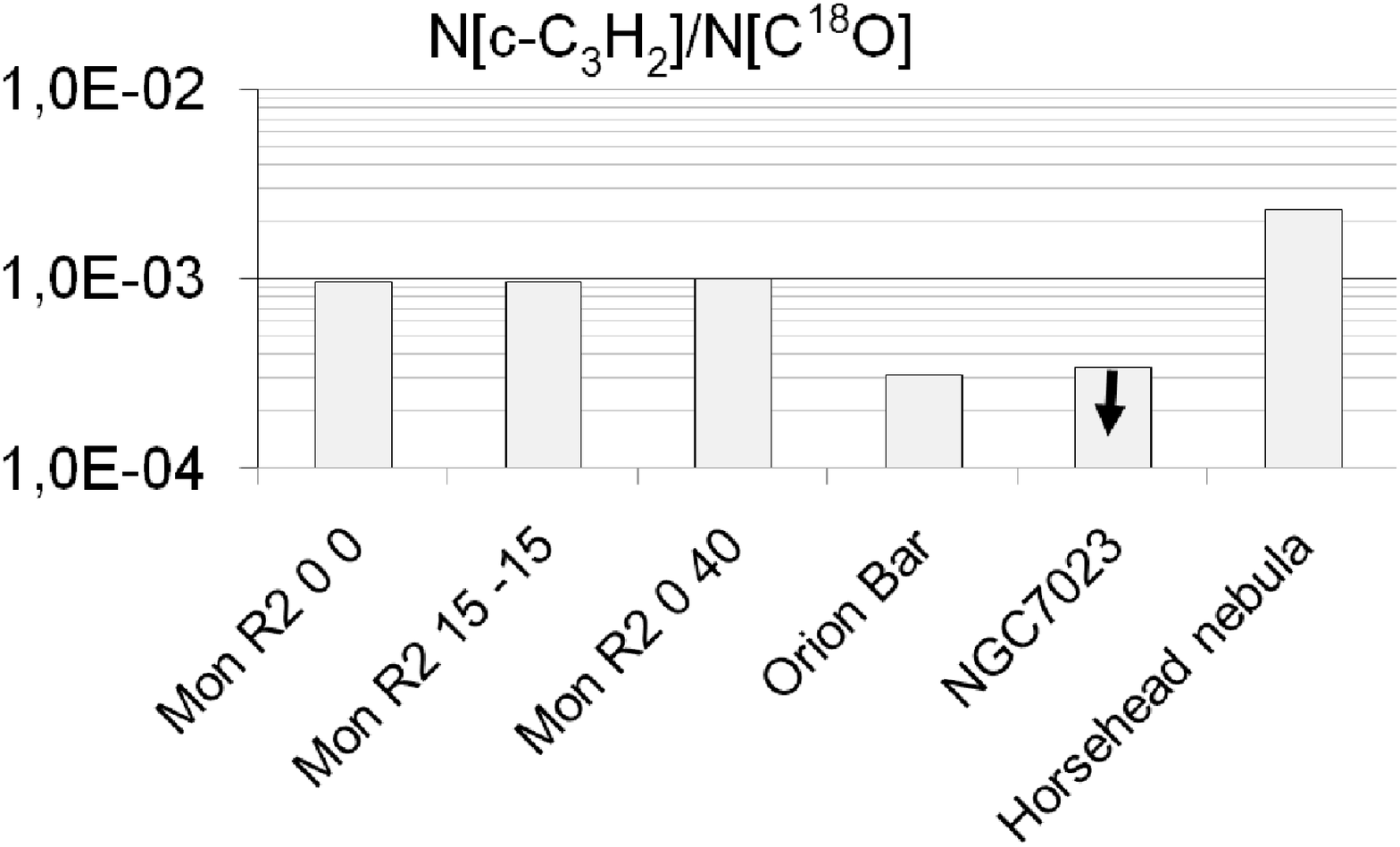}
\includegraphics[width=0.33\textwidth]{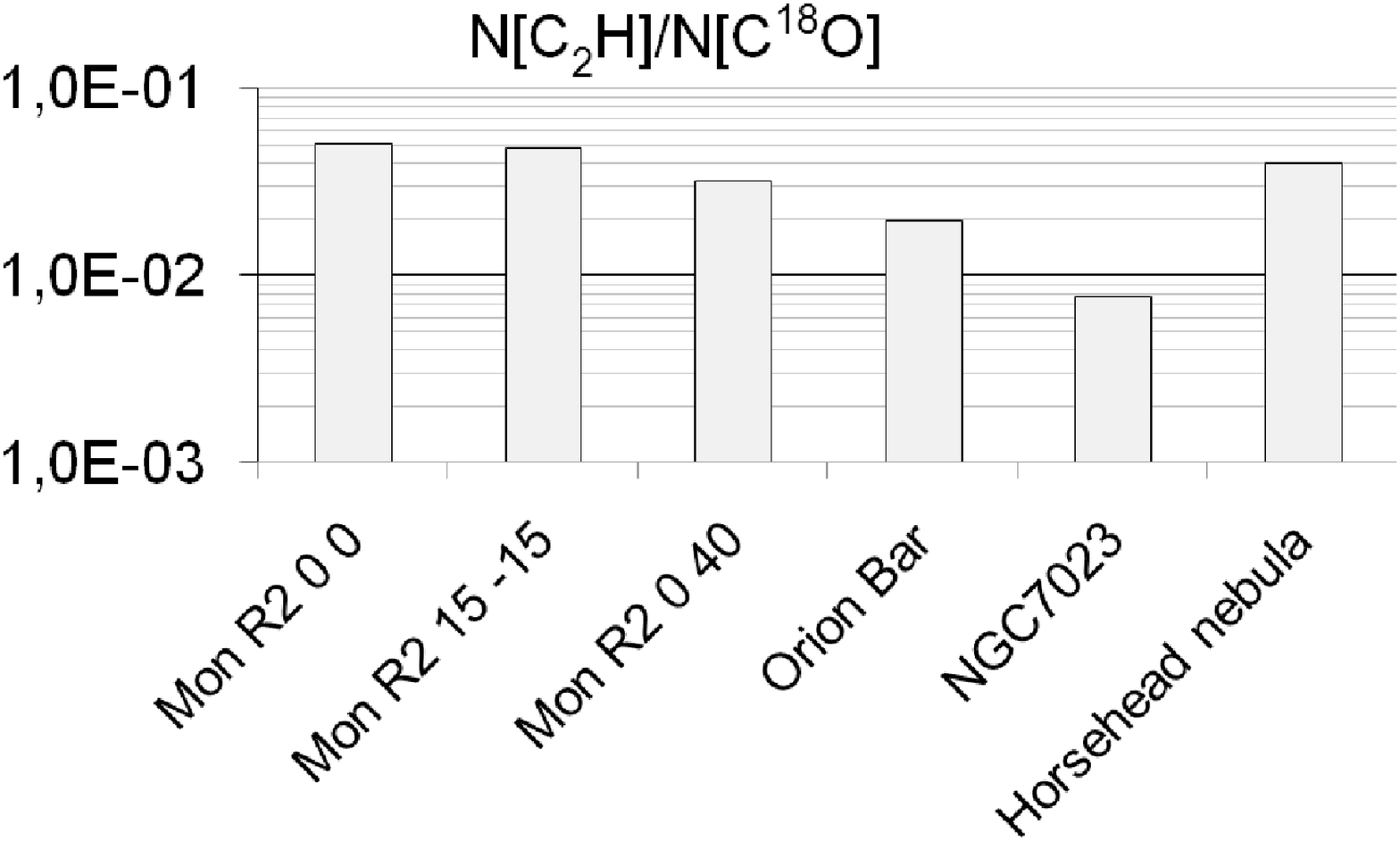}
\includegraphics[width=0.33\textwidth]{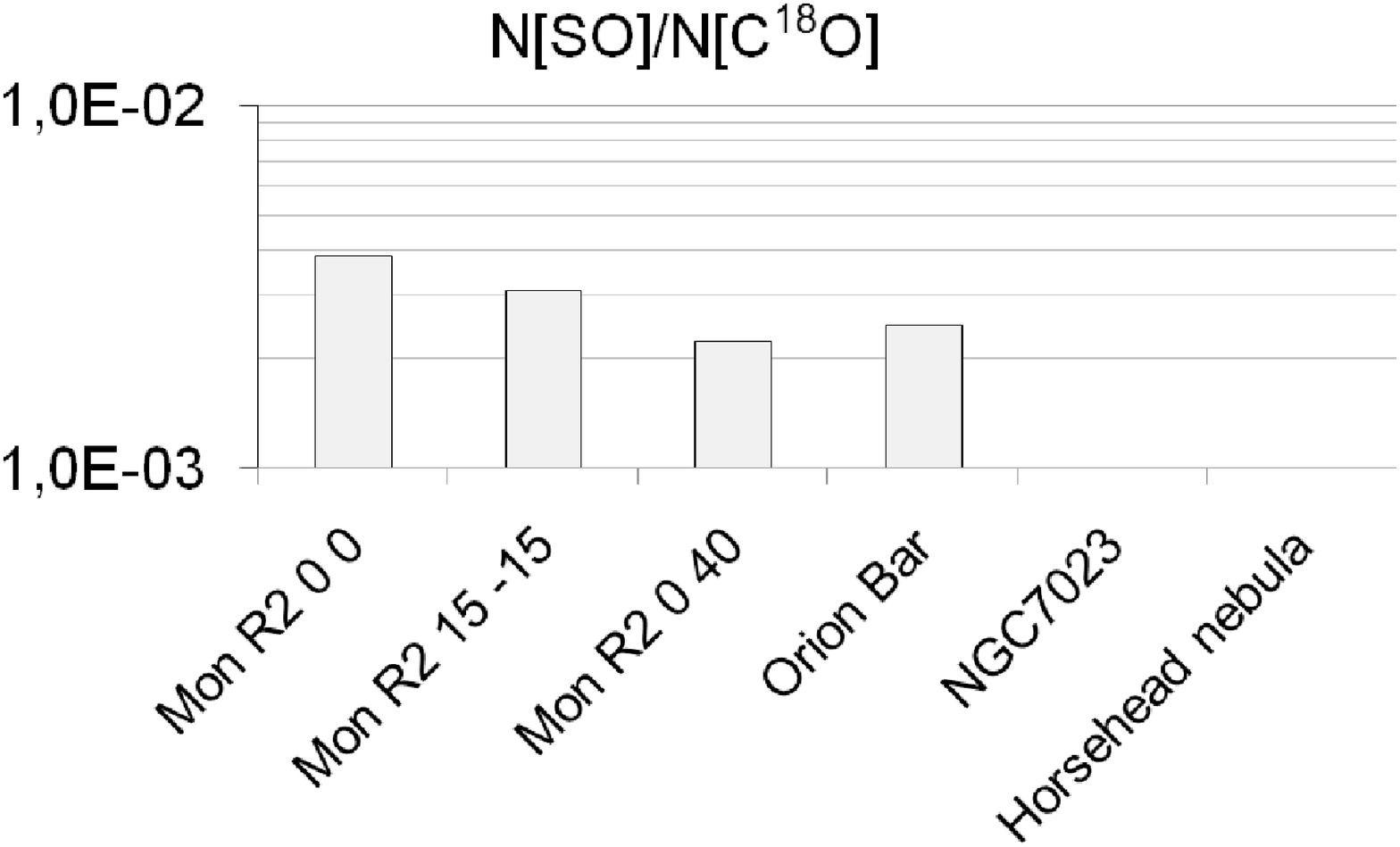}
\includegraphics[width=0.33\textwidth]{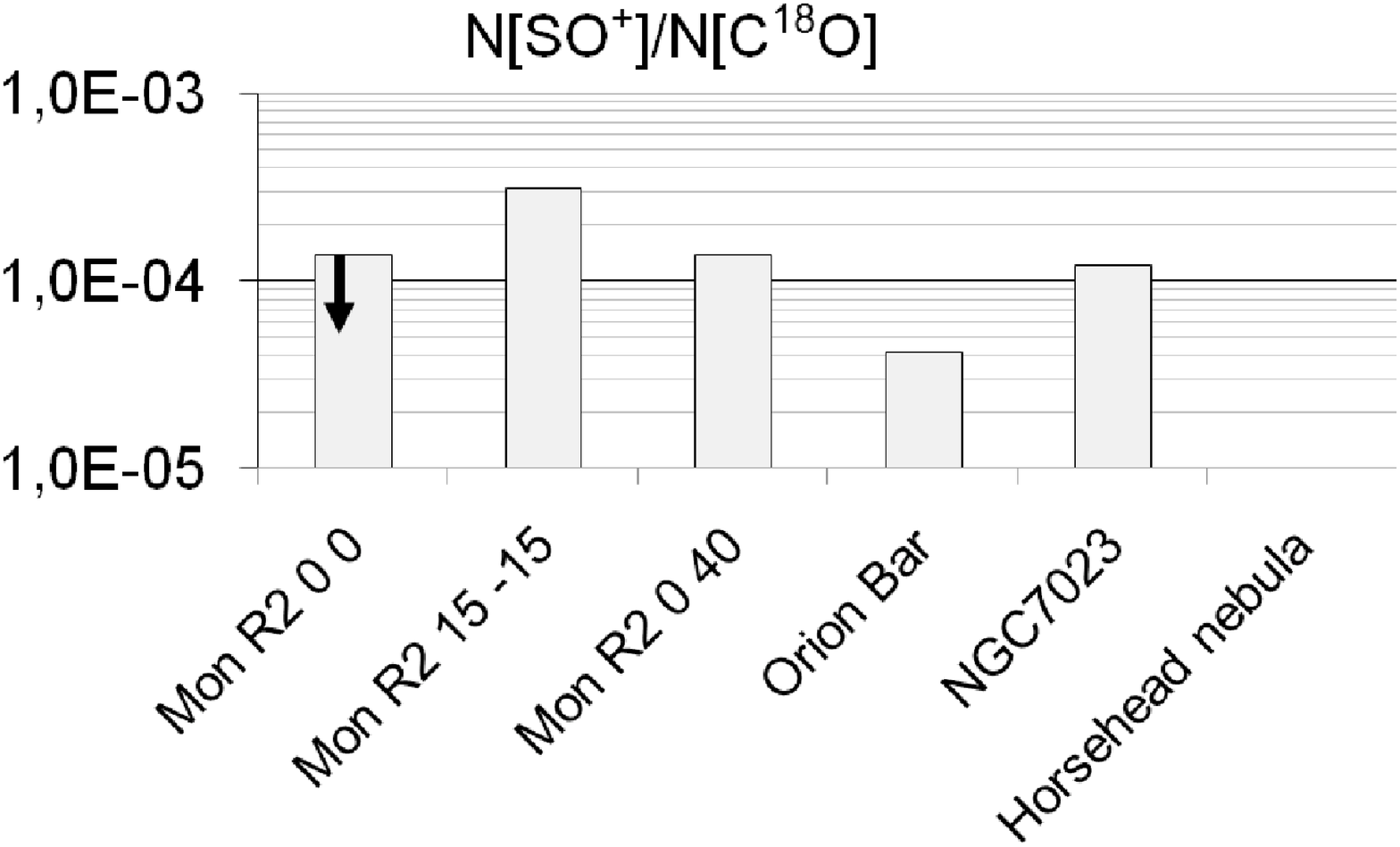}
\includegraphics[width=0.33\textwidth]{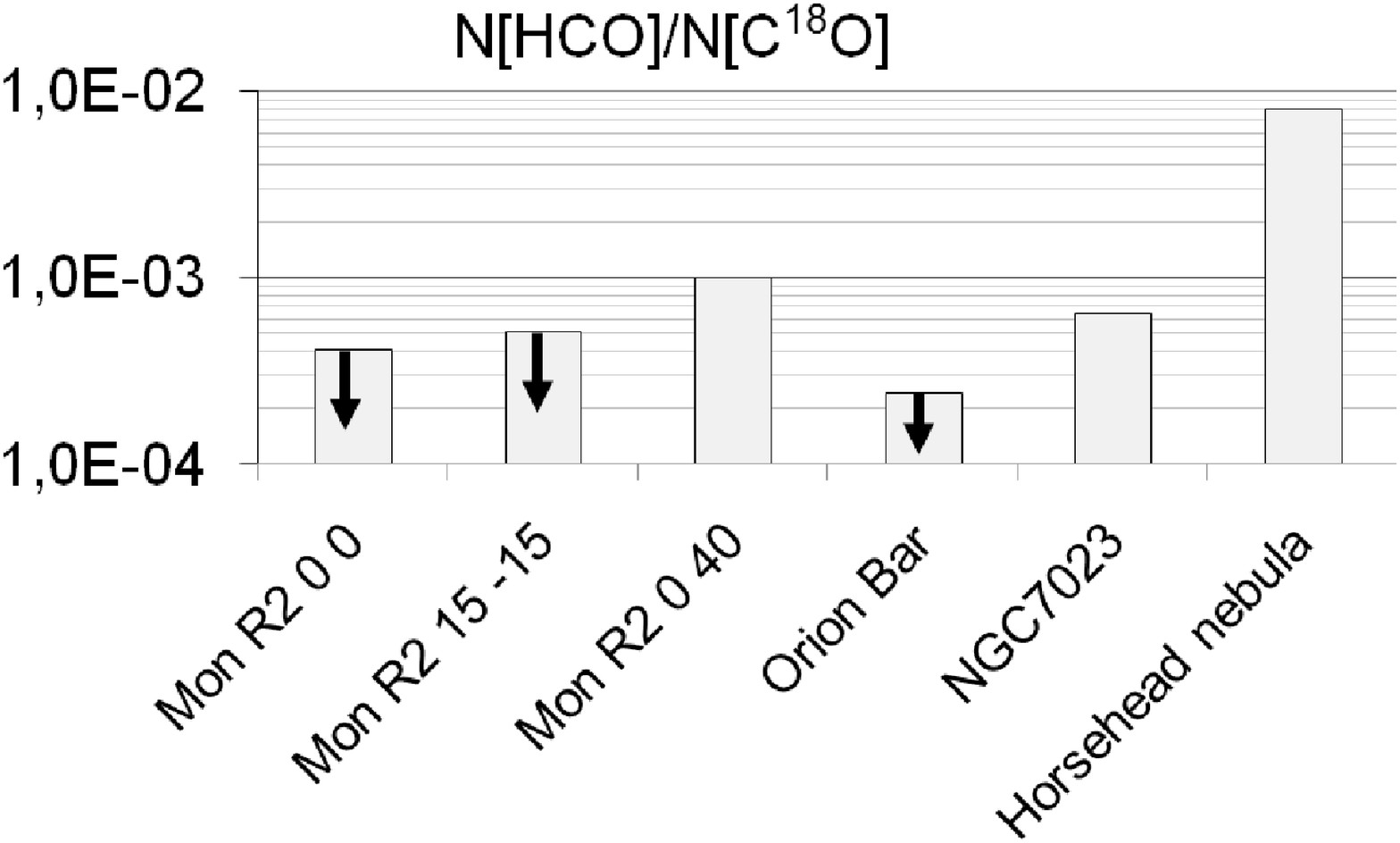}
\includegraphics[width=0.33\textwidth]{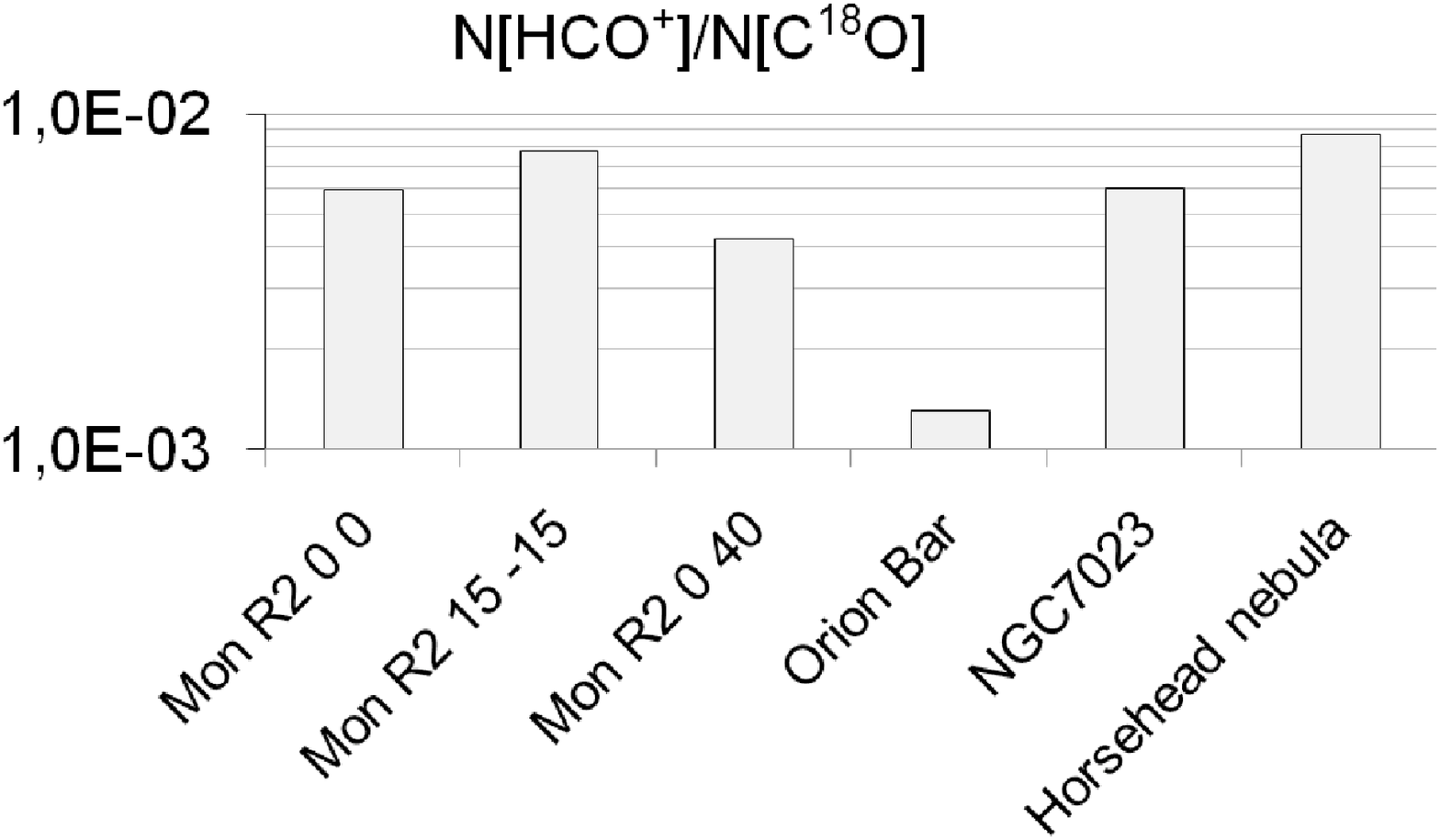}
\includegraphics[width=0.33\textwidth]{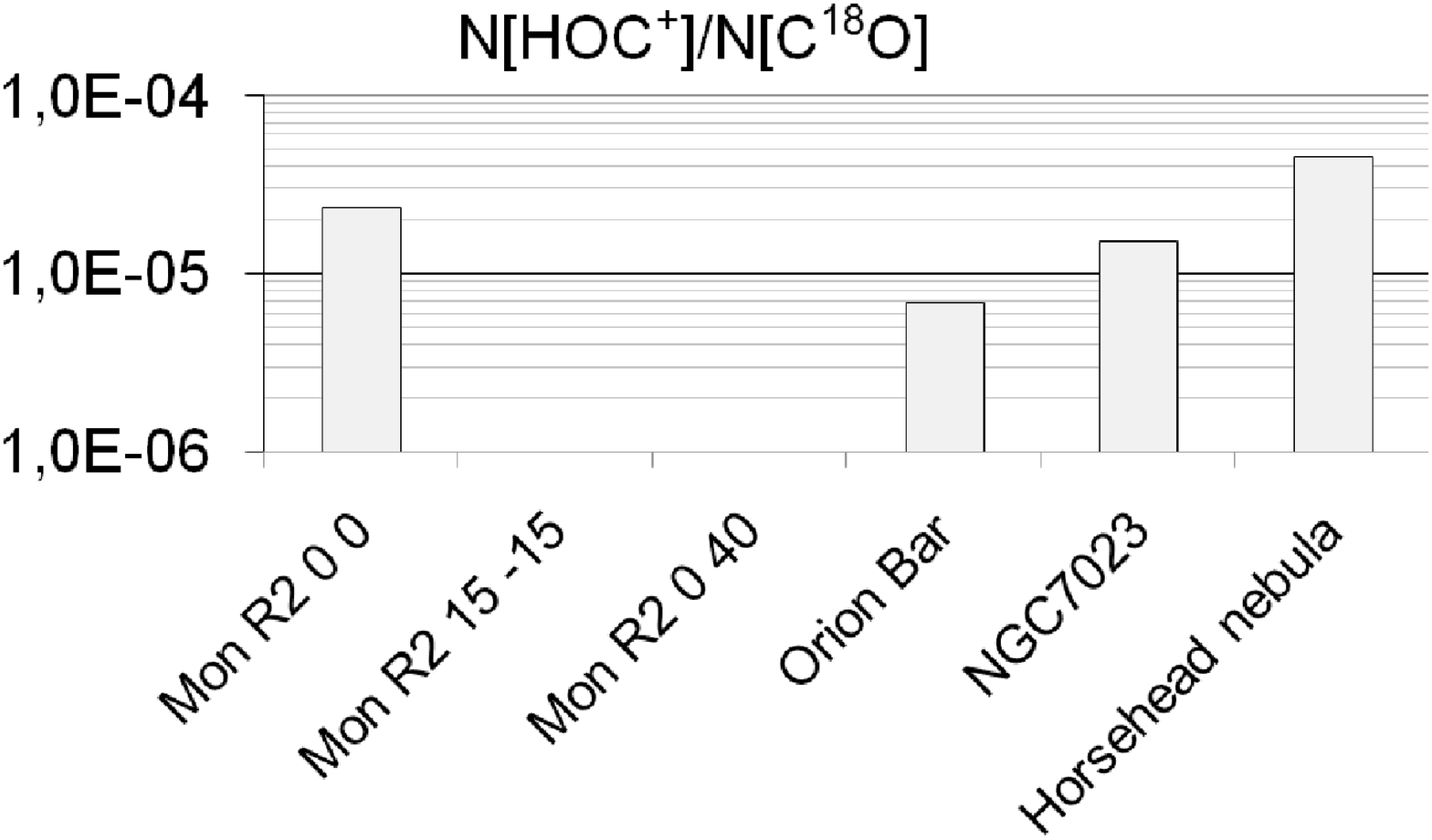}
\includegraphics[width=0.33\textwidth]{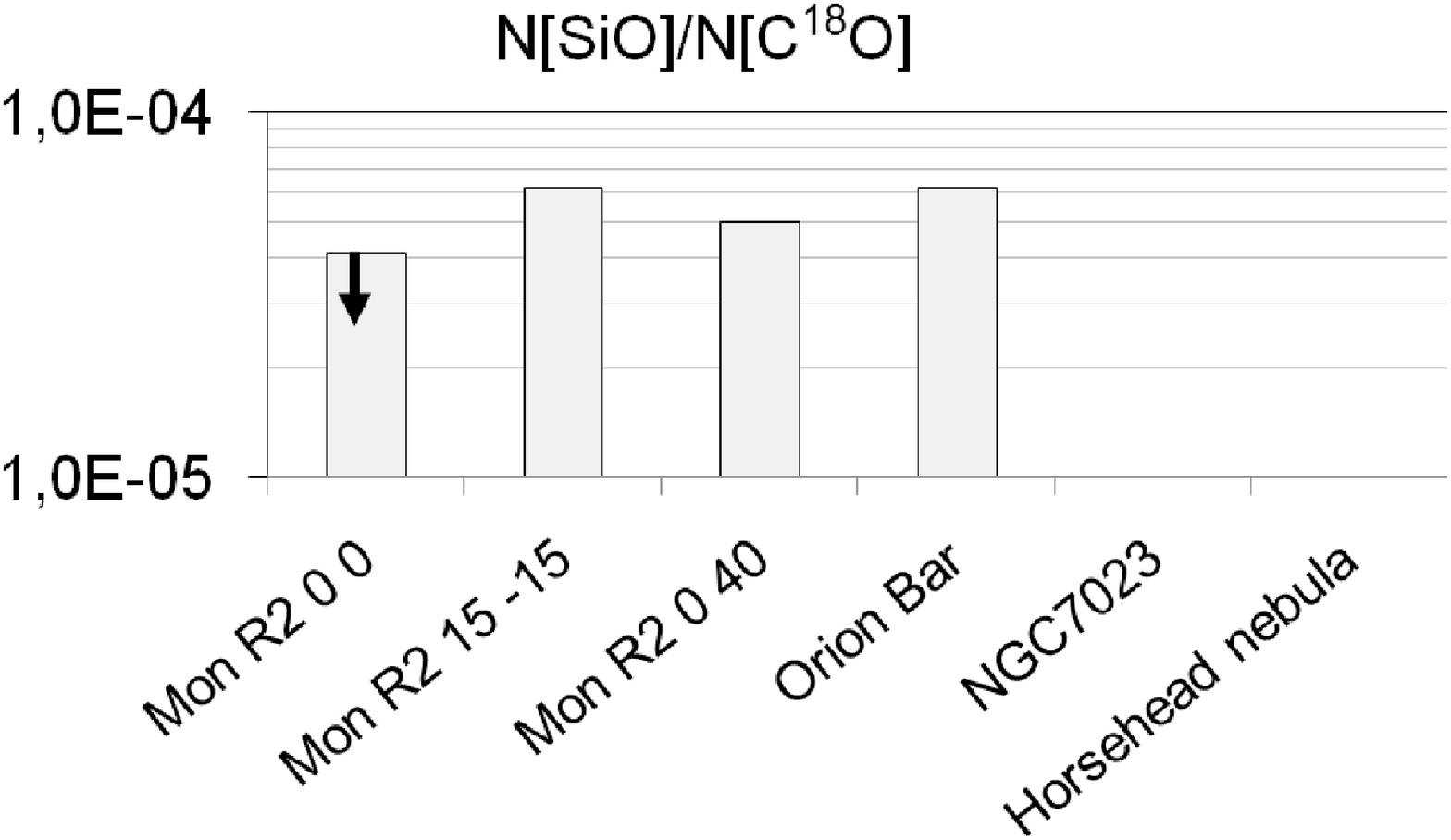}
\includegraphics[width=0.33\textwidth]{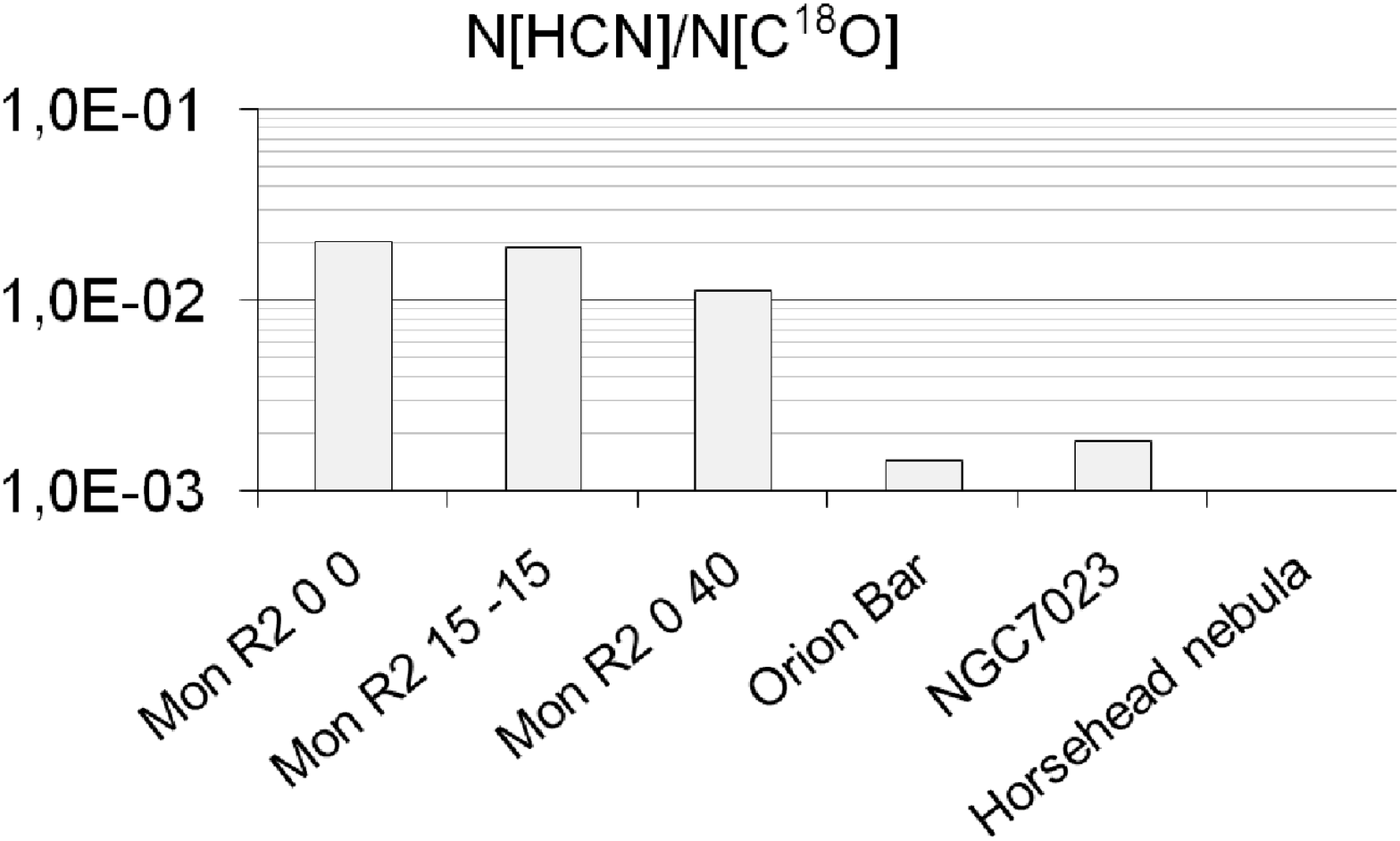}
\includegraphics[width=0.33\textwidth]{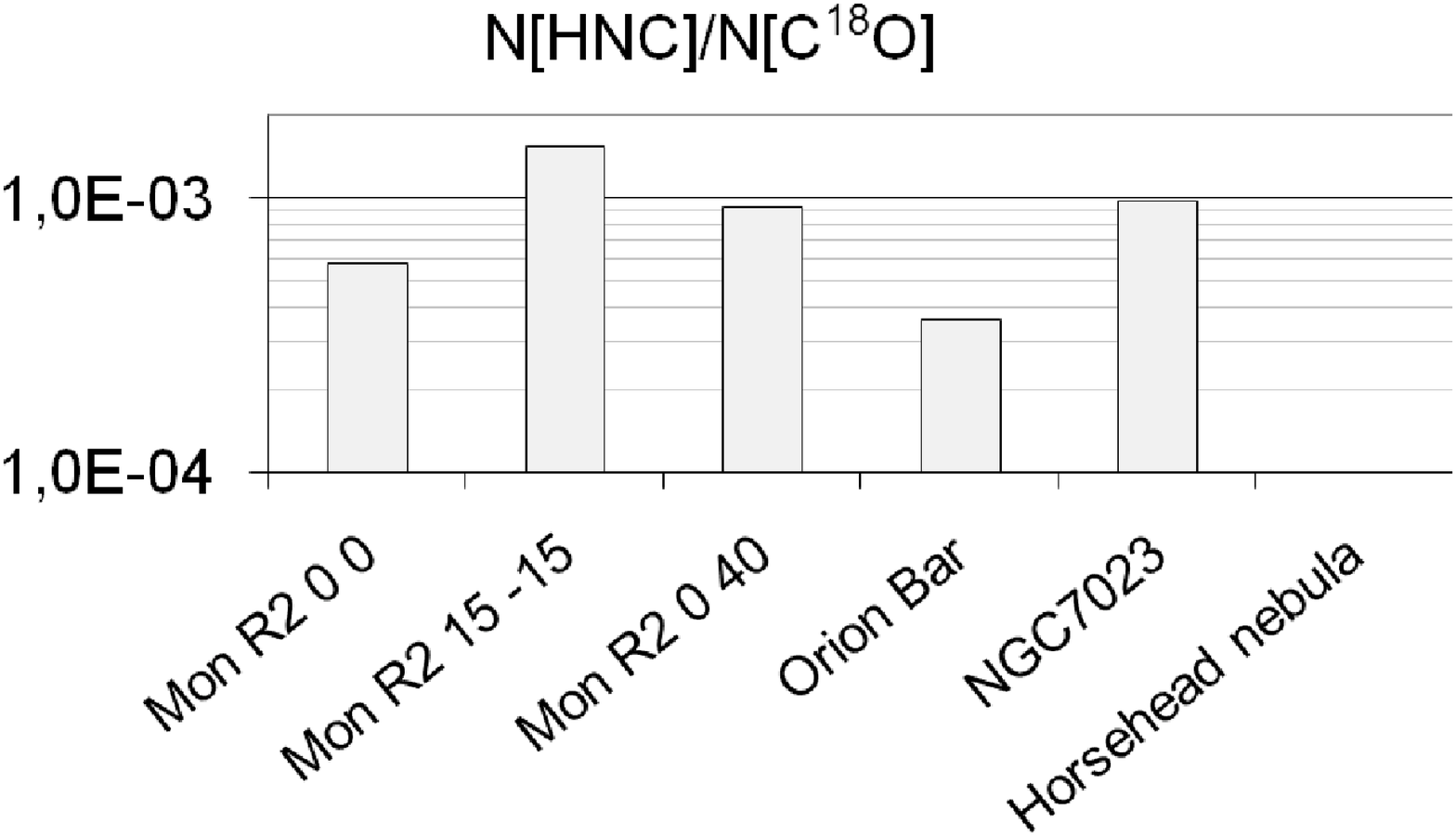}
\includegraphics[width=0.33\textwidth]{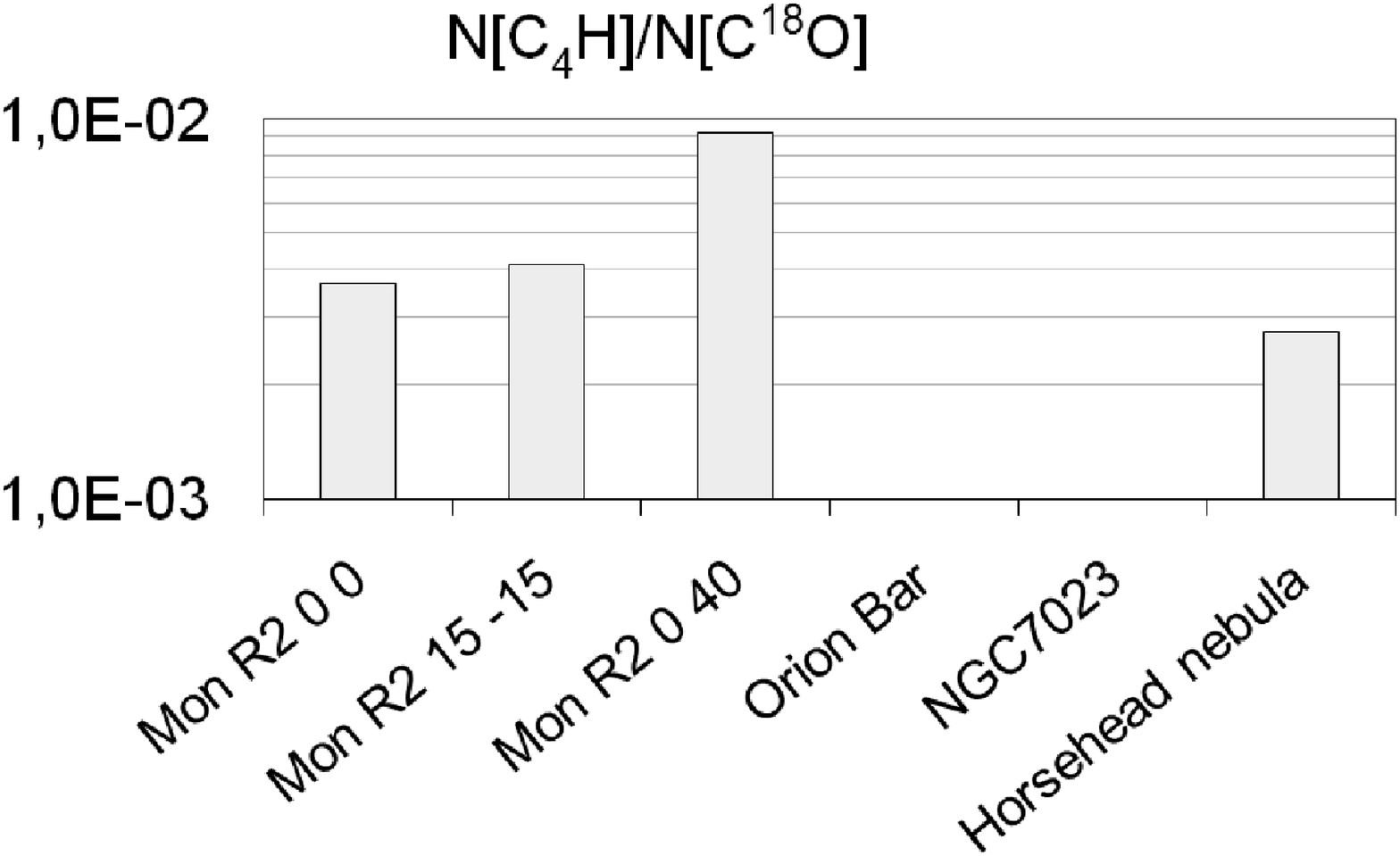}
      \caption{Comparison of the abundance of the most representative species among the prototypical PDRs:  Mon~R2, NGC 7023, Orion Bar and the Horsehead. }
 \end{figure*}

\subsection{HCO}
The formyl radical, HCO, was detected in the interstellar medium by Snyder et al. (1976). Pioneering studies by Schenewerk et al. (1988) pointed out to an association of
large HCO abundances with the gas around {\sc Hii} regions. Schilke et al. (2001) detected HCO towards NGC 2023, the offset (-30$\arcsec$,80$\arcsec$) in NGC 7023, the 
position referred to as Bar(CO) in the Orion Bar and the IF in S140. From a limited mapping
they concluded that HCO is associated with the PDR component and explained the enhancement of HCO abundance as the consequence of the
photodissociation of the H$_2$CO  molecules released from the grain mantles by UV photons. Garc\'{\i}a-Burillo et al. (2002) mapped this radical in the 
starburst M~82 and also interpreted it as arising in the giant PDR in the disk of this galaxy. Gerin et al. (2009a) carried out a high angular resolution mapping of the
Horsehead in this radical. The maximum of the HCO emission was found towards a position where the PAH band emission is particularly intense.
The HCO emission becomes fainter in the more shielded molecular gas. After updating some reaction rates, in particular
O+CH$_2$  $\rightarrow$ HCO + H, they found that gas-phase chemistry could account for the high HCO abundance
without invoking the photodesorption of H$_2$CO from the ice mantles. In Mon~R2, 
we have detected HCO towards MP2 with a fractional abundance of $\sim$1.7$\times$10$^{-10}$. 
%This abundance is 10 times lower than that measuredtowards the HCO peak in the Horsehead nebula. However this can be due to a geometrical effect, HCO can arise in a thin layer and its emission is diluted in a 29$\arcsec$ beam. 
Towards  the IF and the MP1, the HCO abundance 
is at least a factor of $\sim$2 lower than towards the MP2. Note that Schilke et al. (2001) did not detect HCO towards the IF 
but only in the position referred to as Bar(CO)  in their paper that could correspond to a lower UV radiation field. 
% peak, which corroborates that HCO  is preferably found in low UV PDRs.

\subsection{H$_2$CO}
We have derived a H$_2$CO abundance of $\approx$1$-$2$\times$10$^{-9}$ in Mon~R2.
This abundance is close to that found in hot cores (see e.g. Sutton et al 1995) and higher than that expected in a PDR. 
Guzm\'an et al. (2011) studied the formation of H$_2$CO in the Horsehead. They derived an  H$_2$CO abundance
of 4$-$6$\times$10$^{-10}$ in the PDR peak, similar to that found in the shielded core position. They could not explain the H$_2$CO abundance in this PDR 
only with gas phase chemistry. However, they obtained a good agreement between model predictions and observations with a gas phase+surface chemistry model in which H$_2$CO can 
be formed on the grain surfaces and photo-desorbed to the gas phase, in addition to the gas phase route.
In Mon~R2, the emission of H$_2$CO is ubiquitous (Giannakopoulou et al. 1997) and the low spatial resolution of our observations makes it difficult to discern
its origin: the PDR, the more shielded molecular gas or the bipolar outflows.

\subsection{Complex molecules: CH$_3$OH, CH$_3$CN, HC$_3$N}
In addition to PDR-like molecules, we have detected in our survey other complex species that are not expected to be abundant in PDRs and are usually associated with the chemistry of warm clouds. 
This is the case of  CH$_3$OH, CH$_3$CN and HC$_3$N. 
In the case of methanol, our fractional abundance estimates are very uncertain. The abundance of CH$_3$CN is also quite constant and equal to $\approx$7$\times$10$^{-11}$. For HC$_3$N
we measured abundances of 7$\times$10$^{-11}$$-$10$^{-10}$. These values are lower than those found in hot cores, but very similar to those found in warm molecular clouds such
as the Orion Ridge and in the envelopes of massive protostars.

Complex molecules are rapidly destroyed by UV radiation. Assuming a typical photodissociation rate of 10$^{-10}$ s$^{-1}$, the lifetime of complex molecules
would be $>$10$^5$~yr only for  A$_V$$>$10 mag. 
Fuente et al. (2010) and Pilleri et al. (2012a) estimated an expansion velocity of $\lesssim$1.0 km s$^{-1}$ for this UC {\sc Hii} which implies a  dynamical time of  
$\sim$10$^5$~yr.
This means that complex molecules of the initial core could have survived as long as they  were within clumps in which the visual extinction is $>$10 mag. 
High spatial resolution observations are required to
determine the region from which the emission of complex molecules arises and put additional constraints to their formation mechanism.

\subsection{Deuteration: DCN, C$_2$D}
We have detected DCN and C$_2$D towards the MP1 and MP2. The DCN abundances are  X(DCN)=1.0$\times$10$^{-10}$ (MP1) and
X(DCN)=4.2$\times$10$^{-11}$ (MP2). The C$_2$D abundances are X(C$_2$D)=2.3$\times$10$^{-10}$ (MP1) and X(C$_2$D)=2.6$\times$10$^{-10}$ (MP2).
These abundances correspond to deuterium fractionation values of $\sim$0.03$-$0.05 for both HCN and C$_2$H. The deuterium fractionation
of HCN is a factor of $\sim$10 larger than
that measured by Parise et al. (2009) in the position Bar(HCN) of the Orion Bar. The deuterated species C$_2$D was not detected in Orion with a lower
limit to the C$_2$D/C$_2$H ratio of $<$0.04. Pety et al. (2007) detected DCO$^+$ in a cold clump (T$_k$$\sim$10$-$20~K) towards 
the Horsehead and measured [DCO$^+$]/[H$^{13}$CO$^+$] =0.02. There is not any intense transition of DCO$^+$ in the frequency range covered
by our survey. Therefore, we could not obtain an estimate  of the DCO$^+$ abundance.

Several mechanisms have been proposed to explain the high values of deuterium fractionation (higher than the deuterium abundance in the Universe,  D/H$\sim$10$^{-5}$)
observed in the interstellar medium. Deuterated isotopologues of methanol and formaldehyde have been detected
in hot cores and corinos (see e.g. Parise et al. 2002, 2004; Fuente et al. 2005) where the gas kinetic temperatures is $>$100 K. In these cases, the fractionation is thought
to occur on grain surfaces. The deuterated compounds are released to the gas phase when the ice is evaporated, producing a transient  deuterium enhancement
in gas phase. In Mon~R2, we have detected the deuterated compounds of HCN and C$_2$H. Since these molecules are not mainly formed on grain surfaces, 
it is more plausible to interpret that the deuteration has occurred in gas phase. 

In molecular clouds, deuterium is mainly locked into HD. Efficient transfer of deuterium from this reservoir to other species occurs by means of ion-molecule
reactions. In cold clumps, T$_k$$\approx$10$-$20 K, deuteration is usually transferred via reactions with H$_2$D$^+$. This mechanism is very efficient, for instance,
in pre-stellar cores where molecules are highly depleted (see e.g. Caselli et al. 2003). For slightly higher temperatures of T$_k$$\approx$30$-$50 K, 
the transfer is more efficient via CH$_2$D$^+$ (Roueff et al. 2007, Parise et al. 2009). The high temperatures measured in Mon~R2
favor a deuteration mechanism based on CH$_2$D$^+$.  Our measurements are
in agreement with gas-phase model predictions by Roueff et al. (2007) and Parise et al. (2009) for T$_k$$\sim$30$-$50~K and 
densities $\sim$a few 10$^6$~cm$^{-3}$. These densities are not unrealistic in this region (Rizzo et al. 2005). 
One possibility is that C$_2$D comes from the densest part of this PDR. Another possibility is that a fraction of 
these molecules are formed on grain surfaces or their 
chemistry is related with some of the evaporated species.

\section{Chemical diagnostics}
%Our goal here is to investigate the influence of the incident UV radiation and the density on the molecular abundance ratios and explore the possibility of using them as chemical diagnostics.
From the discussion in Sect. 5 it seems clear that the emission of some of the detected species comes from the extended PDR around this {\sc Hii} region. In the following we try to
understand the differences between MP1 and MP2 in terms of PDR chemistry.
We have run a grid of isochoric models using the updated version (1.4.2) of the PDR Meudon code (Le Petit et al. 2006, Goicoechea et al. 2007) and the 
parameters listed in Table~7. Our grid of models range in density between 10$^4$~cm$^{-3}$ and 
10$^7$~cm$^{-3}$, and in FUV field between 10 and 10$^6$ Habing fields. 
We have run the same grid of models using the standard galactic and the Orion extinction curves but 
the results are essentially the same for the molecular abundance ratios discussed below.
For each model we represent the
cumulative column densities summed up to 10 mag. For higher visual extinctions, UV radiation has a negligible effect on the physical conditions and chemistry of the molecular
gas. Our goal is to investigate the influence of the incident UV radiation and the density on the studied molecular abundance
ratios and explore the possibility of using them as chemical diagnostics. These models are useful to interpret molecular observations, although to reproduce the results on a 
particular PDR it is required a good knowledge of its geometry.

\begin{table}
\caption{Input parameters for the Meudon PDR code.}
\begin{center}
\begin{tabular}{p{1.2cm} p{4.8cm} p{1.7cm}} 
\hline
Parameter 	&   & Value \\ \hline
$G_0$  		& Radiation~field~intensity (Habing)           & 10 to 10$^6$  \\
$G_0^{ext}$ 	& External Radiation field intensity (Habing)  &  1 \\
$n$            	& Hydrogen nuclei density             &  10$^4$~to~10$^7$~cm$^{-3}$ \\
$A_V$               & Cloud depth          & 10 mag\\
Extinction           & Standard Galactic        &    \\
$R_V$		& $A_V/E(B-V)$               & 3.1 \\
$\zeta$        &  Cosmic ray ionization rate & 5$\times$10$^{-17}$~s$^{-1}$\\
$a_{min}$         & dust minimum radius        & 3$\times$10$^{-7}$ cm\\
$a_{max}$               & dust maximum radius       & 3$\times$10$^{-5}$ cm\\
$\alpha$                &  MRN dust size distribution index      & 3.5 \\
He/H                    &  Helium abundance                   & 0.1\\
O/H                     & Oxygen abundance                   & 3.2$\times$10$^{-4}$ \\
C/H                     & Carbon abundance                   & 1.3$\times$10$^{-4}$ \\
N/H                     &  Nitrogen abundance                 & 7.5$\times$10$^{-5}$\\
S/H                     &  Sulphur abundance                  &  1.9$\times$10$^{-5}$\\
Fe/H                    & Metal abundance                    &  1.5$\times$10$^{-8}$ \\ 
\hline
\end{tabular}
\end{center}
\end{table}

Some representative cumulative column density ratios have been selected as candidates for chemical diagnostics.
Fig. 4 shows these ratios as a function of the incident UV field and the hydrogen nuclei density for the standard interstellar
extinction curve.  Since we are using a gas phase model, we have avoided those species for which the formation on grain surfaces could be important. 
This is the case of H$_2$CO, CH$_3$CN and CH$_3$OH. Another important reason to neglect these complex molecules is that they could
come from a shielded component of dense gas instead of the PDRs. We are aware, however,
that  the injection of complex molecules in gas phase by photo-desorption changes the gas chemical composition and could influence the chemistry of more simple
species. Consider, for instance, HCO, which is the photodissociation product of H$_2$CO. The abundance of HCO in the PDR would increase if the evaporation of H$_2$CO 
from the grain surfaces were included improving the agreement with the observations. As discussed below, PAHs and grain destruction are also required to account for 
the observed abundances of small hydrocarbons.

 Our chemical modeling shows that for the range of parameters considered, the CN/HCN ratio is not dependent on the incident UV field but on the density.
 Therefore, the CN/HCN ratio is a good probe of the existence of PDRs but it is not a good diagnostic to differentiate between
 PDRs with different UV fields for $G_{\mathrm 0}$$>$100.
 The value measured in Mon~R2, CN/HCN=2$-$12, is consistent with a hydrogen nuclei density of  $\approx$10$^6$ cm$^{-3}$ (n(H$_2$)$\approx$5$\times$10$^5$~cm$^{-3}$)
in agreement with our density estimates. As commented in Sect. 5, the HCN and CN abundances in this region are larger than in other prototypical PDRs. We cannot discard
that the existence of bipolar outflows could contribute to enhance the abundance of these molecules. However, the measured CN/HCN=2$-$12 is more consistent with a PDR
origin.
%The HCN/HNC ratio in Mon~R2,  HCN/HNC$\approx$5, is 
%higher than the value predicted for the physical conditions of this dense PDR. But, as discussed in \textbf{Sect.~} 5.1, our estimate involves important uncertainties.

According with our models, the HCO$^+$/HCN ratio must be a good tracer of the UV field and density. Fig. 4b shows that the HCO$^+$/HCN ratio is lower 
than 1 for a wide range of physical conditions but increases to larger values for high values of G$_0$ ($>$10$^3$) and of the density ($\sim$10$^6$ cm$^{-3}$).
Taking into account the extreme values of $G_0$ and $n$ in Mon~R2, one would expect a higher ratio than that observed. As mentioned above, the detailed geometry of the PDR 
has to be taken into account when comparing with chemical models. In Mon~R2 several gas components 
lie along the line of sight. The dense PDR is surrounded by a lower density envelope that very likely contributes to the emission of the low J rotational lines of HCO$^+$
and HCN. Although we use the rare isotopologues, H$^{13}$CO$^+$ and H$^{13}$CN, to compute this abundance ratio, optical depth effects could still be important.

\begin{figure*}
\includegraphics[width=0.9\textwidth]{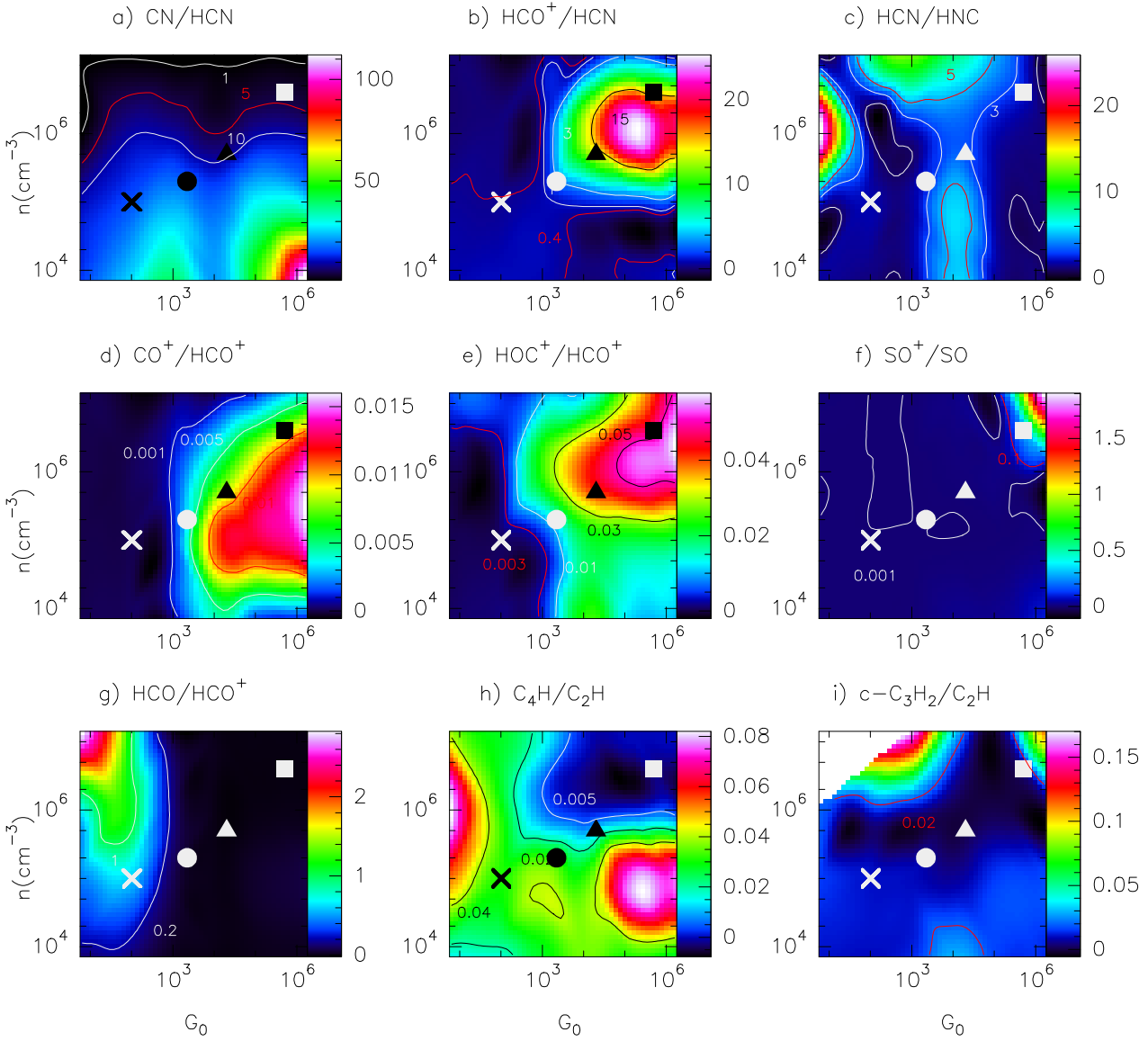}
     \caption{Cumulative column density ratios in a plane slab of A$_V$=10 mag illuminated by the left side for a grid of UV fields and hydrogen nuclei densities. The ratios have
     been calculated using the Meudon code (Le Petit et al. 2006, Goicoechea et al. 2007). G$_0$ is given in unit of the Habing field.
     Levels drawn in red correspond to the values measured in Mon R2. Symbols indicate the physical 
     conditions of the protypical PDRs: Mon~R2 (square), Orion Bar (triangle), NGC~7023 (circle) and Horsehead (cross). Note that the
     CO$^+$/HCO$^+$ and HCO/HCO$^+$ ratios are excellent diagnostics of the UV incident field. The IF and MP positions in Mon~R2 present abundances characteristics of
     dense PDRs with high UV field while the chemistry of the MP2 position is more typical of a low UV PDR.}
 \end{figure*}

One of the best diagnostic of the UV incident field is
the CO$^+$/HCO$^+$ ratio. This ratio is $>$0.05 only when the UV field is higher than 10$^3$. This is consistent with the lack of detection of CO$^+$ in the Horsehead nebula
and in the MP2 position of Mon~R2. The same dependency on the UV incident field stands also for the HOC$^+$/HCO$^+$ ratio. On the contrary, the formyl radical  is
abundant only in PDRs with moderate or low UV radiation field (see also Gerin et al. 2009ab). The HCO/HCO$^+$ ratio takes values $>$0.2 only for UV fields $<$10$^3$. 
%Although the exact value could be
%wrong because of the non-inclusion of the formation of H$_2$CO on grain surfaces, the qualitative behavior is correct (Guzm\'an et al. 2011). 
The CO$^+$/HCO$^+$
and HCO/HCO$^+$ abundance ratios are excellent diagnostics to differentiate between low and high UV PDRs. There is a good qualitative agreement between models and observations 
although they still fail to quantitatively predict these ratios.

Observations show that the SO$^+$/SO ratio is also higher for dense and highly ionized regions and the value towards Mon~R2 is well predicted by our gas-phase
chemical models. However, observationally this ratio does not follow the same trend that the CO$^+$/HCO$^+$ 
(see e.g. Fuente et al. 2003). This ratio is influenced by the amount of sulphur in gas phase that could be
changed, for instance, by the existence of bipolar outflows and slow shocks. Therefore, we do not propose it as a good diagnostic of PDRs. 

Small hydrocarbons  can be good tracers of PDRs, and they are specially abundant in low UV PDRs. 
%However the abundance ratios among the different species do not
%seem to be good chemical diagnostics. 
According with our gas phase calculations, the c-C$_3$H$_2$/C$_2$H ratio presents a fairly uniform value in a wide region of the parameter space. 
Although our models can explain the value
measured in MP2, they fall short to account for the value measured in the Horsehead. In the case of the C$_4$H/C$_2$H ratio, our models fall short to 
reproduce the observed values in both, MP2 and the Horsehead. Different authors have pointed out that the observed abundances of these hydrocarbons in PDRs requires of another production
mechanism than gas-phase reactions, possibly the destruction of PAHs and/or very small grains (VSGs).
It is a matter of a lower value of the UV field that prevents too fast destruction of hydrocarbons, but having
destruction of PAHs/VSGs at a reasonable rate. Going further requires a full modelling of formation/destruction
processes and it is beyond the scope of this paper.

Summarizing,  the CO$^+$/HCO$^+$ and HCO/HCO$^+$ ratios are excellent chemical diagnostic of the UV field, while the CN/HCN and HCN/HNC ratios are more dependent on the
density. Applying these chemical diagnostics to the three positions observed in Mon~R2, we can conclude that the incident UV radiation field in the IF and MP1  are different, being
the highest in IF, with a value consistent with that estimated by Rizzo et al. (2003). Towards the MP2, the incident UV field must be lower. The morphology of the 8~$\mu$m emission 
(see Fig. 1) suggests that this PDR is the consequence of the border of the molecular cloud being illuminated by IRS~1.

%This is also compatible with
%the highest C$_4$H abundance toward the PAH position, although our gas phase chemical model cannot account for the high C$_4$H/C$_2$H ratio measured at this position.

\section{Summary and Conclusions}

 We have carried out a mm survey towards three positions of the PDR that represent different physical
 and chemical environments: (i) the ionization front (IF); two positions in the molecular cloud, (ii) MP1 and (iii) MP2. 
 Our goal has been to investigate the chemistry of the molecular gas around the UC {\sc Hii} region and the possible variations due
  to the different local physical conditions. Our results can be summarized as follows:

  \begin{itemize}   
   \item We have detected more than thirty different species  (including isotopologues and deuterated compounds) in Mon~R2.
  In particular, we have  detected SO$^+$ and C$_4$H towards this region, which are well known tracers of PDRs. 
  In addition to SO$^+$ and C$_4$H, the list of identified species 
  includes typical tracers of PDRs like CN, HCN,HCO, C$_2$H, and c-C$_3$H$_2$ but also other complex molecules more
  common in warm molecular clouds such as CH$_3$CN, H$_2$CO, HC$_3$N, CH$_3$OH or CH$_3$C$_2$H. The origin of these complex species, dense and well shielded
 clumps within the molecular cloud or PDR, is still to be investigated. The existence of high velocity molecular outflows in the region could
also affect the abundance of some molecules such as HCN and the sulfuretted species. 
  
  \item Within the PDR component, the comparison of the fractional abundances measured in Mon~R2 with those in the prototypical PDRs shows that the 
 positions IF and MP1 present a chemistry similar to that in high-UV PDRs (G$_0$$>$10$^3$ Habing fields)
while the chemistry in the position MP2 resembles that of the Horsehead ( $<$10$^3$ Habing field). Chemical models predict that  [CO$^+$]/[HCO$^+$] and 
[HCO]/[HCO$^+$] ratios are  good diagnostics to differentiate between these two types of PDRs. 
   
 \item The deuterated species   DCN and C$_2$D are detected in our spectral survey.  The observed deuteration fractionations,
  [DCN]/[HCN]$\sim$0.03 and [C$_2$D]/[C$_2$H]$\sim$0.05, are among the highest in warm regions.

   \end{itemize}

One important question is whether these stationary models are adequate for PDRs associated with rapidly evolving star forming regions. The time to reach the equilibrium for each
species at a given A$_V$ depends
on the density and the UV radiation (see e.g Bayet et al. 2009). The higher the density and incident UV radiation, the fastest the equilibrium is reached at a given A$_v$. 
For the physical conditions in Mon~R2, simple species are expected to have reached the  equilibrium in $\sim$10$^5$~yr which
explains the success of our chemical diagnostics.
%We cannot discard, however, that small departures from 
%equilibrium especially in regions located at high extinctions  and with low incident UV field.
However, we cannot exclude small deviations from equilibrium especially in those regions at high extinction and with low incident UV field. 
In following works, we will investigate the role of grain surface chemistry and time dependent effects on the chemistry of the molecular gas in Mon~R2.

\begin{acknowledgements}
We are grateful to the IRAM staff for their great help during the observations and data reduction. This paper was partially
supported within the programme CONSOLIDER INGENIO 2010, under
grant "Molecular Astrophysics: The Herschel and ALMA Era.- ASTROMOL"
(Ref.: CSD2009-00038). J.C. and J.R.G. thank the Spanish MICINN for funding support through grants
AYA2006-14876 and AYA2009-07304. J.R.G. is supported by a Ram\'on y Cajal research contract. Part of this work was supported by the Deutsche
Forschungsgemeinschaft, project number Os 177/1-1.
\end{acknowledgements}

{}

\Online
\onecolumn
\begin{appendix}
\input{table-paolo.tex}

\begin{table*}
\caption{Summary of recombination lines}
\begin{tabular}{llcccc} 
\hline 
\multicolumn{6}{c}{IF (0$\arcsec$,0$\arcsec$)} \\ 
\multicolumn{1}{c}{Freq(MHz)} & \multicolumn{1}{l}{Line} &
\multicolumn{1}{c}{I(K km~s$^{-1}$)} & \multicolumn{1}{c}{v$_{lsr}$(km~s$^{-1}$)} & 
\multicolumn{1}{c}{$\Delta$v(km~s$^{-1}$)} & \multicolumn{1}{c}{T$_{MB}$(K)} \\ \hline
83801.829 & H(66)$\delta$ & 1.23 (0.11) & 13.3(1.3) & 27.8 (2.9) & 0.04 \\
84914.394 & H(60)$\gamma$ & 2.26 (0.12) & 11.8 (0.8) & 30.7 (1.8) & 0.07 \\
85688.390 & H(42)$\alpha$  & 15.97 (0.13) & 12.4 (0.1)  &  30.9 (0.28) & 0.48 \\
85731.144 & C(42)$\alpha$  & 0.23 ( 0.06)   & 12.2 (1.6)  &  9.0 ( 2.9)  &  0.02 \\
86488.417 & H(70)$\epsilon$ & 0.95 (0.12) & 14.8 (2.6) & 36.9 (5.5) & 0.02 \\
86689.621 & H(74)$\zeta$ & 0.99 (0.17) & 8.2 (2.7) & 32.5 (8.0) & 0.03 \\
87614.996 & H(65)$\delta$ & 1.86 (0.08) & 9.7 (0.7) & 30.3 (1.5) & 0.06 \\
88405.687 & H(52)$\beta$  & 6.77 (0.08) & 9.3 (0.2) & 34.2 (0.5) & 0.19 \\
89198.545 & H(59)$\gamma$ & 2.50 (0.21) & 7.2 (1.1)$^a$ & 26.8 (2.5) & 0.09\\
90164.065 & H(73)$\zeta$$^b$   &  0.63(0.10)   &  9.8(2.3)    &  31.0(7.0)       &  0.02     \\
90174.339 & H(69)$\epsilon$$^b$ & 0.98 (0.12)&  8.4 (1.3)   & 32.0 (3.0)   & 0.03 \\
91663.133 & H(64)$\delta$ & 1.86 (0.11) & 9.0 (1.0) & 33.2 (2.0) & 0.05 \\
92034.434 & H(41)$\alpha$  & 20.45 (0.17) &  9.6 (0.2) & 30.8 (0.3) & 0.62 \\
92080.355 & C(41)$\alpha$$^t$  & 0.38 (0.08) & 7.0 (1.2) & 11.7(3.1) & 0.03 \\
93607.316 & H(51)$\beta$  & 6.04 (0.10) & 9.8 (0.3) & 30.9 (0.6) & 0.18 \\
93775.871 & H(58)$\gamma$ & 2.77 (0.17) & 8.7 (0.9) & 30.5 (2.1)  & 0.08 \\
93826.790 & H(72)$\zeta$   & 0.89 (0.09) & 10.7 (1.5) & 29.6 (3.8) & 0.03 \\
94072.811 & H(68)$\epsilon$ & 1.03 (0.07) & 11.4 (0.9) & 26.8 (2.1) & 0.04 \\
103914.838 & H(56)$\gamma$ & 3.02 (0.07) & 12.6 (0.4) & 33.0 (0.9) & 0.09 \\
105301.857 & H(49)$\beta$ & 6.22 (0.10) & 12.4 (0.3) & 31.5 (0.6) & 0.19 \\
105410.216 & H(61)$\delta$ & 1.70 (0.07) & 12.7 (0.7) & 29.5 (1.4) & 0.05 \\
106079.529 & H(69)$\zeta$  & 0.71 (0.09)) & 14.0 (1.9) & 28.8 (4.1) & 0.02 \\
106737.357 & H(39)$\alpha$ & 18.80 (0.11) & 12.53 (0.09) & 31.0 (0.2) & 0.57 \\
107206.108 & H(65)$\epsilon$ & 1.23 (0.12) & 10.31 (1.8) & 35.5 (3.9) & 0.03 \\
109536.001 & H(55)$\gamma$ & 3.25 (0.10) & 12.5 (0.5) & 33.5 (1.2) & 0.09 \\
110600.675 & H(60)$\delta$ & 1.66 (0.13) & 10.4 (1.3) & 34.0 (3.2) & 0.04 \\
209894.050 & H(44)$\gamma$ & 2.37 (0.16) & 13.3 (1.1) & 31.6 (2.6) & 0.07 \\
210501.771 & H(31)$\alpha$$^c$  & 15.18 (0.20) & 13.7 (0.3) & 33.0 (0.4) & 0.43 \\
\hline
\multicolumn{6}{c}{MP1 (15$\arcsec$,-15$\arcsec$)} \\ 
\multicolumn{1}{c}{Freq(MHz)} & \multicolumn{1}{l}{Line} &
\multicolumn{1}{c}{I(K km~s$^{-1}$)} & \multicolumn{1}{c}{v$_{lsr}$(km~s$^{-1}$)} & 
\multicolumn{1}{c}{$\Delta$v(km~s$^{-1}$)} & \multicolumn{1}{c}{T$_{MB}$(K)} \\  \hline
83801.829 & H(66)$\delta$ & 0.24 (0.06) & 16.1(2.4) & 17.0 (4.9) & 0.01 \\
84914.394 & H(60)$\gamma$ & 0.70 (0.04) & 15.8 (0.9) & 31.5 (1.8) & 0.02\\
85688.390 & H(42)$\alpha$  & 4.08 (0.13) & 15.3 (0.5) & 29.7 (1.2) & 0.13 \\
85731.144 & C(42)$\alpha$ & 0.22 (0.03) & 8.7 (0.7) & 12.0 (1.7) & 0.02\\
87614.996 & H(65)$\delta$ & 0.41 (0.09) & 10.0 (4.2) & 35.7 (8.5) & 0.01\\
88405.687 & H(52)$\beta$  & 1.01 (0.07) & 14.2 (0.9) & 27.2 (2.2) & 0.03\\
91663.133 & H(64)$\delta$ & 0.52 (0.08) & 18.0 (2.3) & 26.7 (4.5) & 0.02\\
92034.434 & H(41)$\alpha$  & 4.01 (0.18) & 12.8 (0.7) & 33.9 (1.8) & 0.11\\
93607.316 & H(51)$\beta$  & 1.06 (0.08) & 14.5 (1.3) & 34.2 (3.0) & 0.03\\
93775.871 & H(58)$\gamma$ & 0.56 (0.09) & 15.6 (2.9) & 33.8 (5.8) & 0.016 \\ 
103914.838 & H(56)$\gamma$ & 0.46 (0.07) & 18.3 (2.7) & 40.1 (7.2) & 0.01 \\
105301.857 & H(49)$\beta$ & 0.66 (0.06) & 17.1 (1.3) & 27.3 (3.3) & 0.02 \\
106737.357 & H(39)$\alpha$ & 2.33 (0.07)& 16.4 (0.5) & 31.5 (1.1) & 0.07\\
\hline
\multicolumn{6}{c}{MP2 (0$\arcsec$,40$\arcsec$)} \\ 
\multicolumn{1}{c}{Freq(MHz)} & \multicolumn{1}{l}{Line} &
\multicolumn{1}{c}{I(K km~s$^{-1}$)} & \multicolumn{1}{c}{v$_{lsr}$(km~s$^{-1}$)} & 
\multicolumn{1}{c}{$\Delta$v(km~s$^{-1}$)} & \multicolumn{1}{c}{T$_{MB}$(K)} \\ \hline
85688.390 & H(42)$\alpha$  & 0.89 (0.16) & 11.0 (2.1) & 23.2 (4.8) & 0.04 \\
88405.687 & H(52)$\beta$  & 0.47 (0.06) & 9.3 (2.2) & 35.0 (5.3) & 0.01 \\
92034.434 & H(41)$\alpha$  & 1.39 (0.05) & 6.7 (0.6) & 33.4 (1.4) & 0.04\\
93607.316 & H(51)$\beta$  & 0.35 (0.08) & 6.6 (2.7) & 29.6 (10.7) & 0.01\\
105301.857 & H(49)$\beta$ & 0.39 (0.05) & 9.4 (2.4) & 33.4 (4.4) & 0.01\\
106737.357 & H(39)$\alpha$ & 0.57 (0.06) & 8.0 (1.5) & 26.6 (3.2) & 0.02\\
209998.777 & C(44)$\gamma$ & 0.51 (0.11) & 19.1 (0.9) & 8.2 (1.8) & 0.06\\
\hline
\end{tabular}

\noindent
Note: $^a$ Overlapped with the HCO$^+$ J=1$\rightarrow$0 line and self-absorbed; $^b$ these
two recombination lines are overlapped and the fit of the two each lines is  uncertain; $^t$ Doubtful detection;
$^c$ This recombination line is contaminated by the emission of C$^{18}$O J=2$\rightarrow$1 from
the image band.

\end{table*}

% 0 0
\clearpage

\begin{figure*}
\includegraphics[angle=-89.9, width=1.00000\textwidth, height=0.000925\textheight, viewport=0 0 460 690]{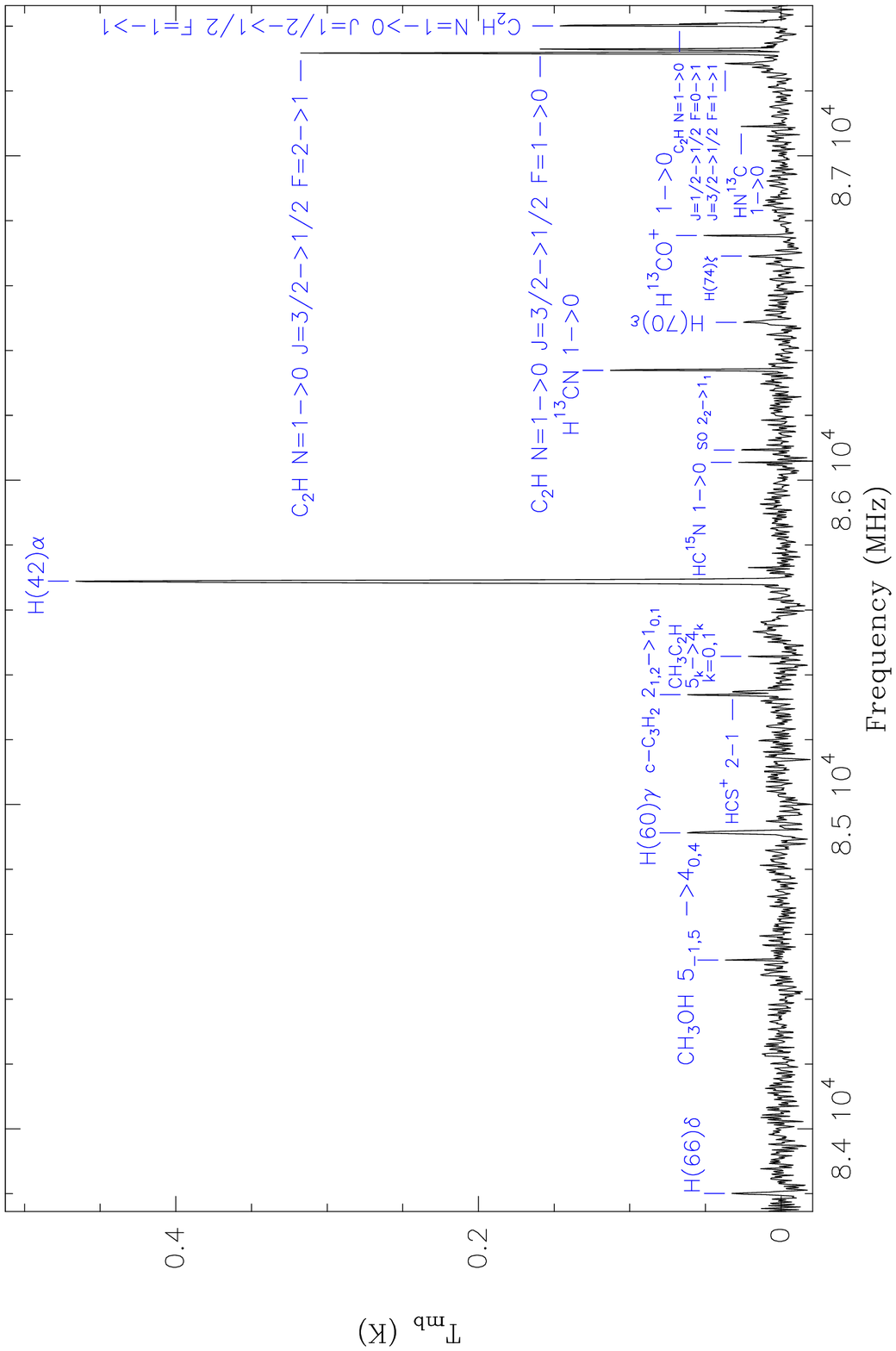}
\includegraphics[angle=-89.9, width=1.00000\textwidth, height=0.000925\textheight, viewport=0 0 460 690]{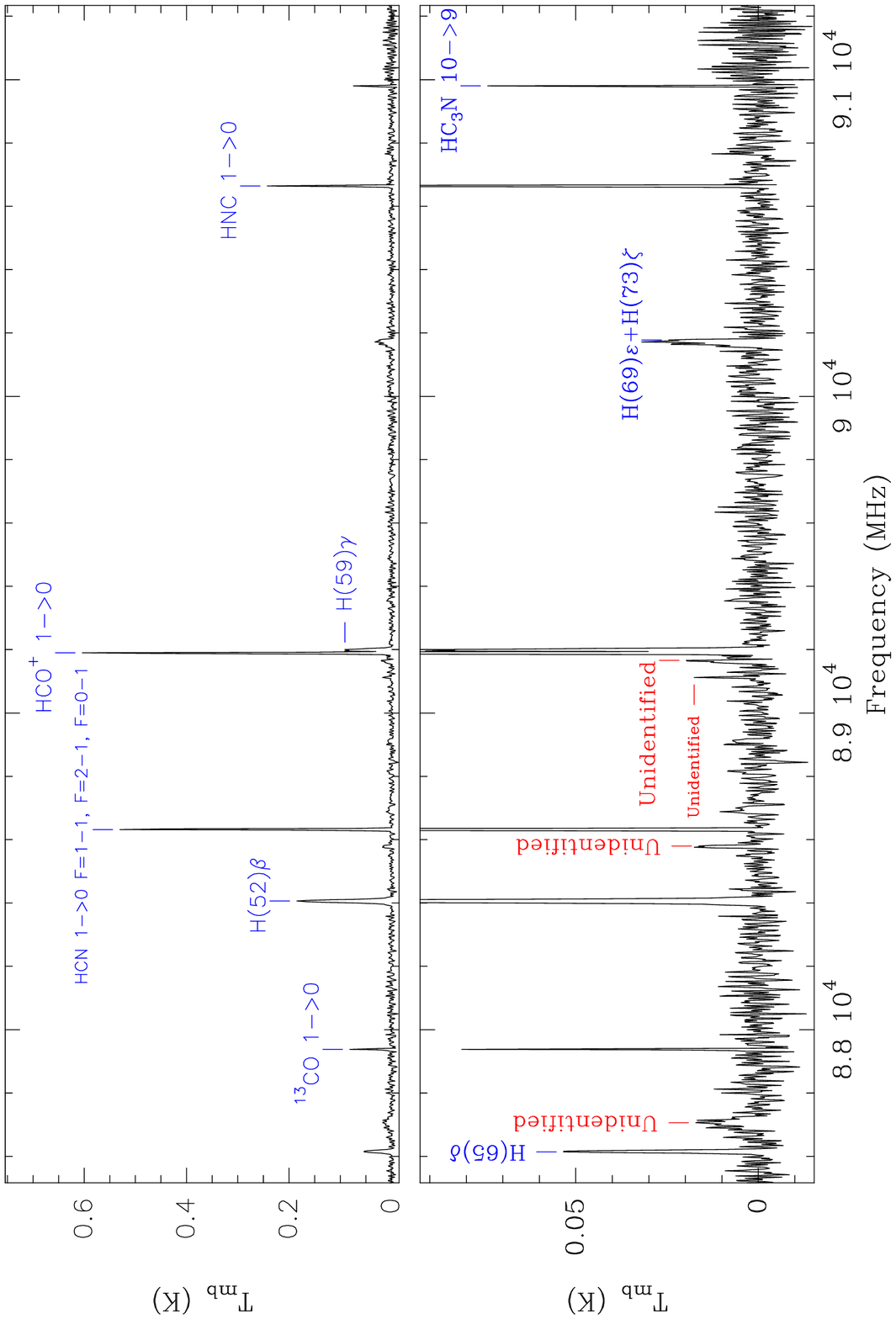}
\includegraphics[angle=-89.9, width=1.00000\textwidth, height=0.000925\textheight, viewport=0 0 460 690]{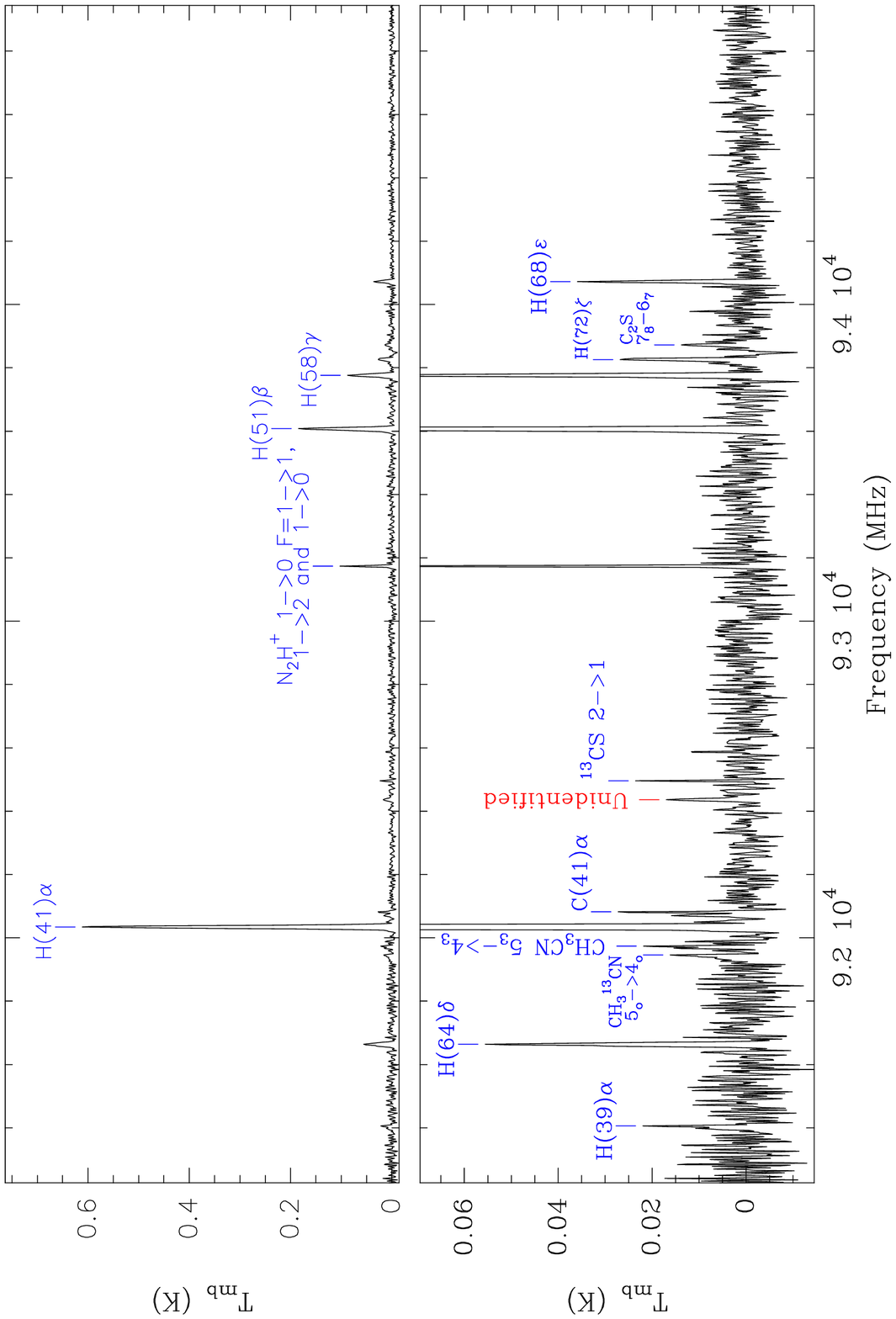}
\end{figure*}

\begin{figure*}
\includegraphics[angle=-89.9, width=1.00000\textwidth, height=0.000925\textheight, viewport=0 0 460 690]{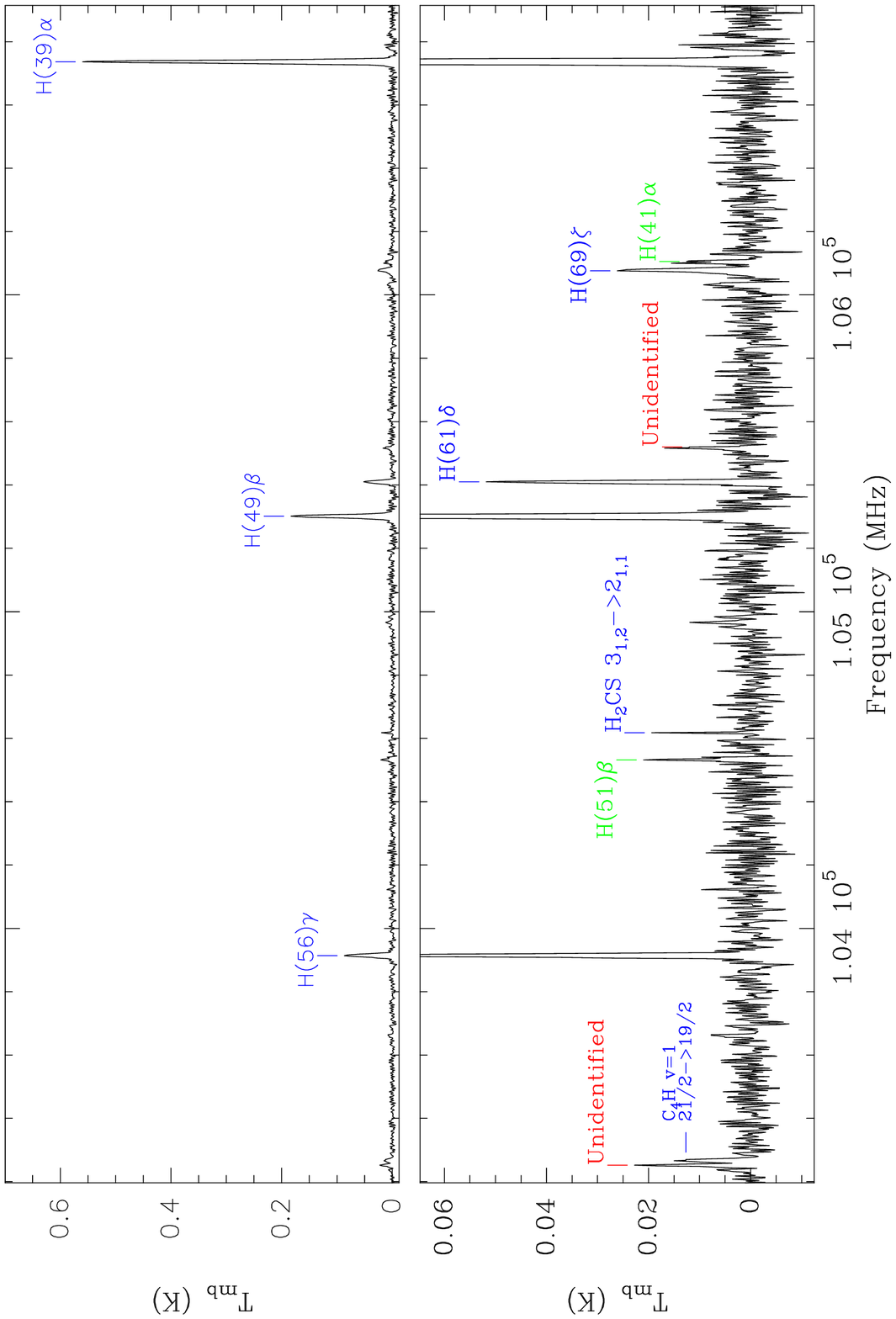}
\includegraphics[angle=-89.9, width=1.00000\textwidth, height=0.000925\textheight, viewport=0 0 460 690]{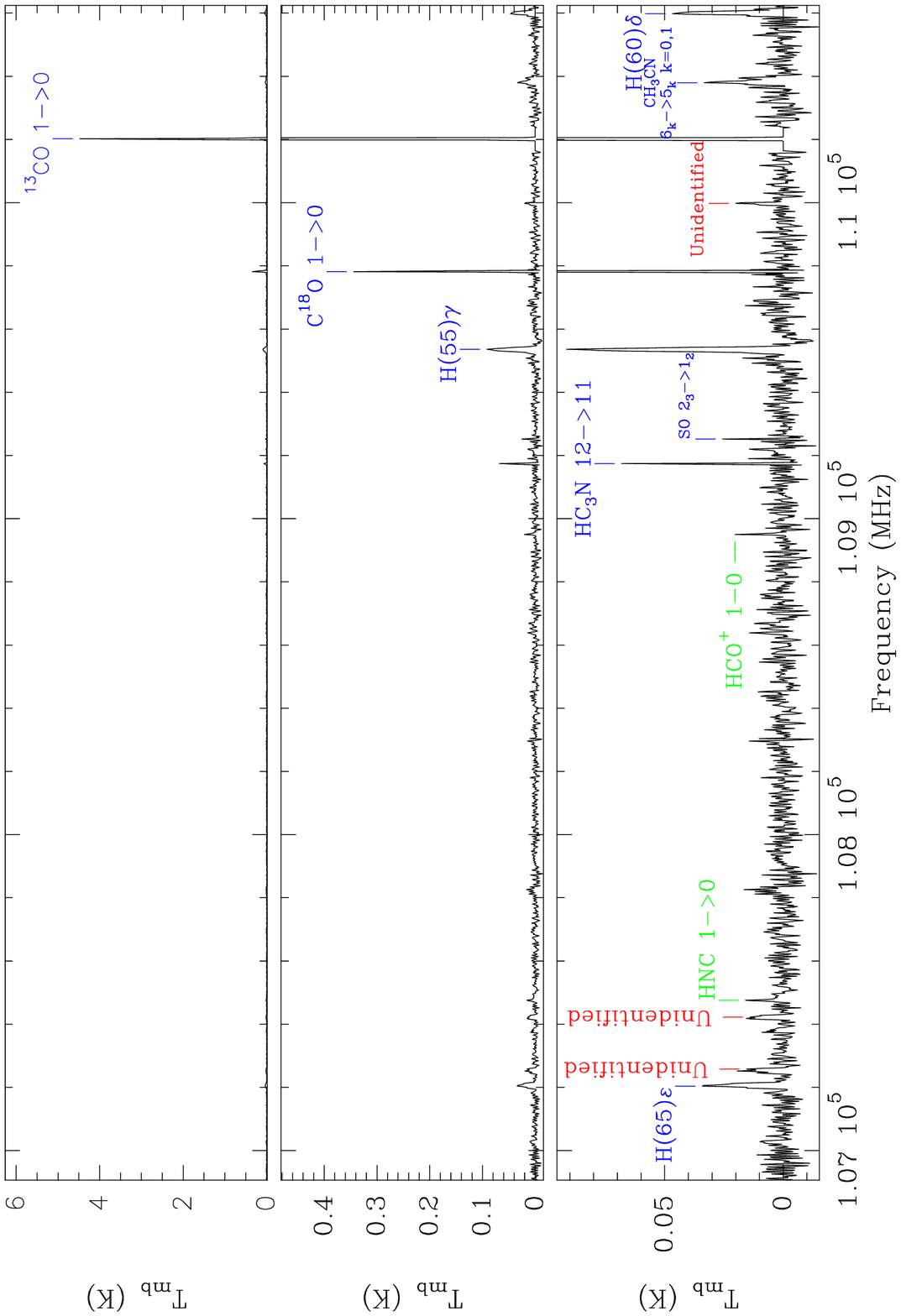}
\includegraphics[angle=-89.9, width=1.00000\textwidth, height=0.000925\textheight, viewport=0 0 460 690]{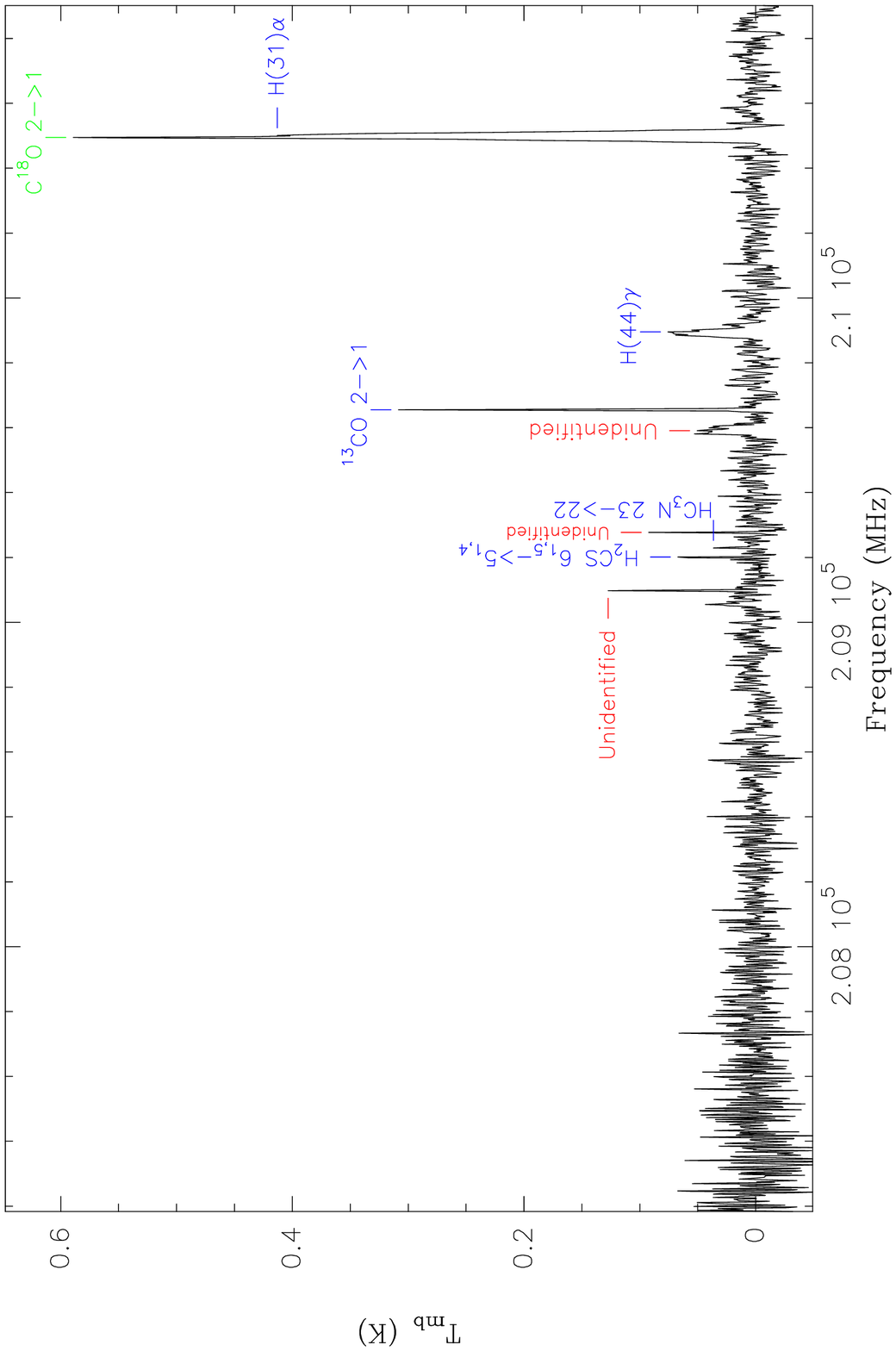}
\end{figure*}

\begin{figure*}
\includegraphics[angle=-89.9, width=1.00000\textwidth, height=0.000925\textheight, viewport=0 0 490 690]{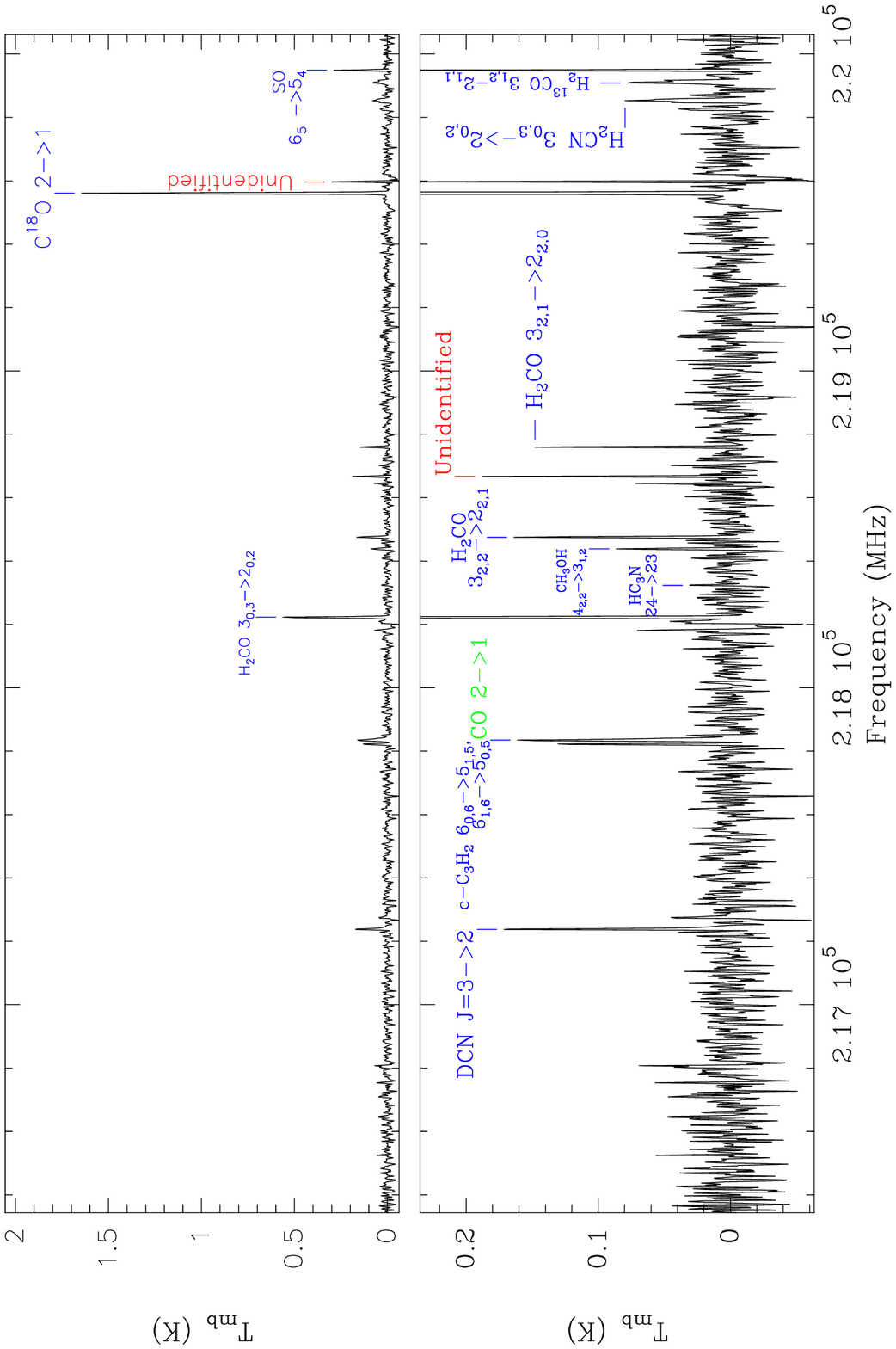}
\caption{Identified lines in offset (0$\arcsec$,0$\arcsec$). In blue lines detected in signal band, in green lines identified in image band, and in red lines detected but not identified.}
\end{figure*}

% 15 -15
\newpage

\begin{figure*}
\includegraphics[angle=-89.9, width=1.00000\textwidth, height=0.000925\textheight, viewport=0 0 460 690]{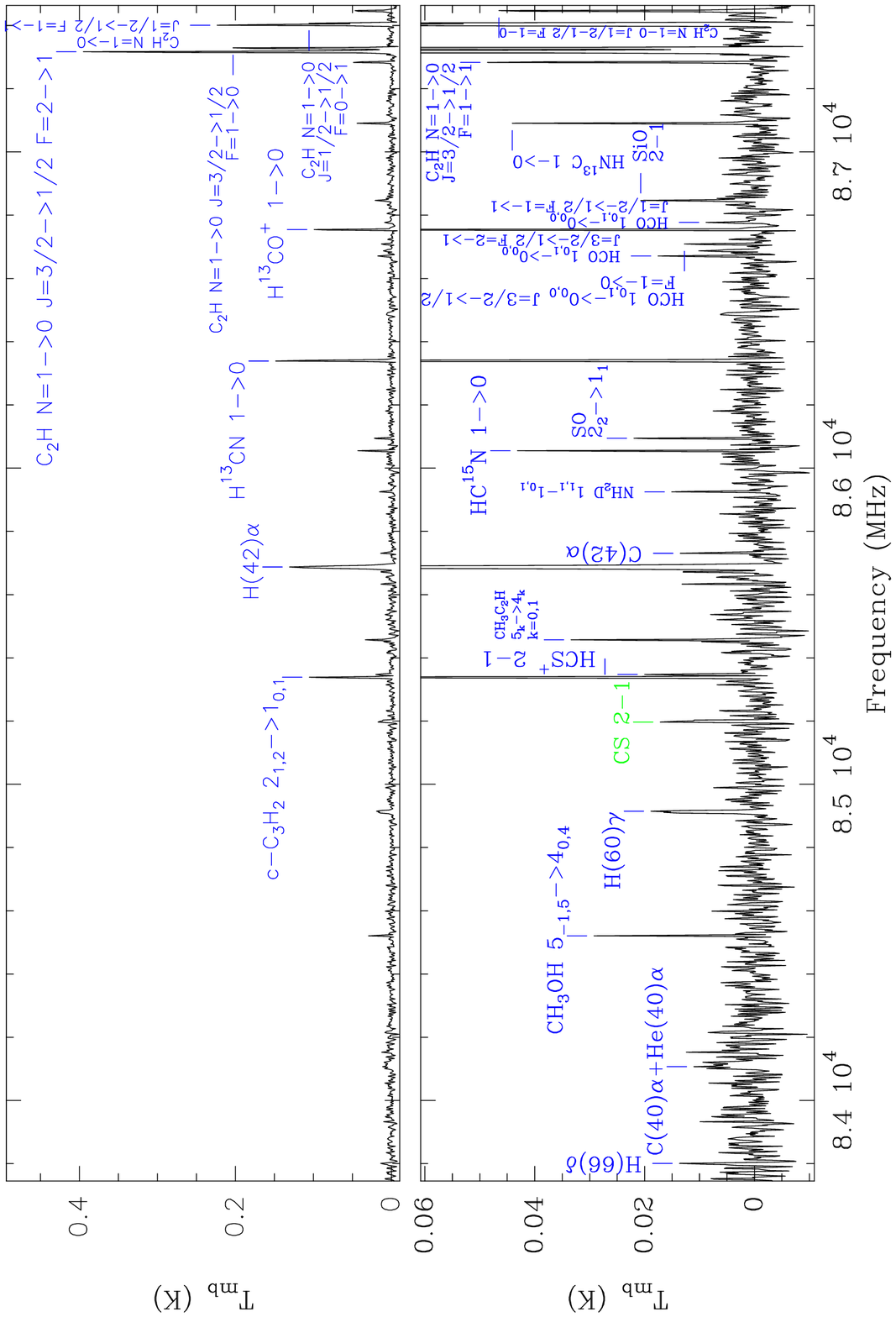}
\includegraphics[angle=-89.9, width=1.00000\textwidth, height=0.000925\textheight, viewport=0 0 460 690]{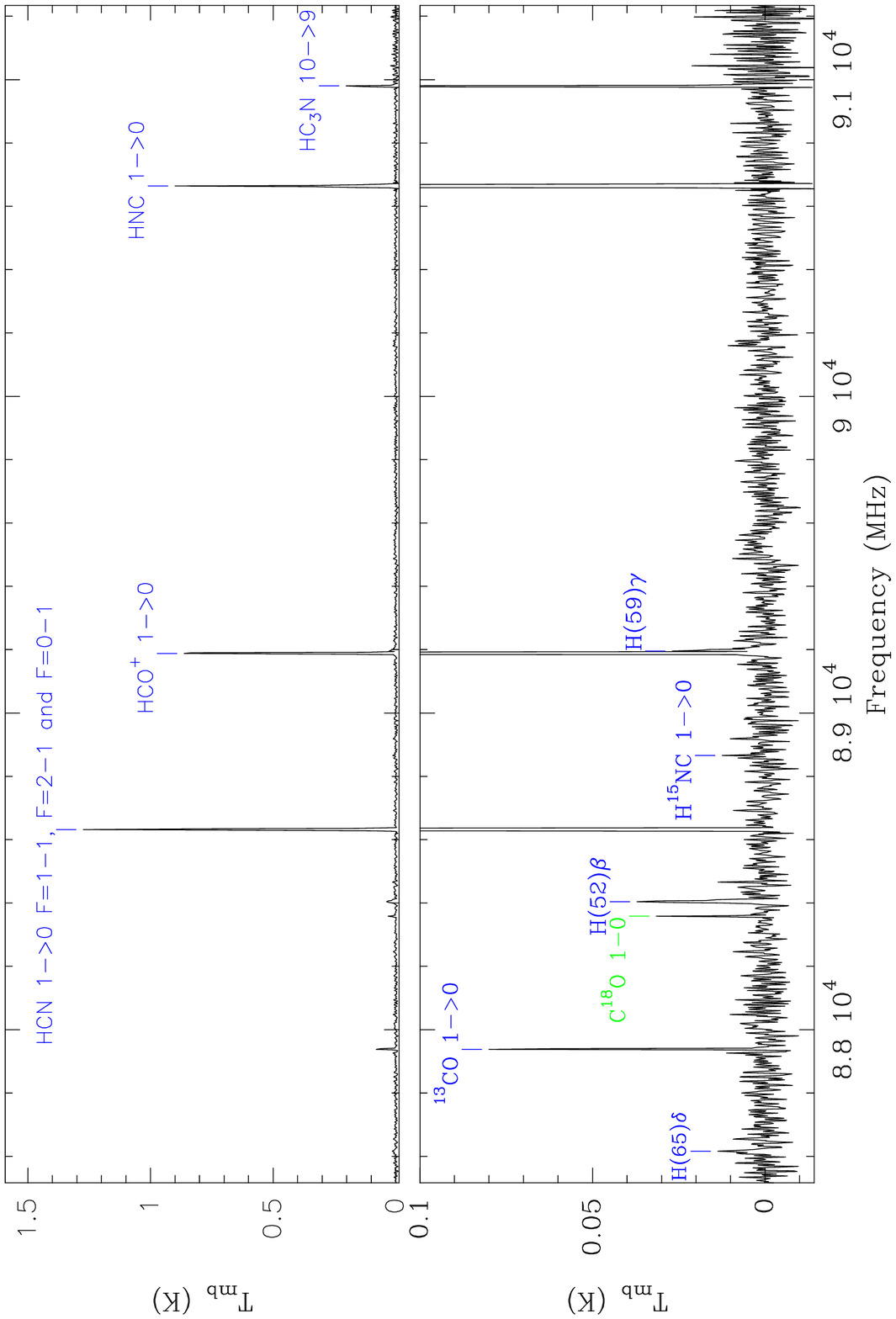}
\includegraphics[angle=-89.9, width=1.00000\textwidth, height=0.000925\textheight, viewport=0 0 460 690]{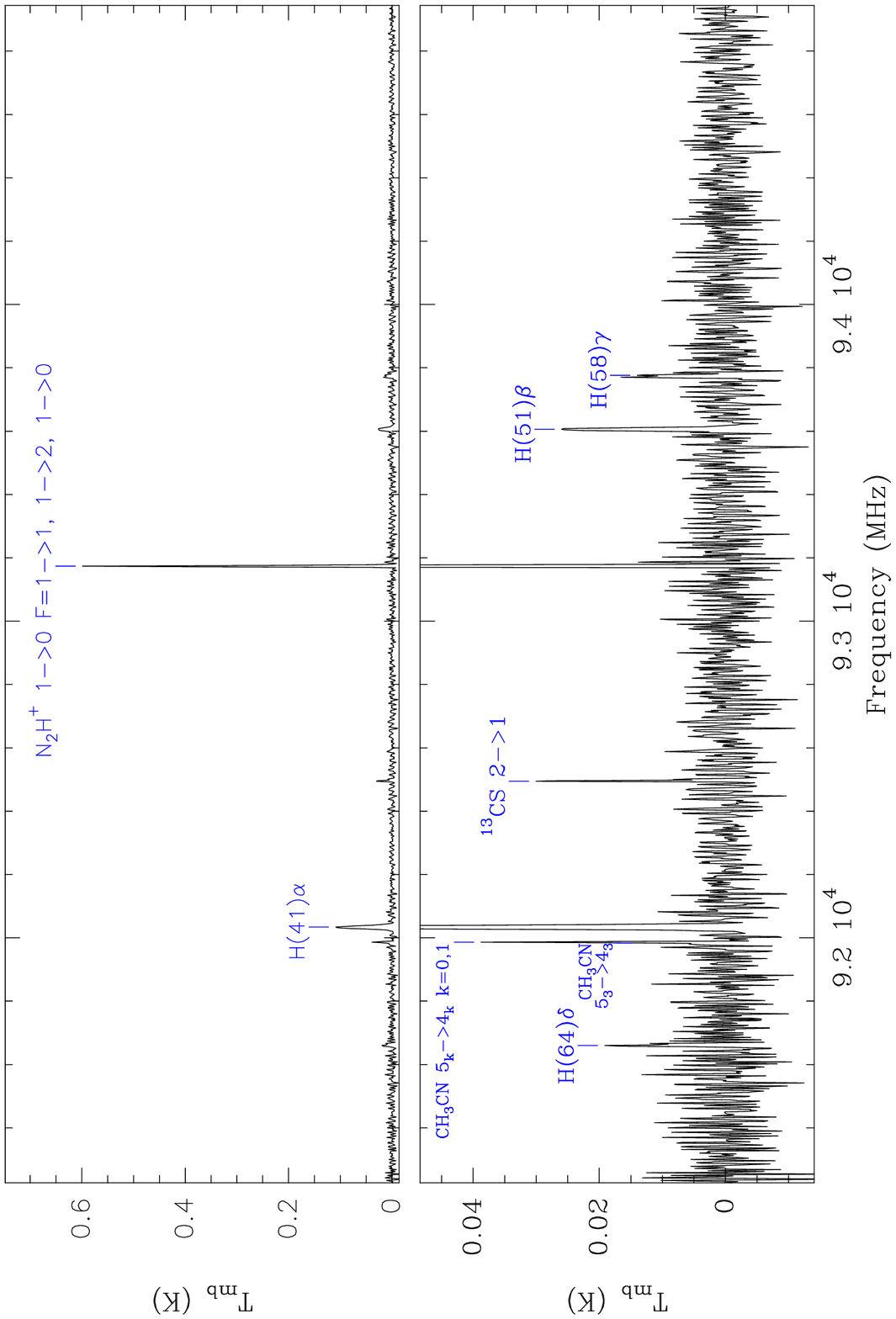}
\end{figure*}

\begin{figure*}
\includegraphics[angle=-89.9, width=1.00000\textwidth, height=0.000925\textheight, viewport=0 0 460 690]{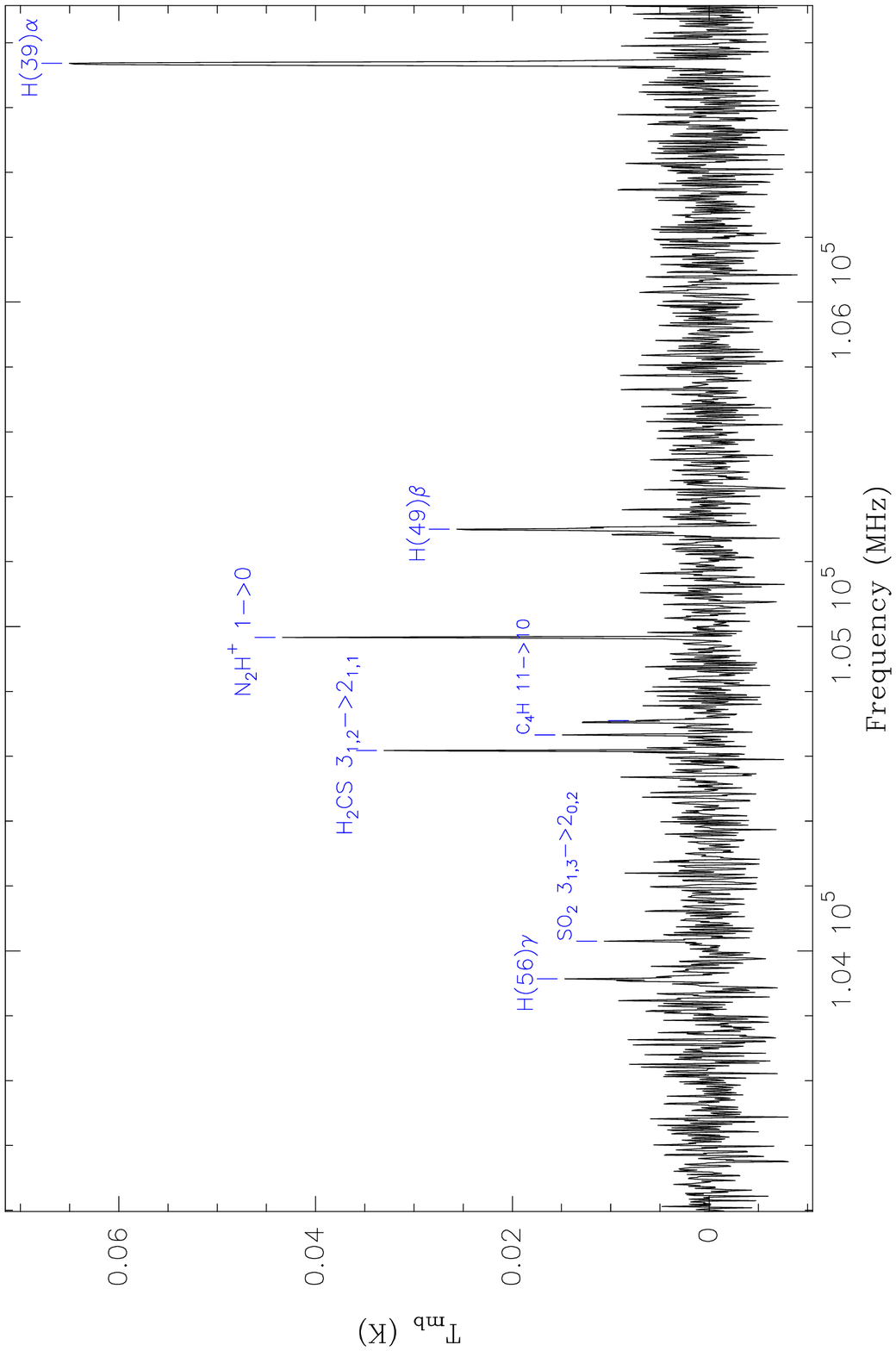}
\includegraphics[angle=-89.9, width=1.00000\textwidth, height=0.000925\textheight, viewport=0 0 460 690]{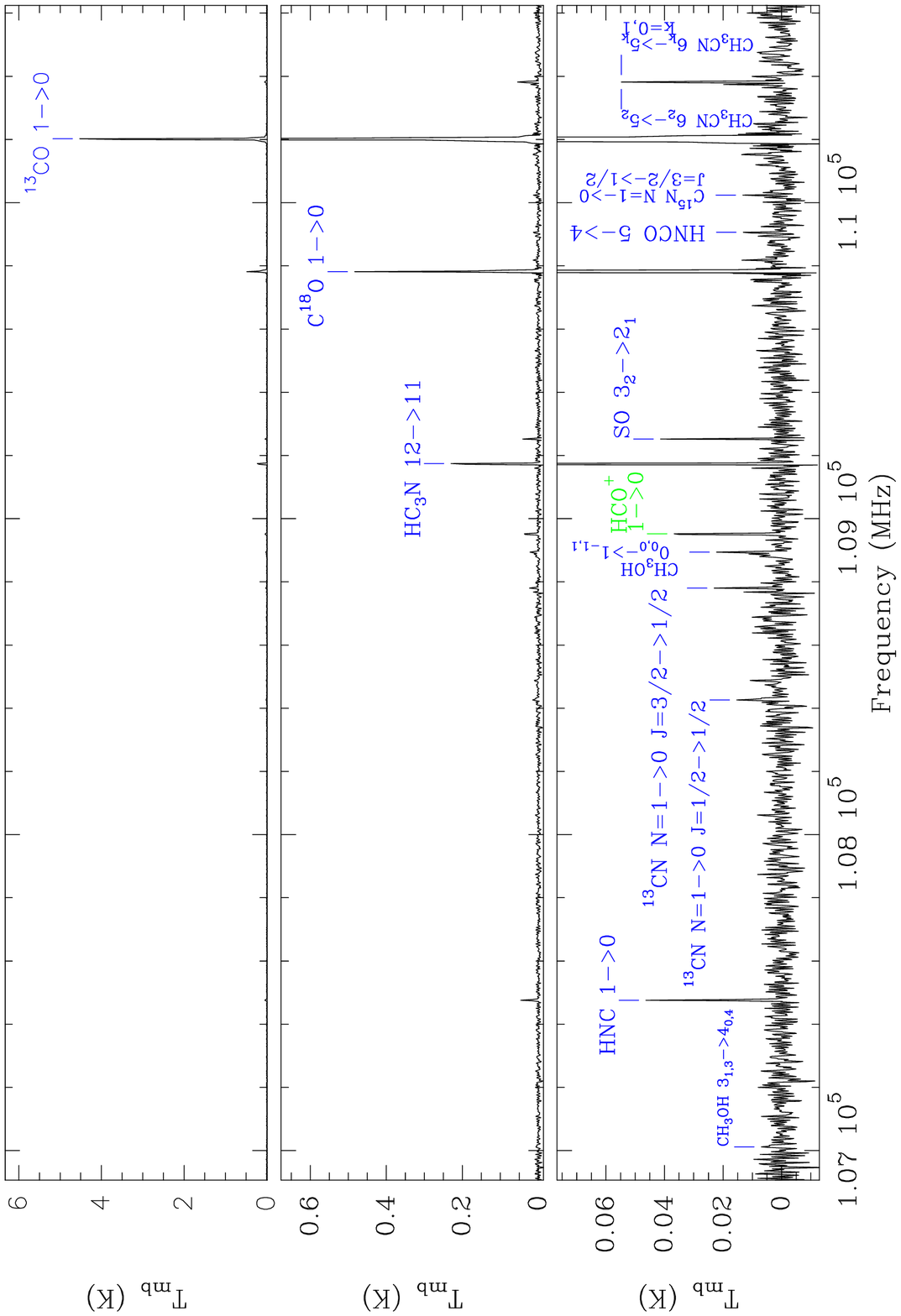}
\includegraphics[angle=-89.9, width=1.00000\textwidth, height=0.000925\textheight, viewport=0 0 460 690]{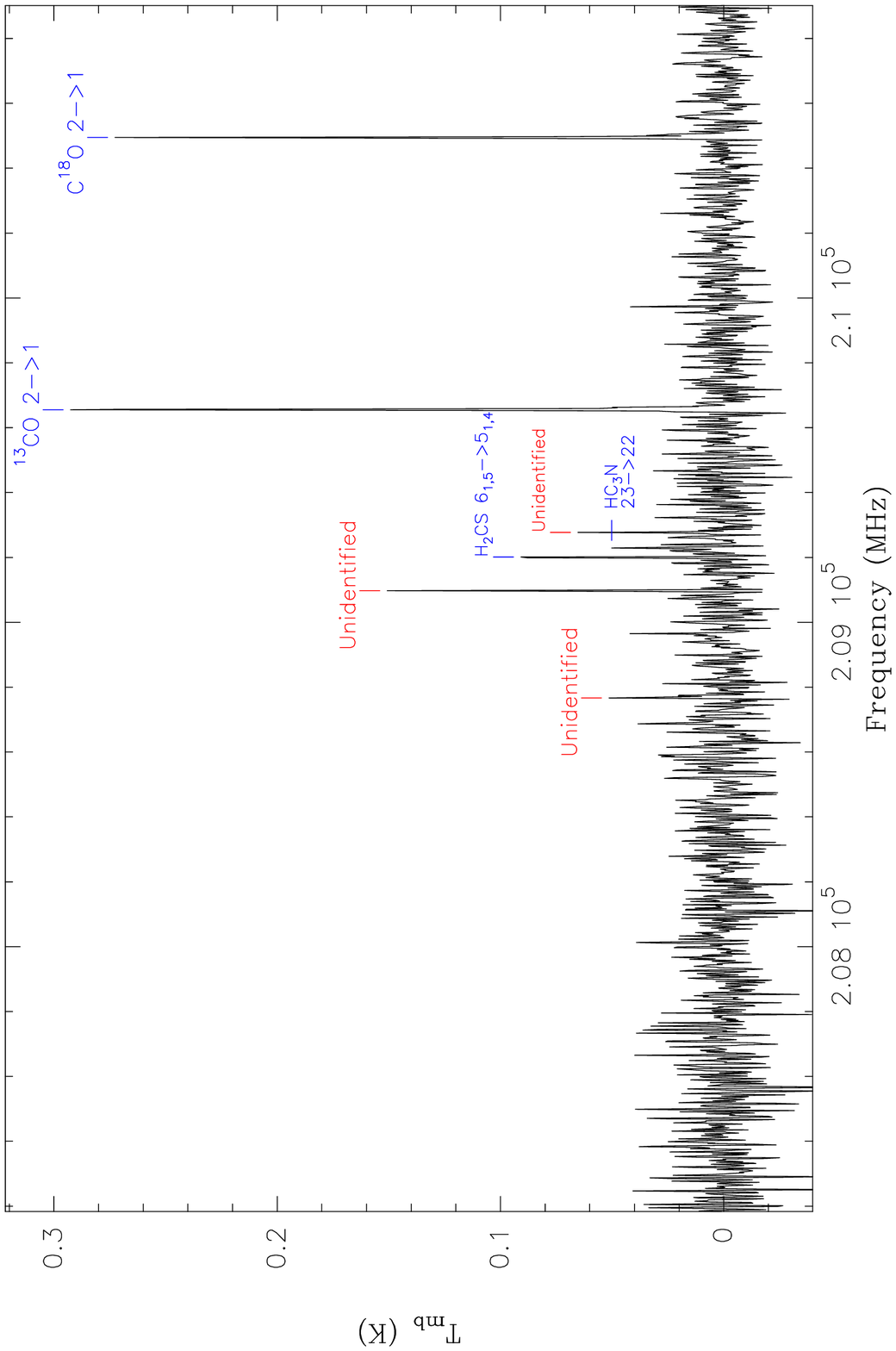}
\end{figure*}

\begin{figure*}
\includegraphics[angle=-89.9, width=1.00000\textwidth, height=0.000925\textheight, viewport=0 0 490 690]{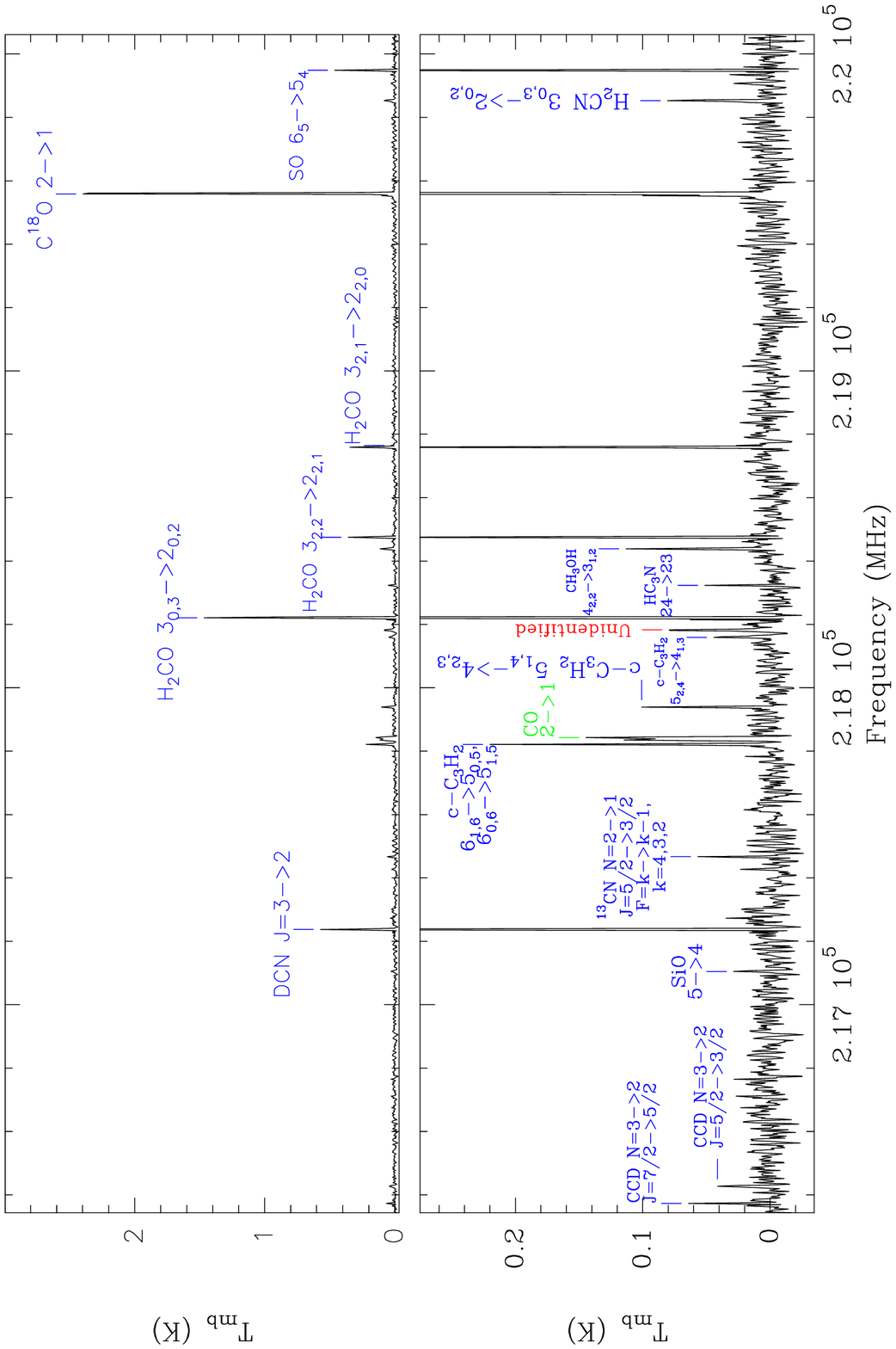}
\caption{Identified lines in offset (+15$\arcsec$,-15$\arcsec$). In blue lines detected in signal band, in green lines identified in image band, and in red lines detected but not identified.}
\end{figure*}

% 0 40
\newpage

\begin{figure*}
\includegraphics[angle=-89.9, width=1.00000\textwidth, height=0.000925\textheight, viewport=0 0 460 690]{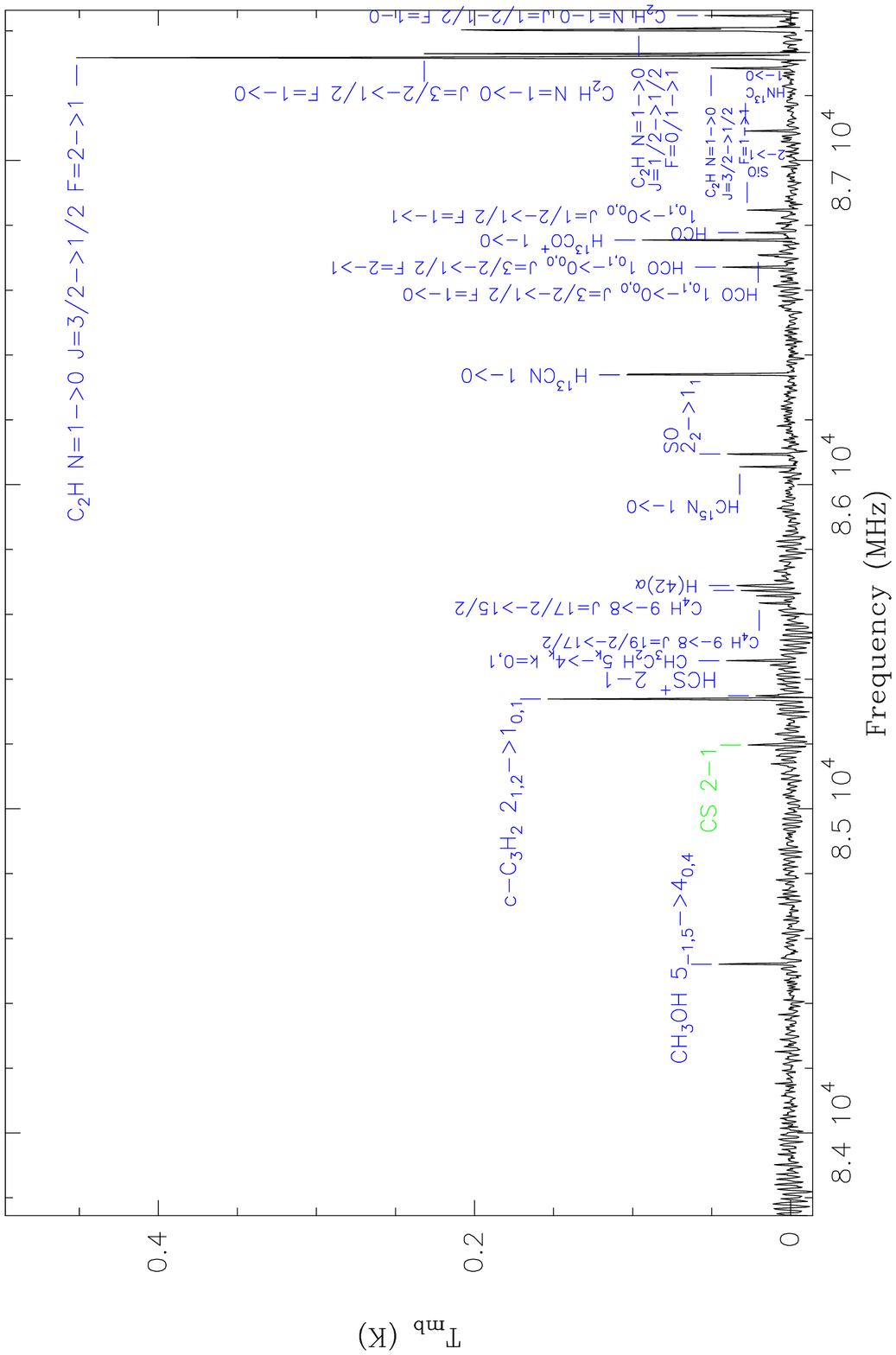}
\includegraphics[angle=-89.9, width=1.00000\textwidth, height=0.000925\textheight, viewport=0 0 460 690]{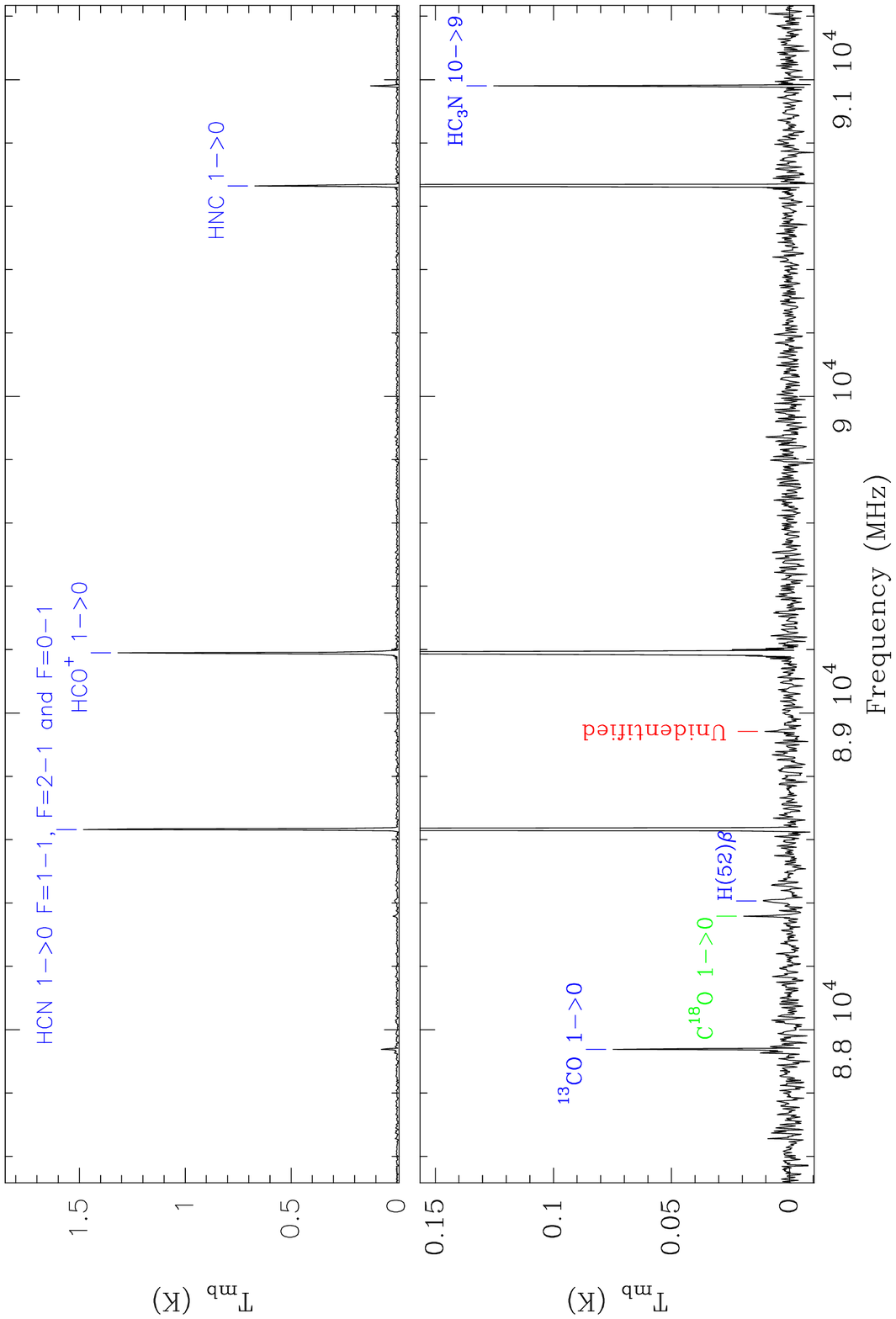}
\includegraphics[angle=-89.9, width=1.00000\textwidth, height=0.000925\textheight, viewport=0 0 460 690]{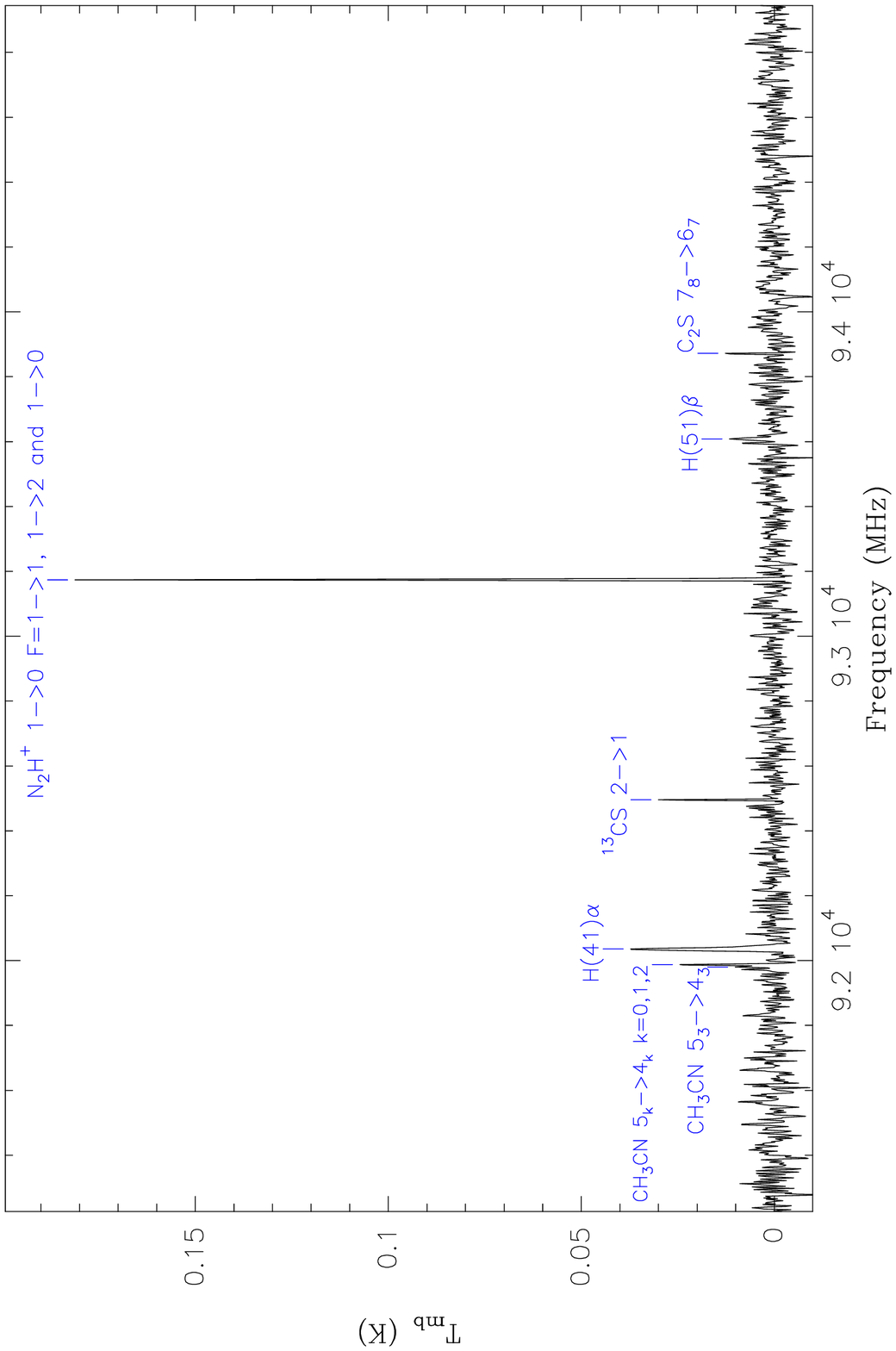}
\end{figure*}

\begin{figure*}
\includegraphics[angle=-89.9, width=1.00000\textwidth, height=0.000925\textheight, viewport=0 0 460 690]{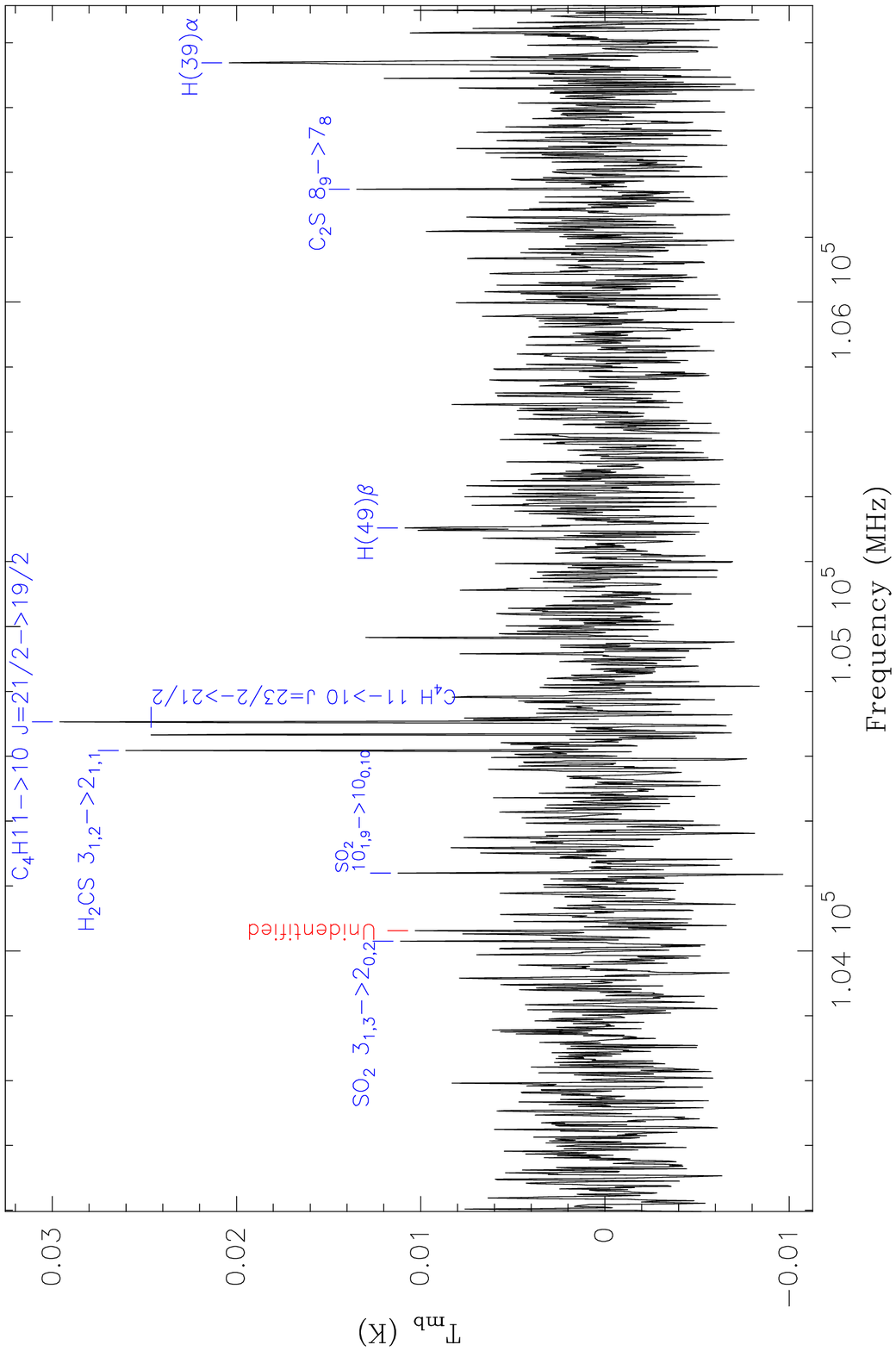}
\includegraphics[angle=-89.9, width=1.00000\textwidth, height=0.000925\textheight, viewport=0 0 460 690]{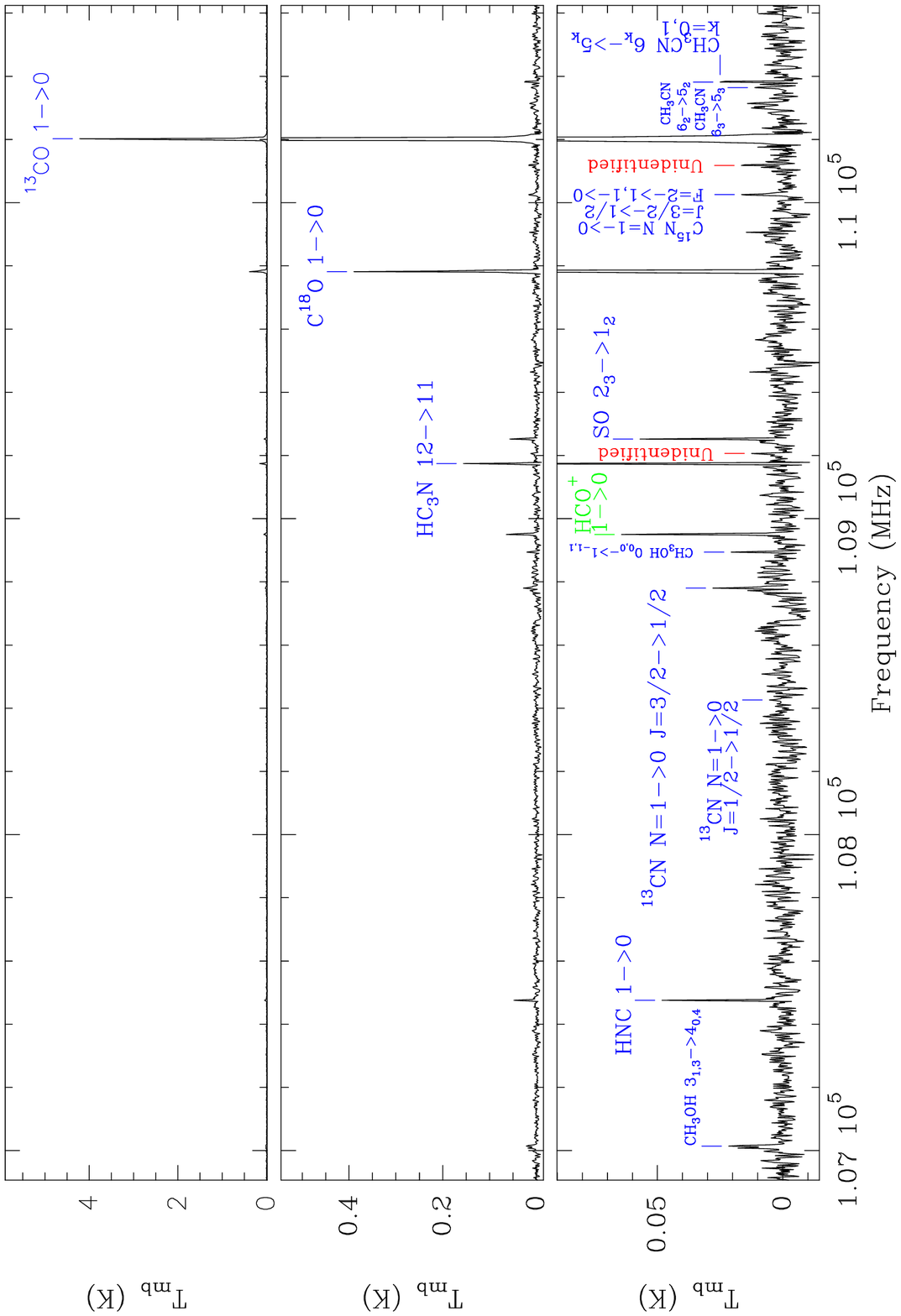}
\includegraphics[angle=-89.9, width=1.00000\textwidth, height=0.000925\textheight, viewport=0 0 460 690]{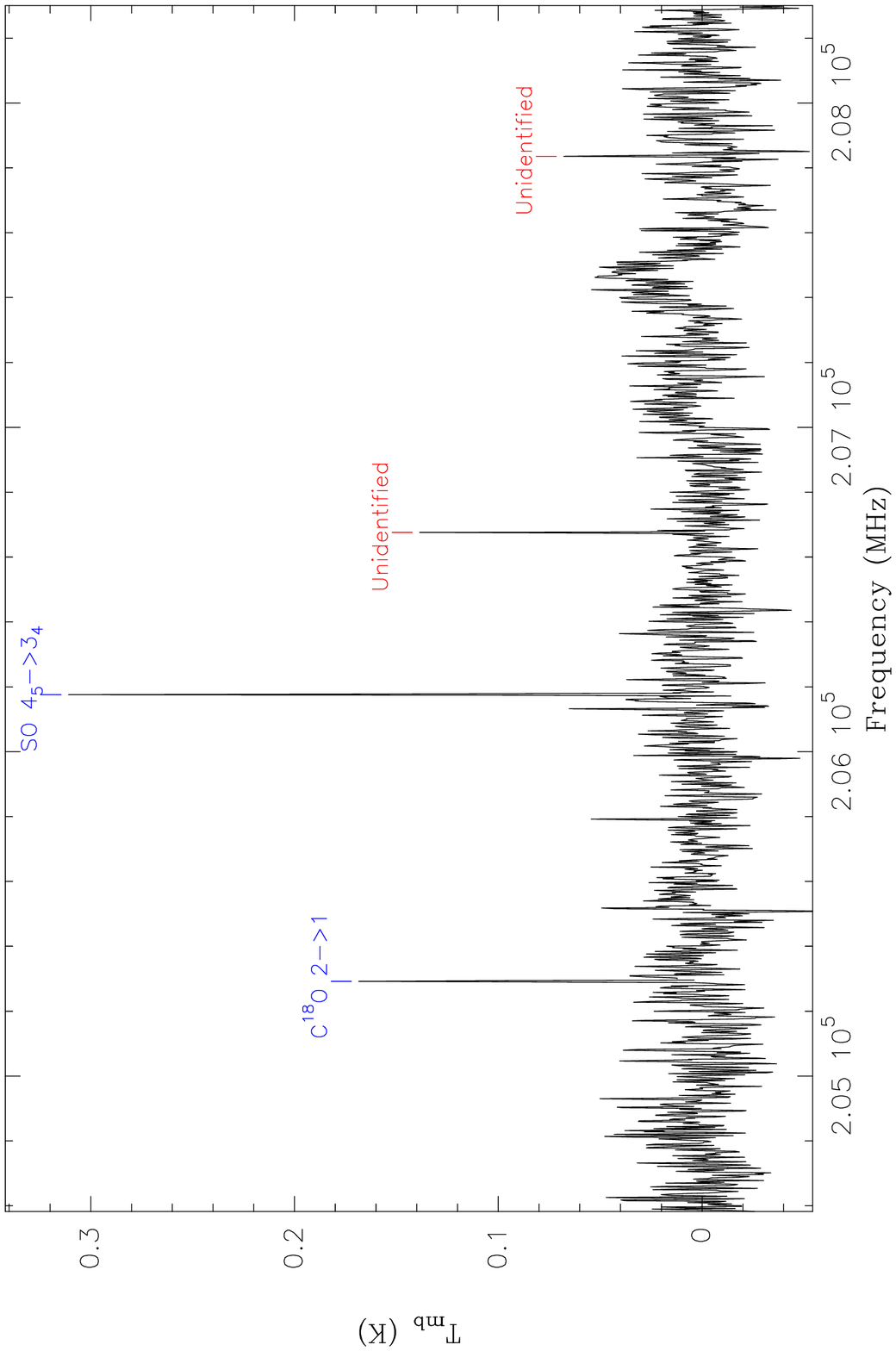}
\end{figure*}

\begin{figure*}
\includegraphics[angle=-89.9, width=1.00000\textwidth, height=0.000925\textheight, viewport=0 0 490 690]{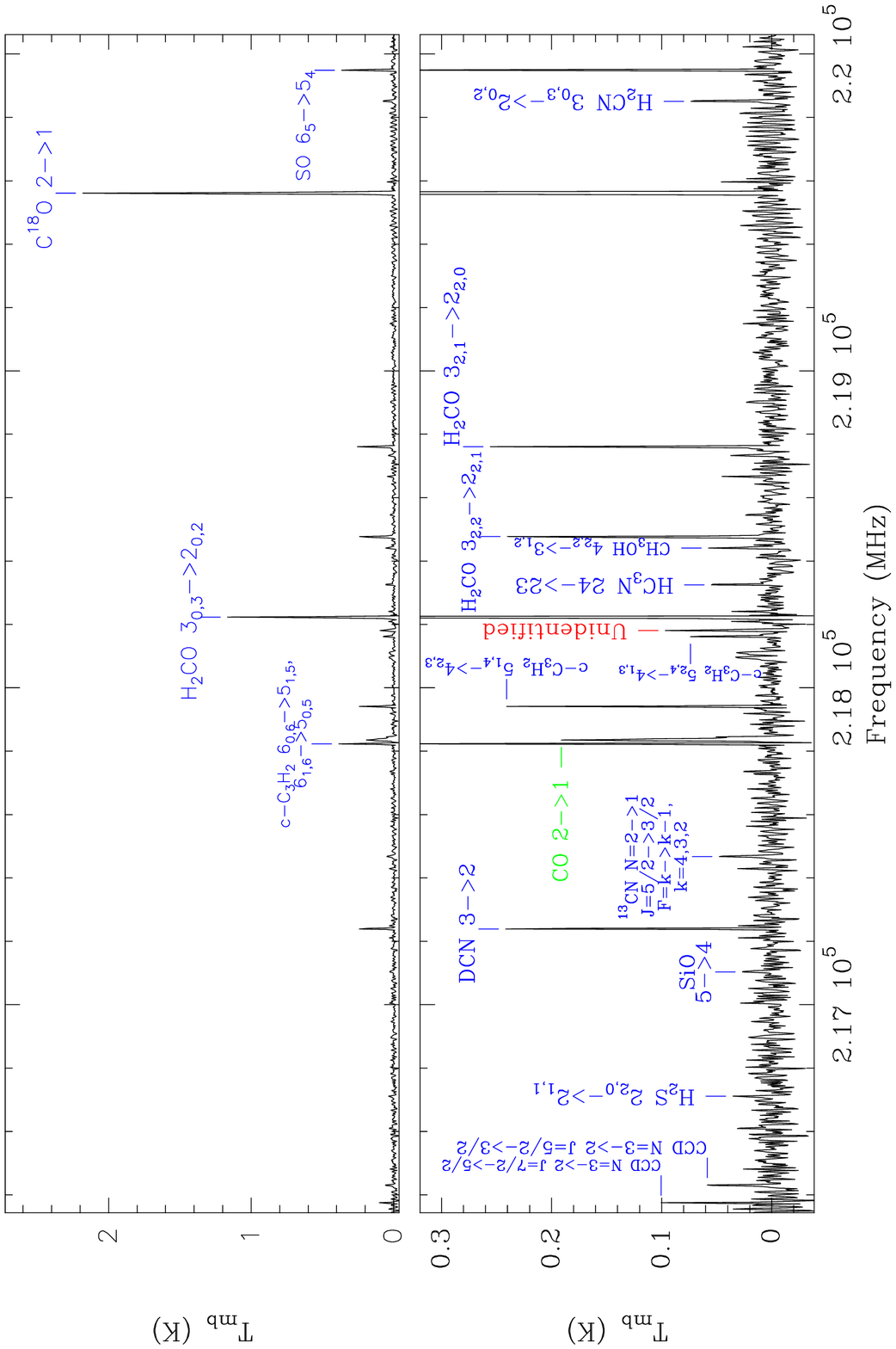}
\caption{Identified lines in offset (0$\arcsec$,40$\arcsec$). In blue lines detected in signal band, in green lines identified in image band, and in red lines detected but not identified.}
\end{figure*}
%\end{appendix}

\clearpage
%\newpage
%\begin{appendix}
\section{Rotational Diagrams}

\begin{figure*}[hb]
\includegraphics[width=0.33\textwidth]{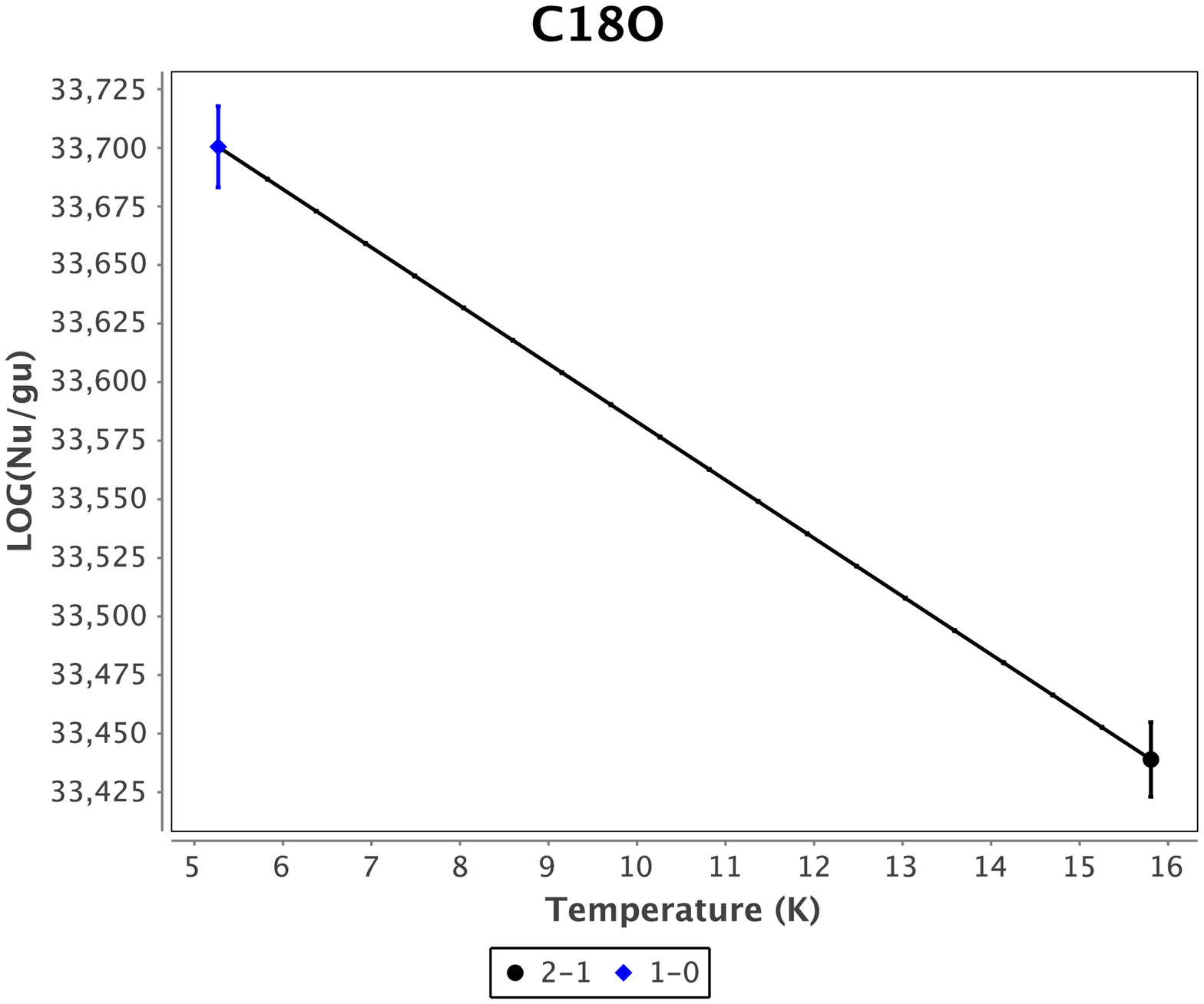}
\includegraphics[width=0.33\textwidth]{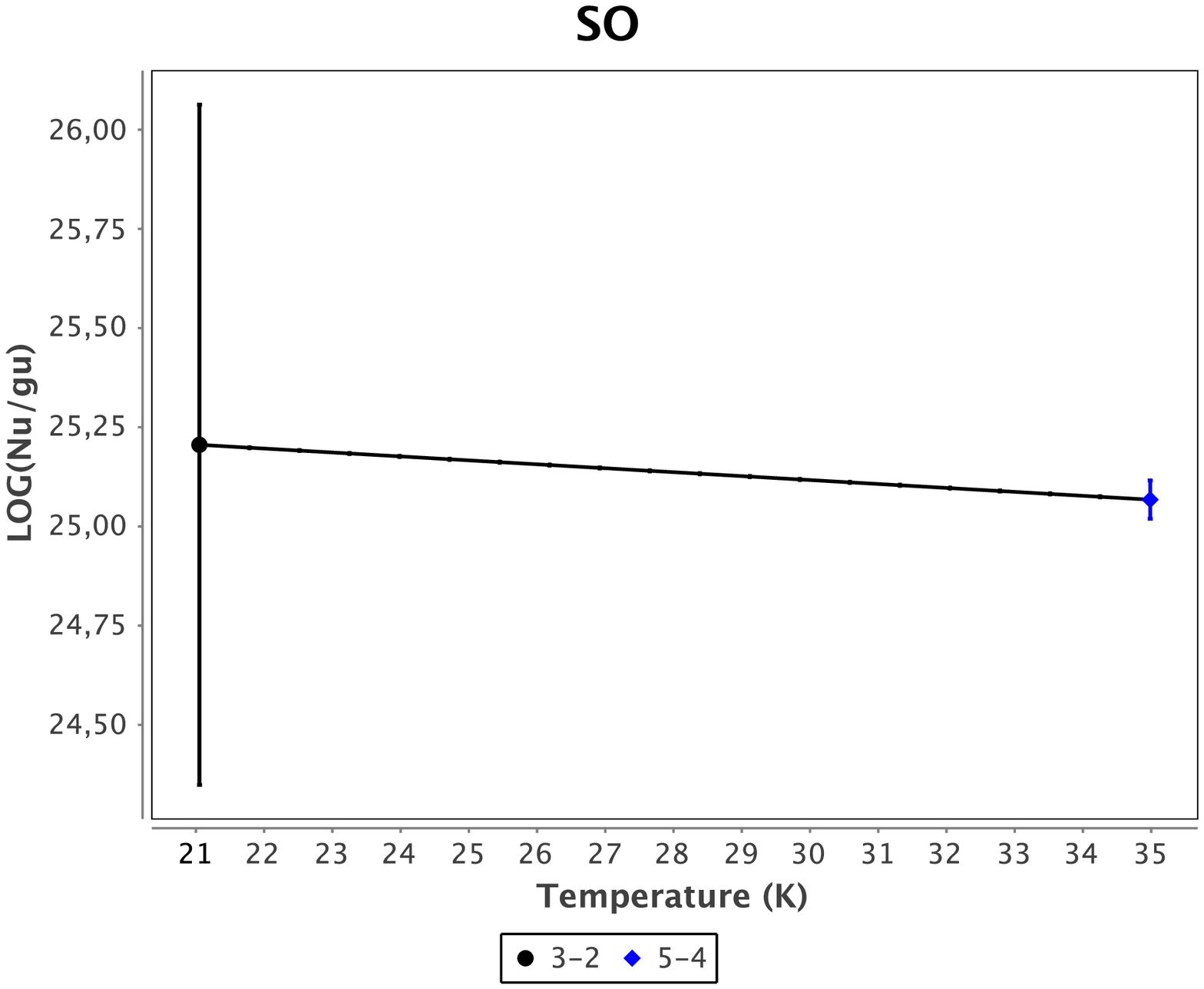}
\includegraphics[width=0.33\textwidth]{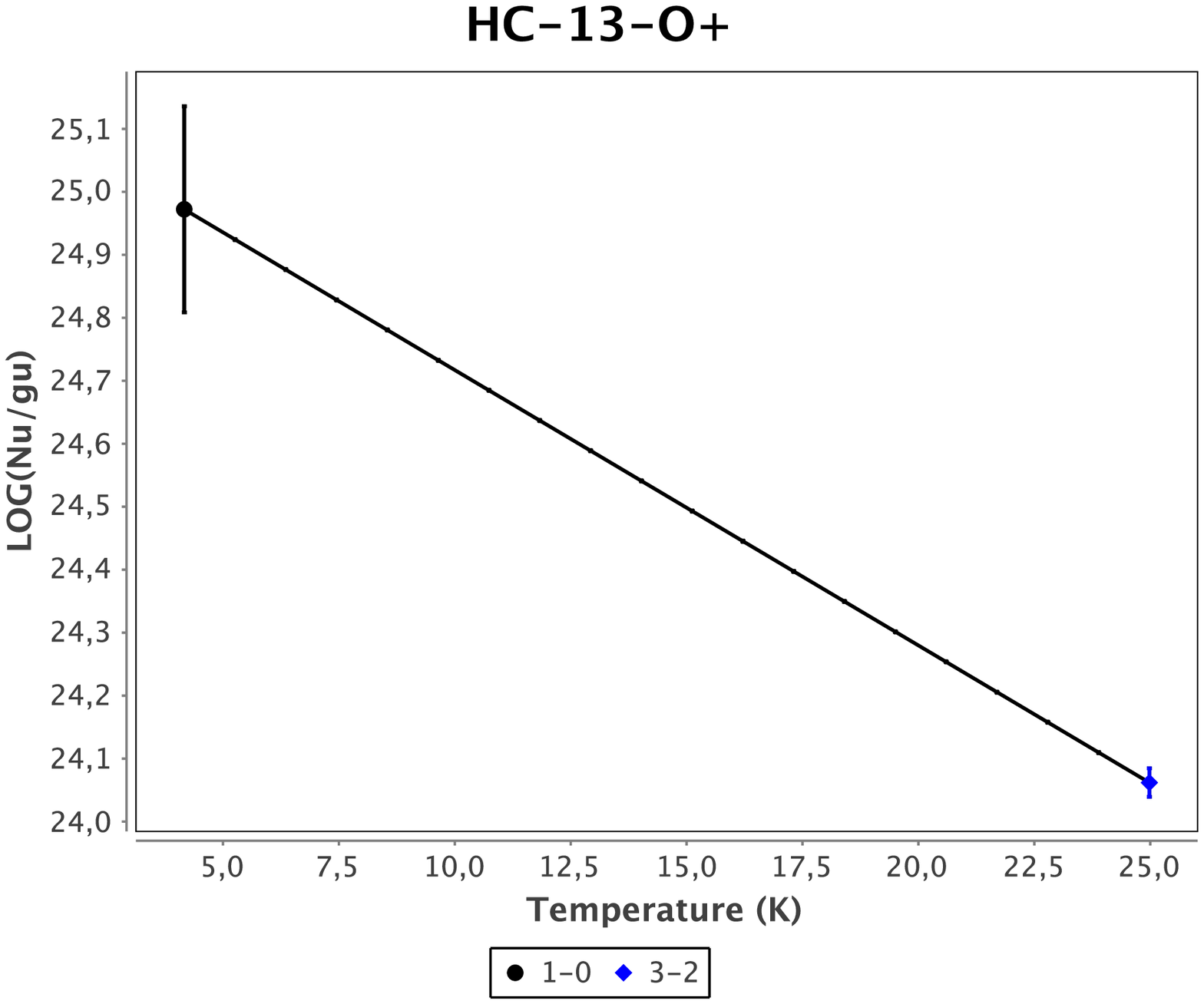}
\includegraphics[width=0.33\textwidth]{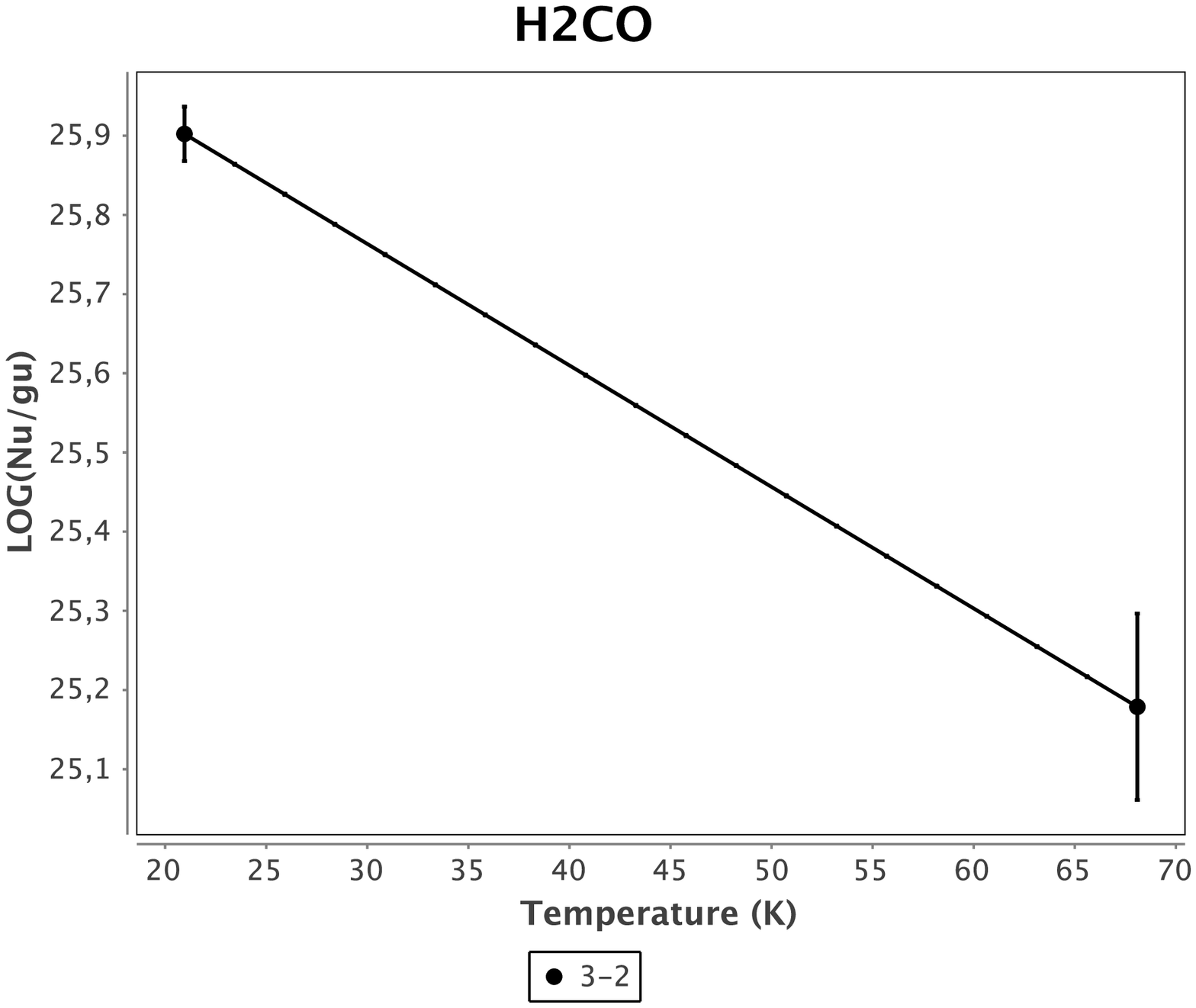}
\includegraphics[width=0.33\textwidth]{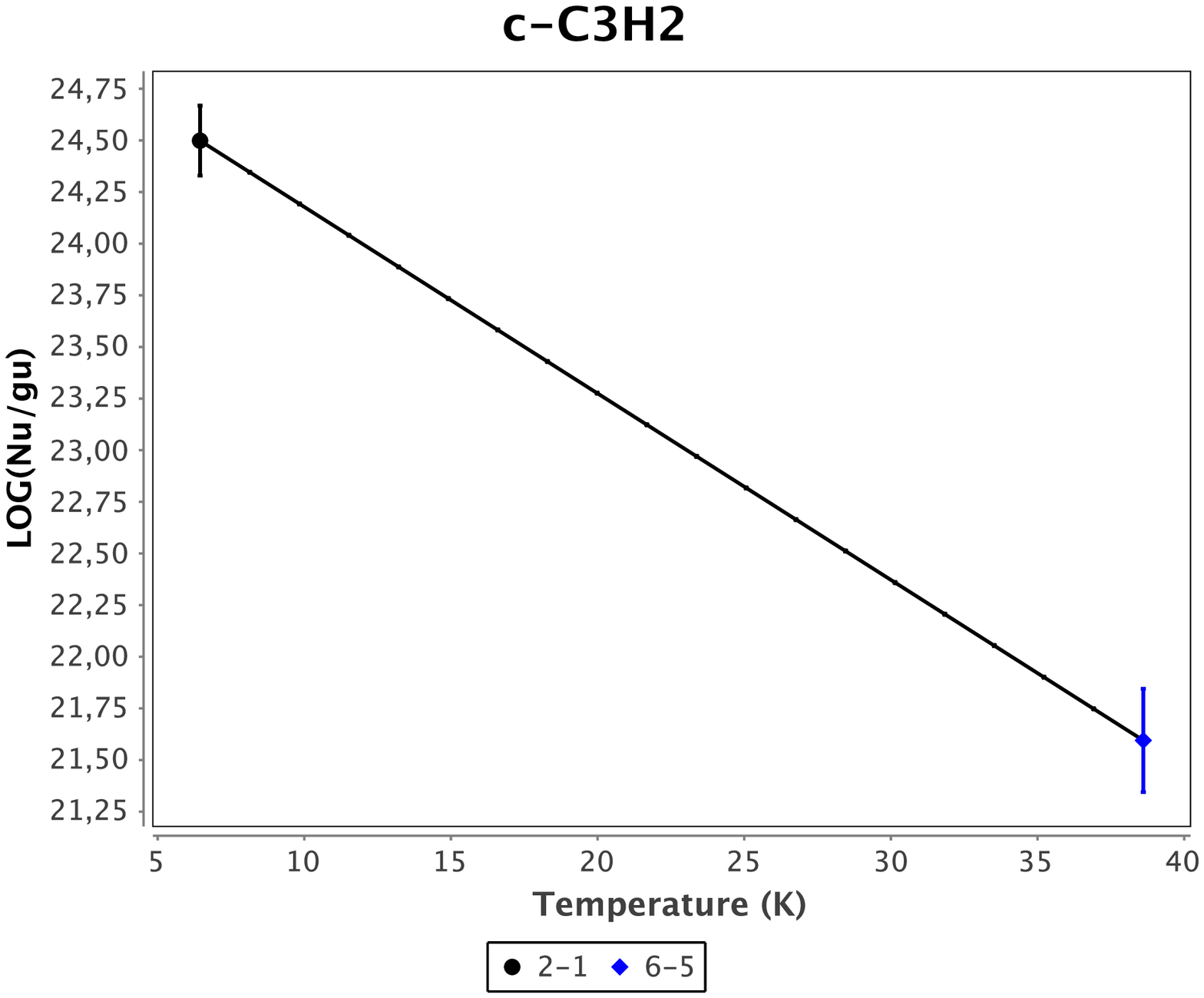}
\includegraphics[width=0.33\textwidth]{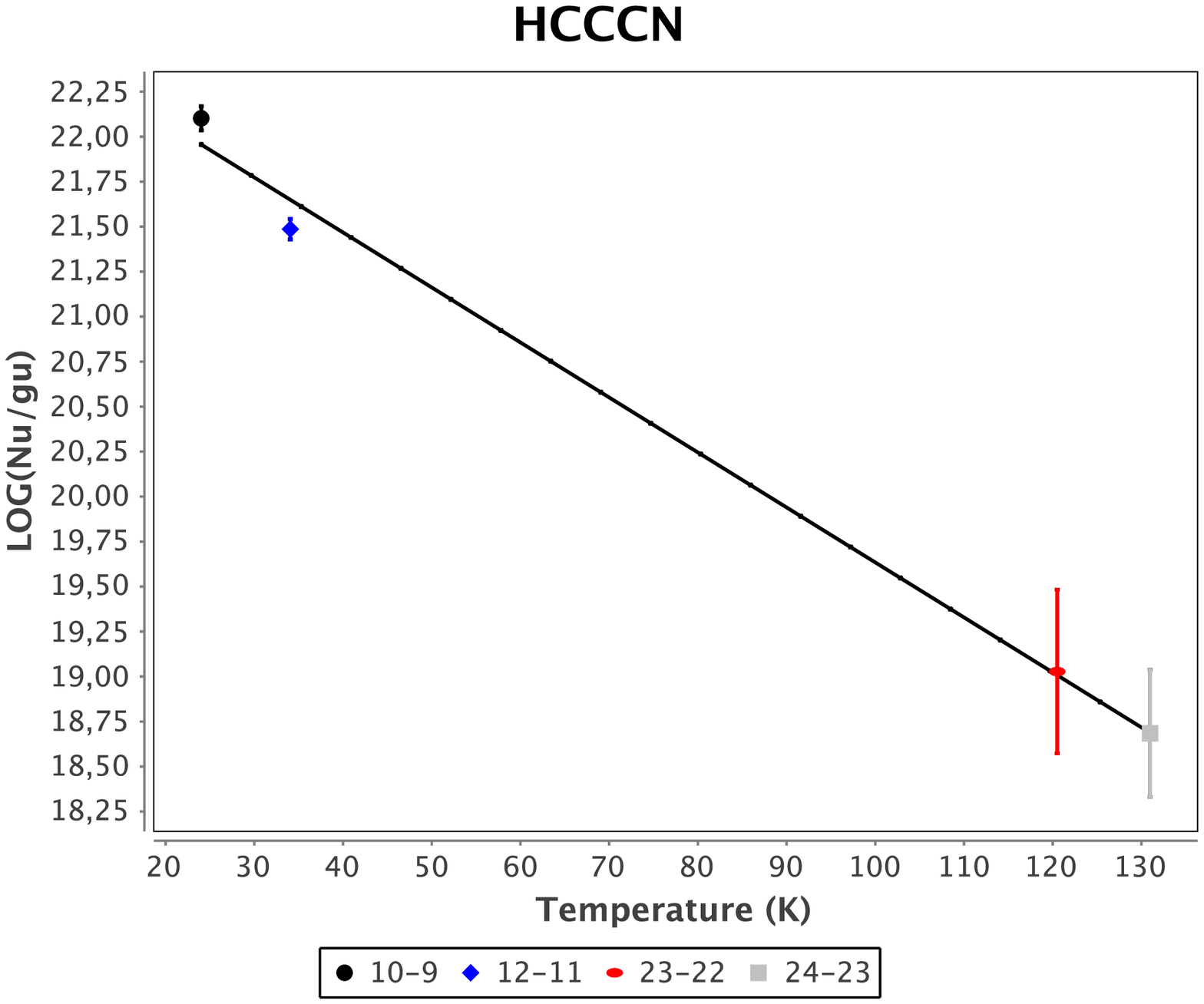}
\includegraphics[width=0.33\textwidth]{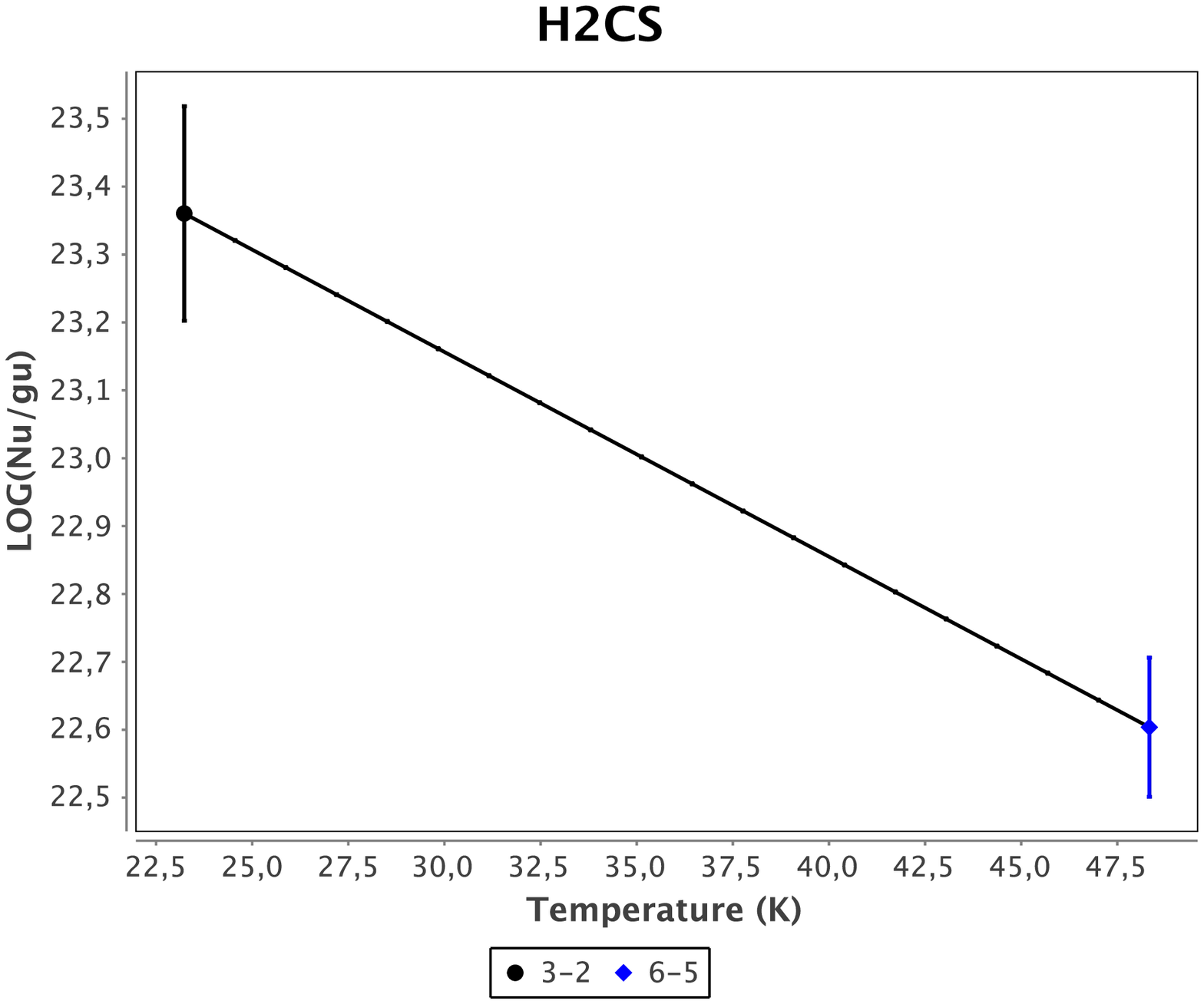}
\includegraphics[width=0.33\textwidth]{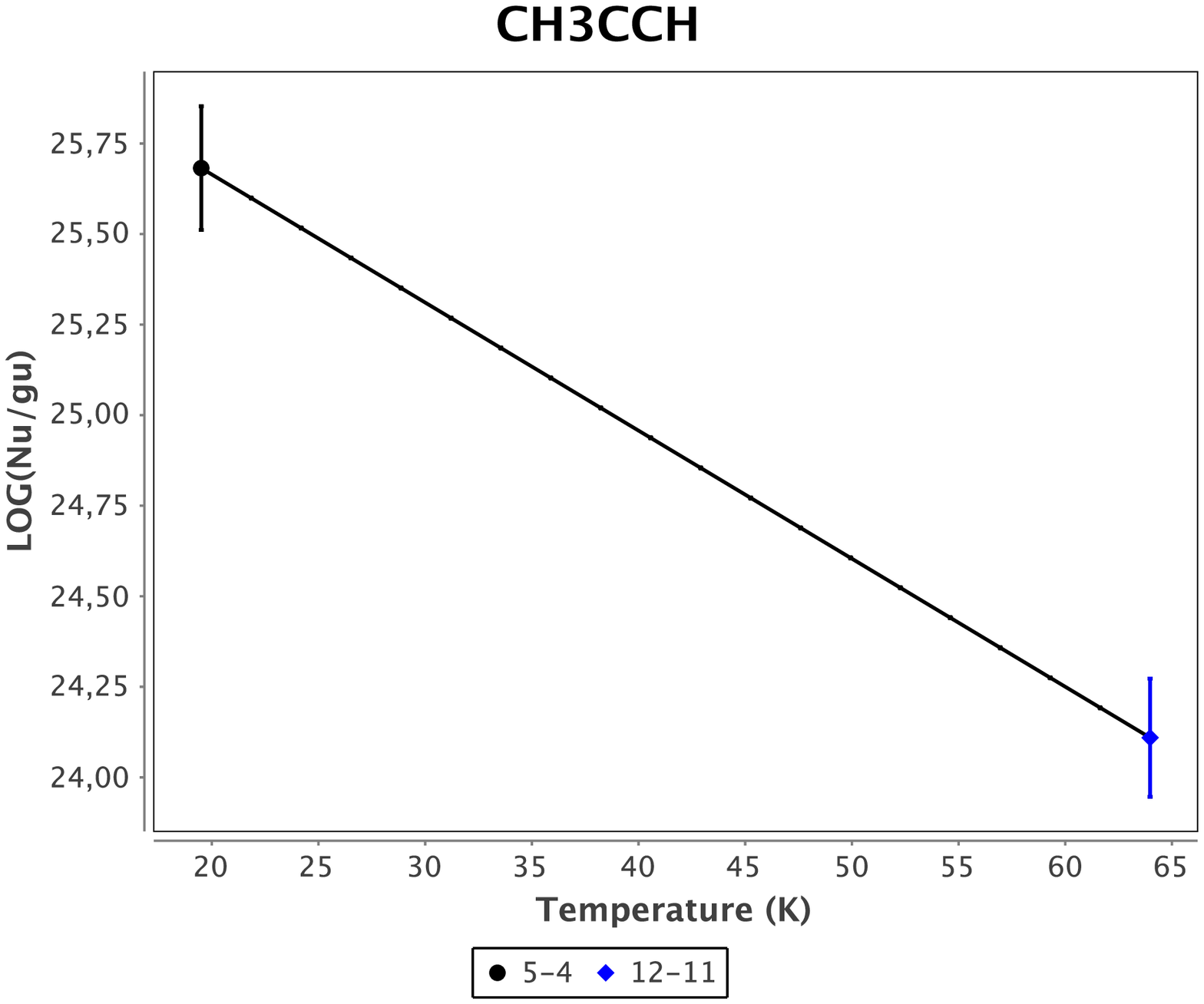}
\includegraphics[width=0.33\textwidth]{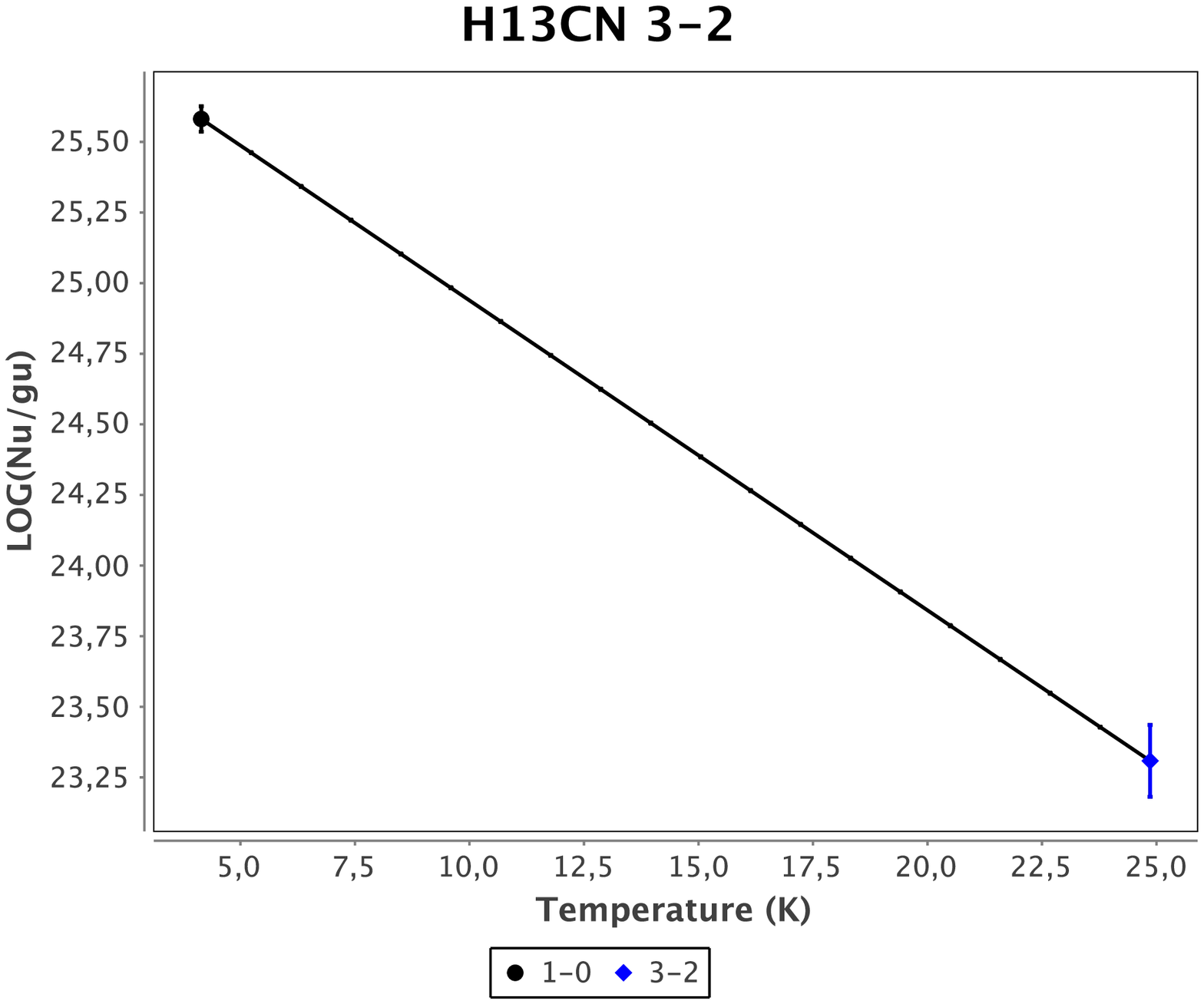}
\includegraphics[width=0.33\textwidth]{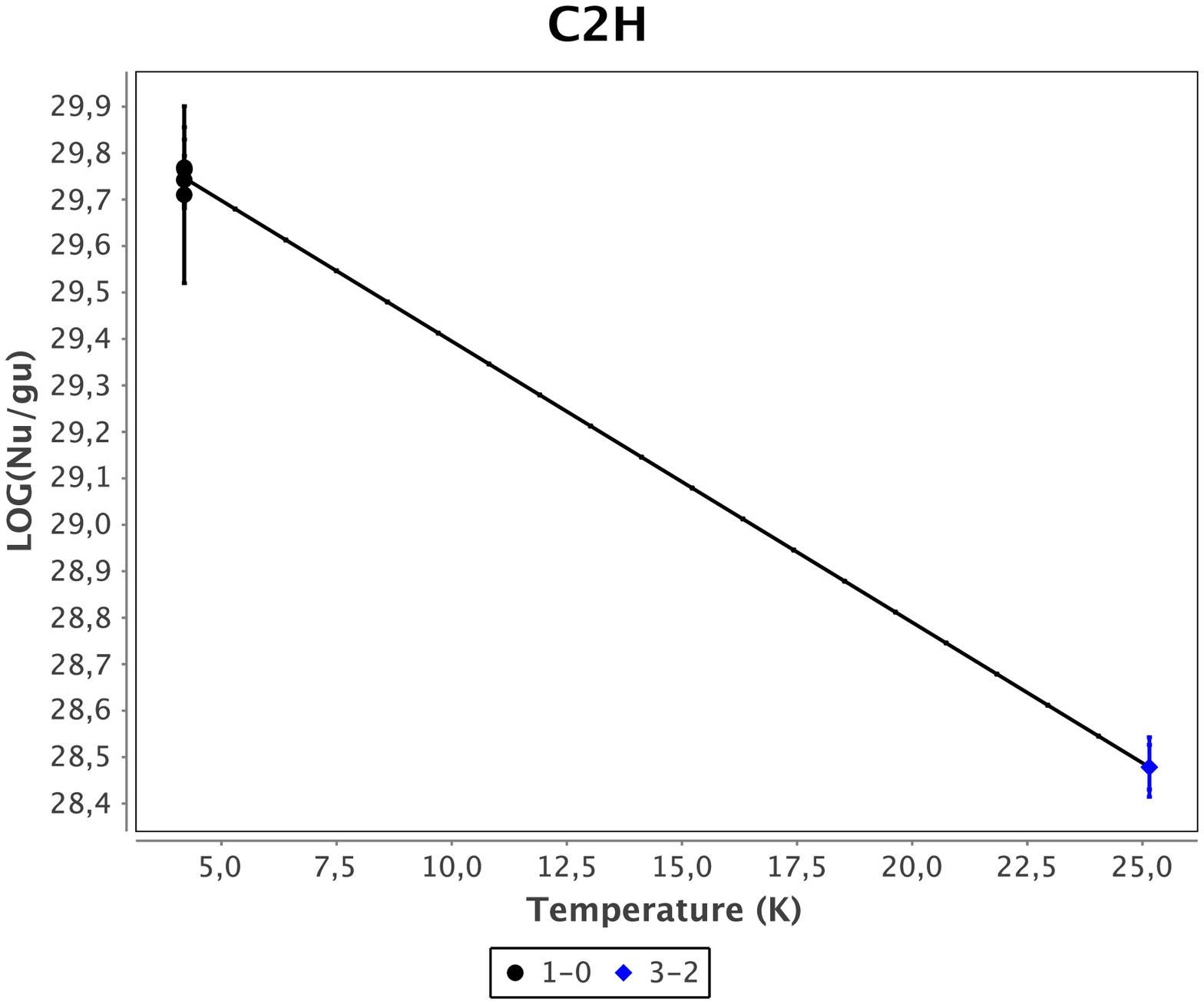}
\includegraphics[width=0.33\textwidth]{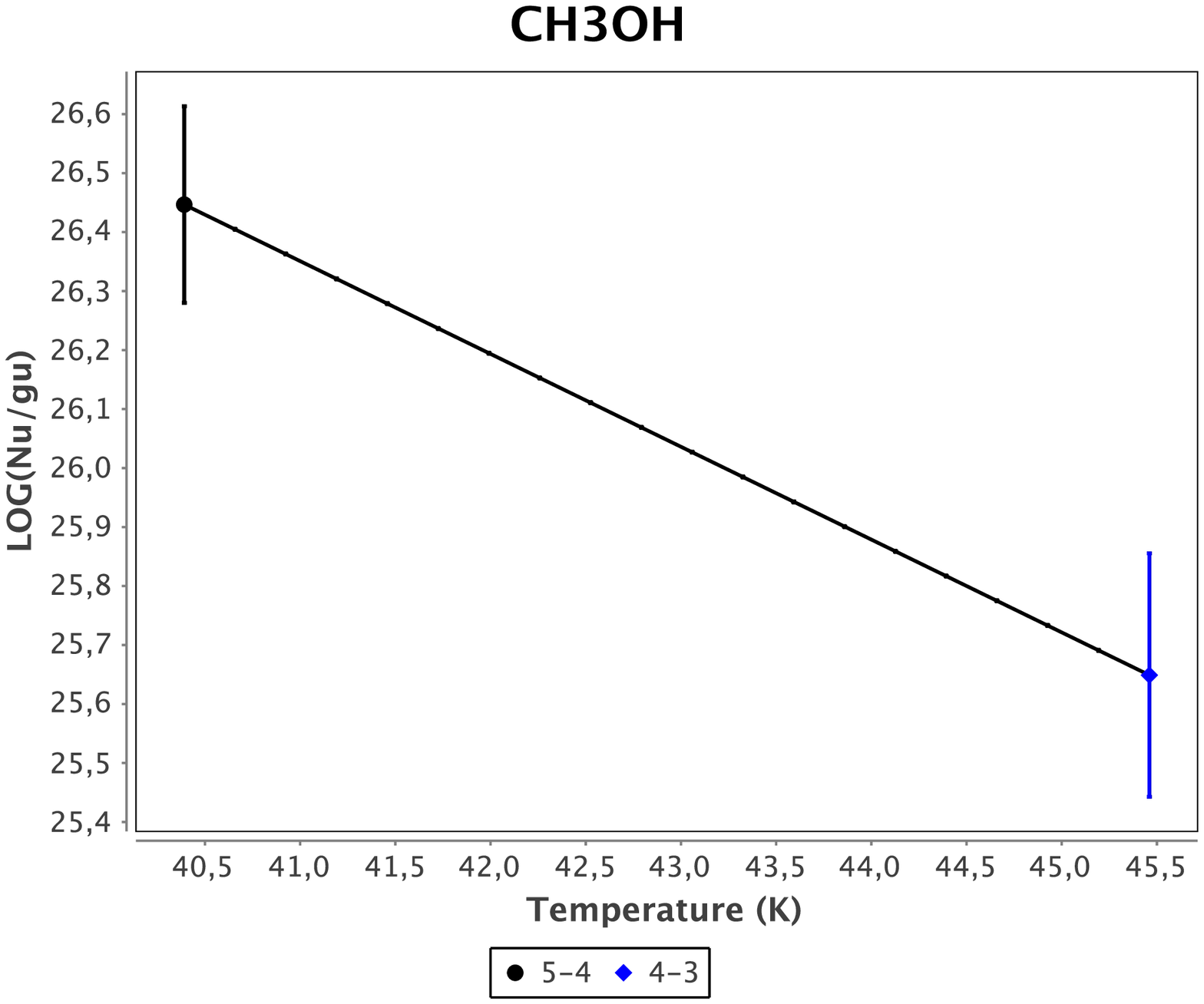}
     \caption{Rotational diagrams of the IF's most representative molecules (Offset (0$\arcsec$, 0$\arcsec$)). }
 \end{figure*}
 
\begin{figure*}
\includegraphics[width=0.33\textwidth]{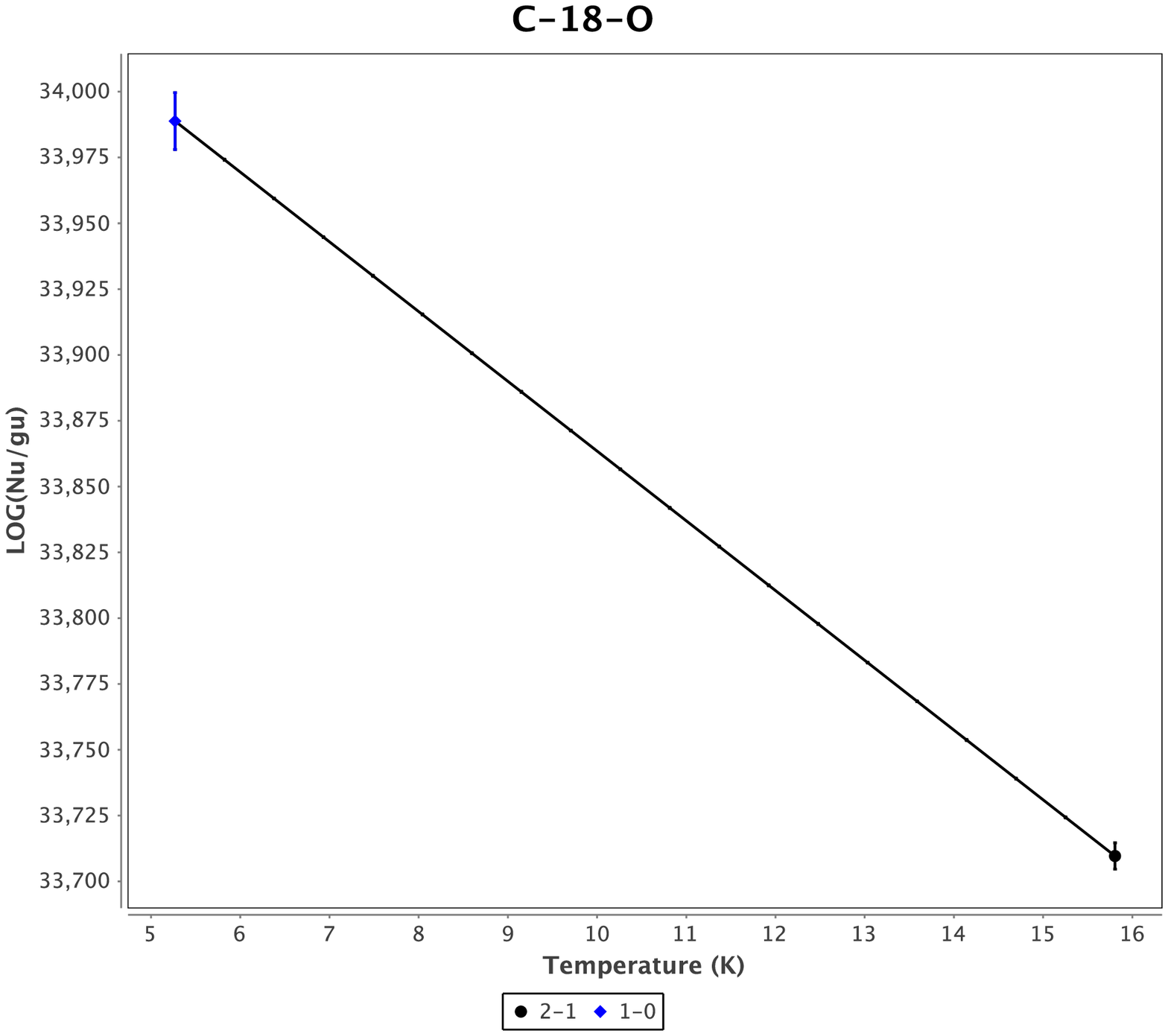}
\includegraphics[width=0.33\textwidth]{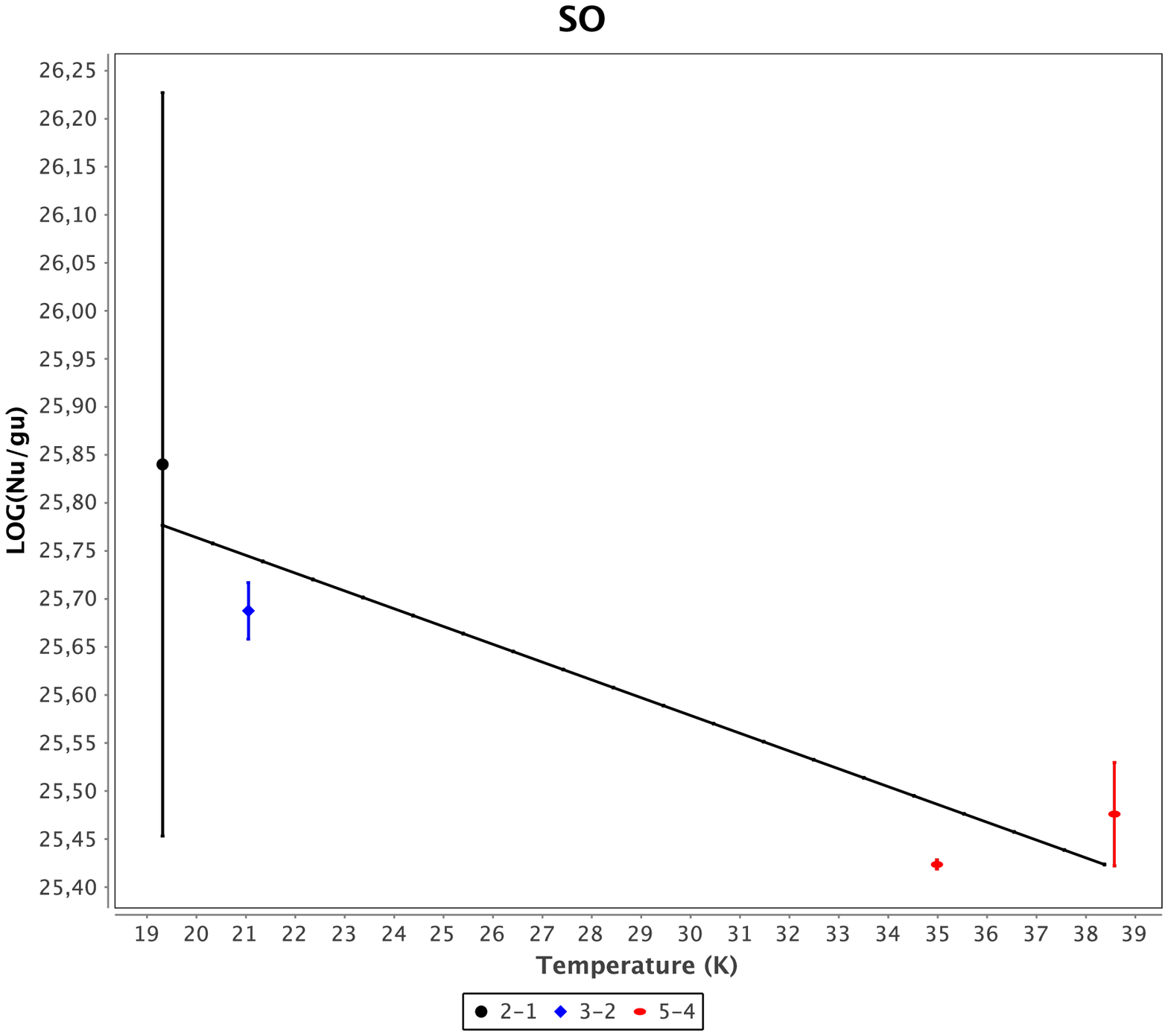}
\includegraphics[width=0.33\textwidth]{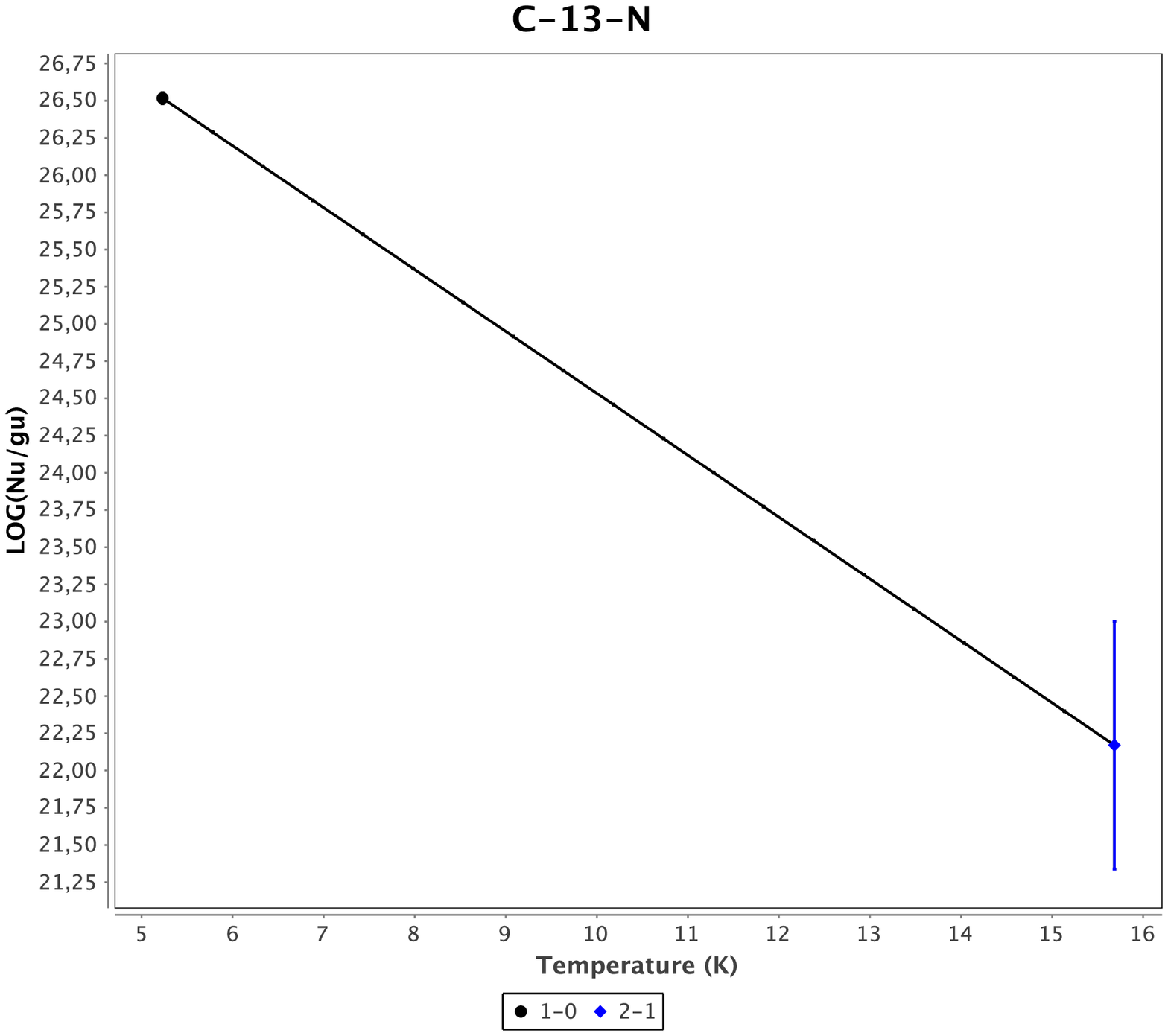}
\includegraphics[width=0.33\textwidth]{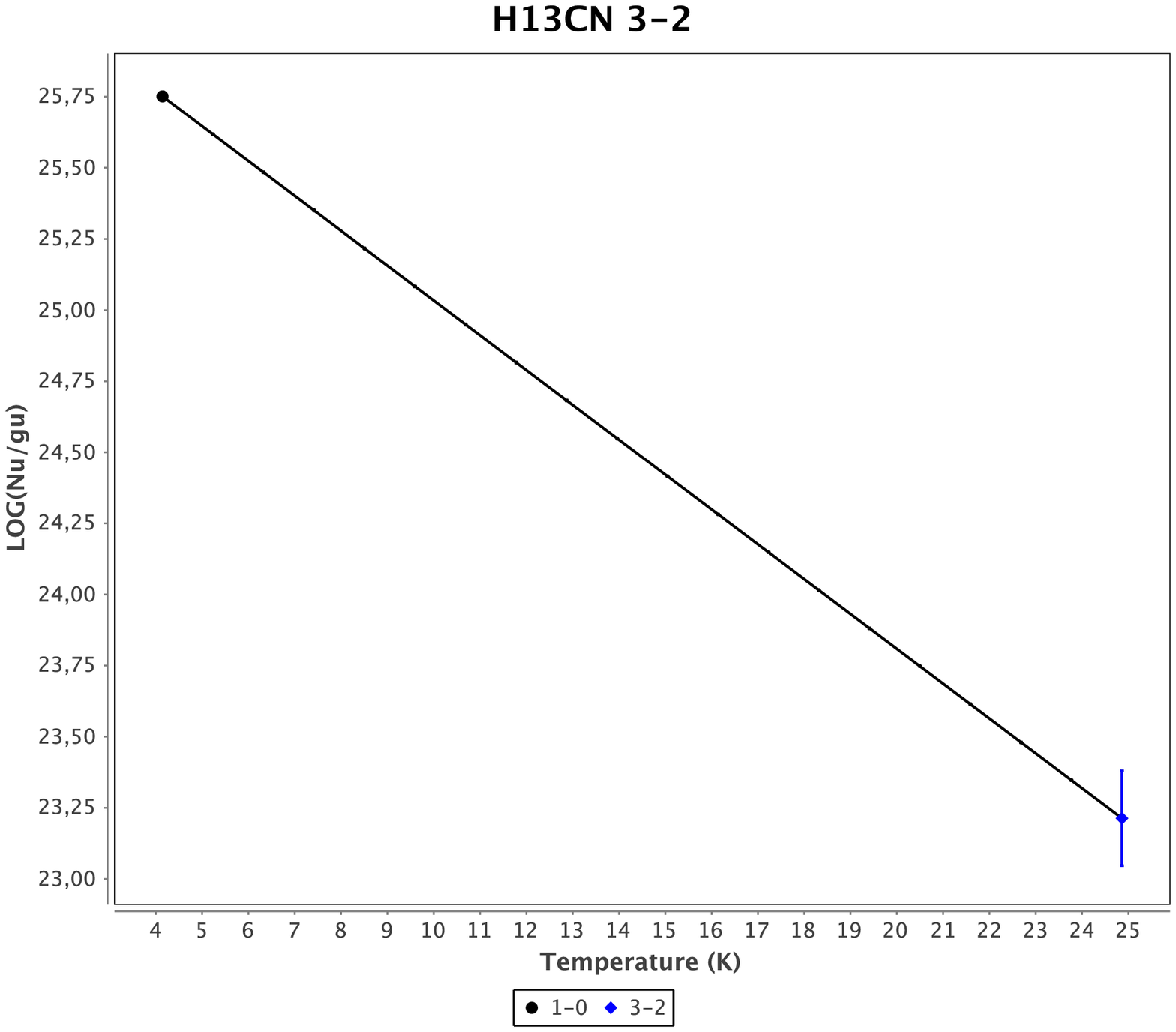}
\includegraphics[width=0.33\textwidth]{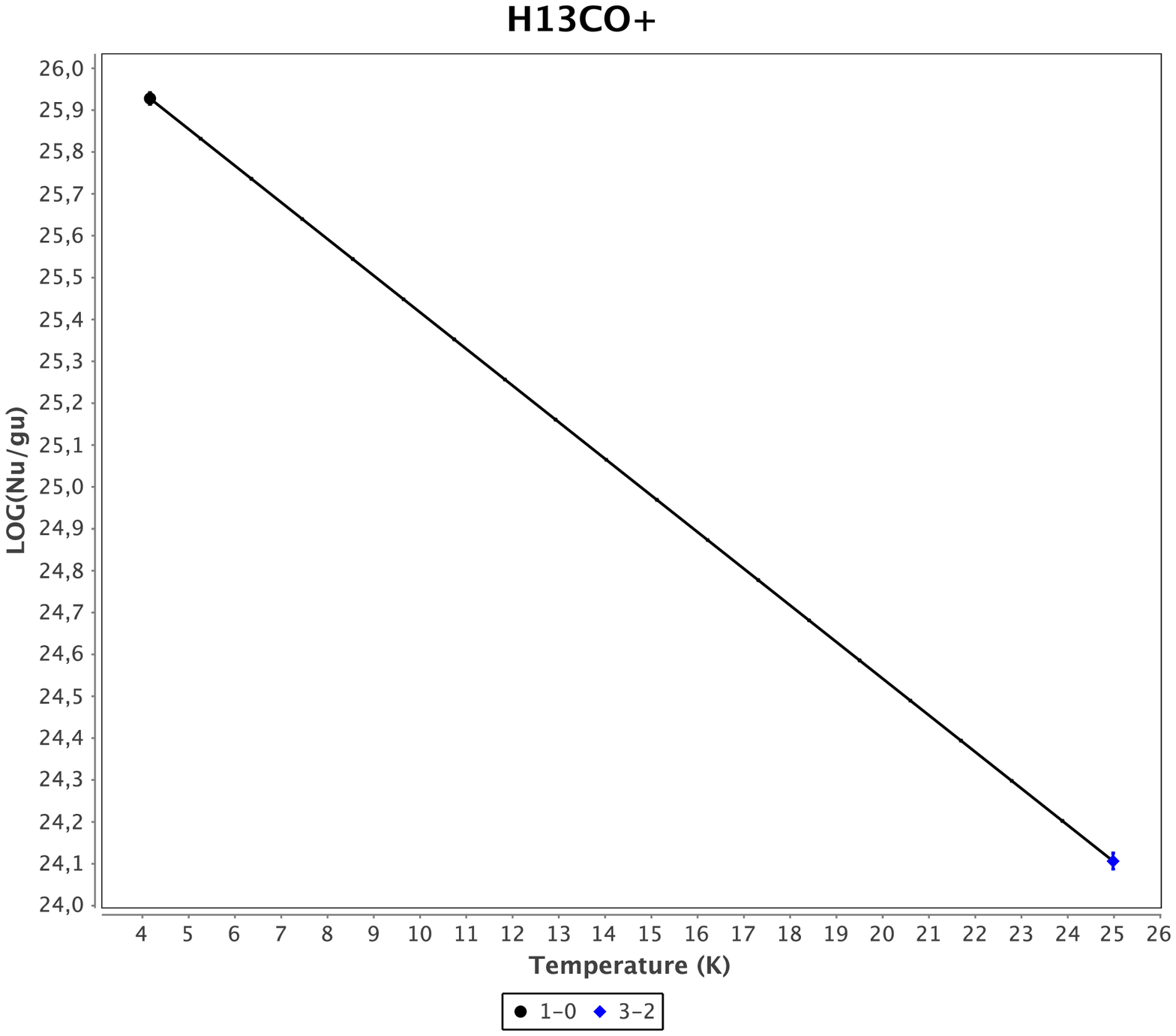}
\includegraphics[width=0.33\textwidth]{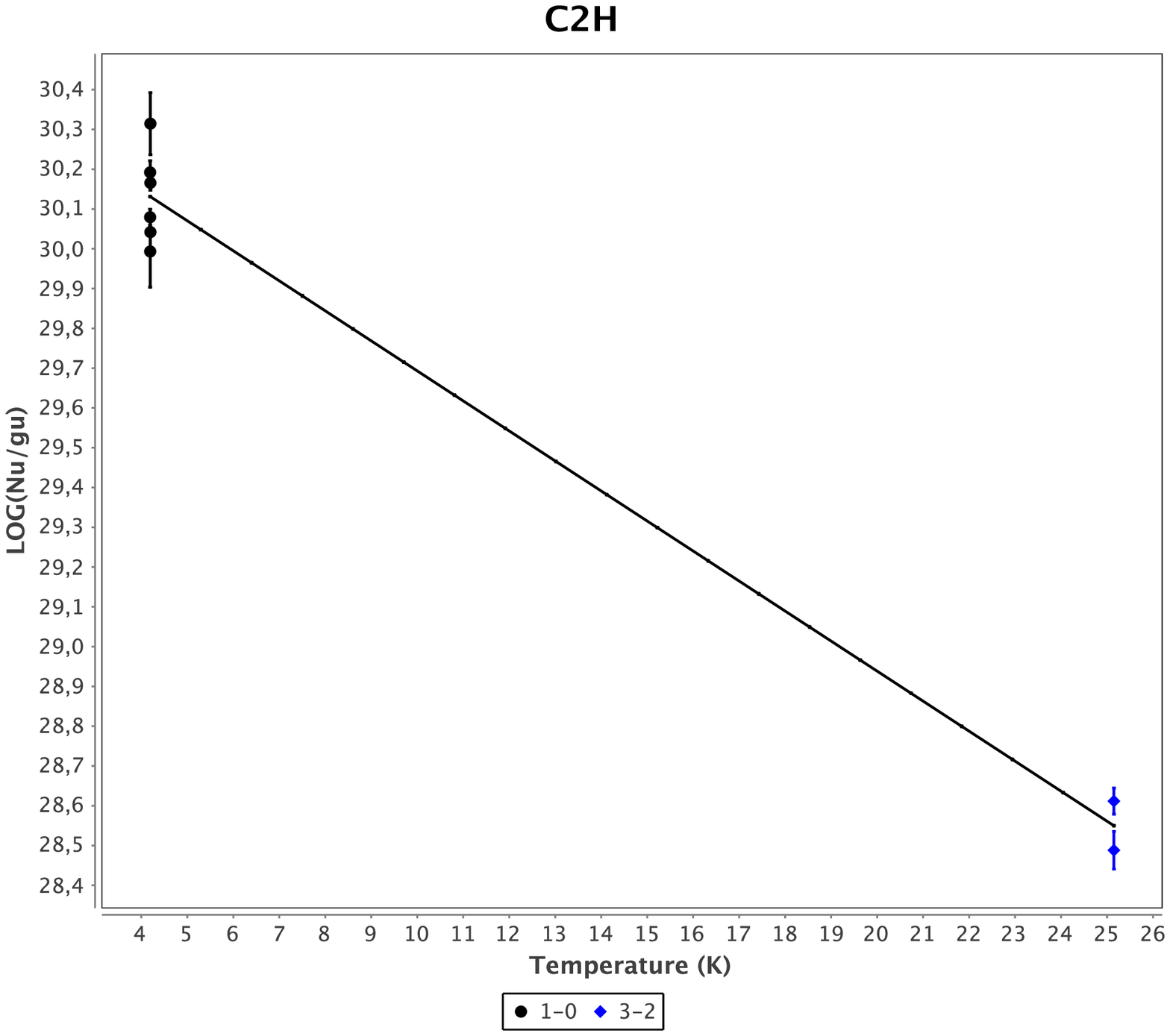}
\includegraphics[width=0.33\textwidth]{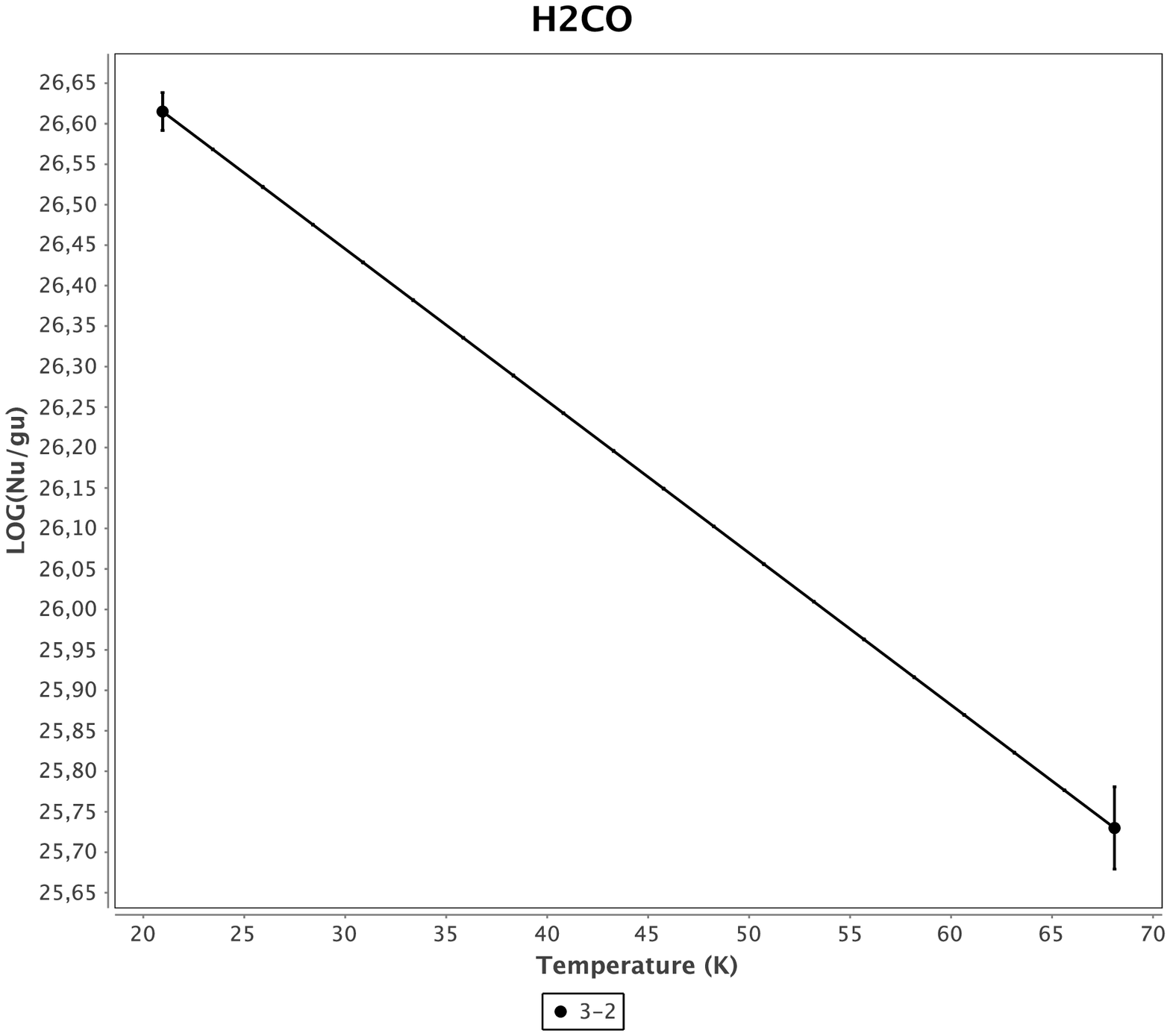}
\includegraphics[width=0.33\textwidth]{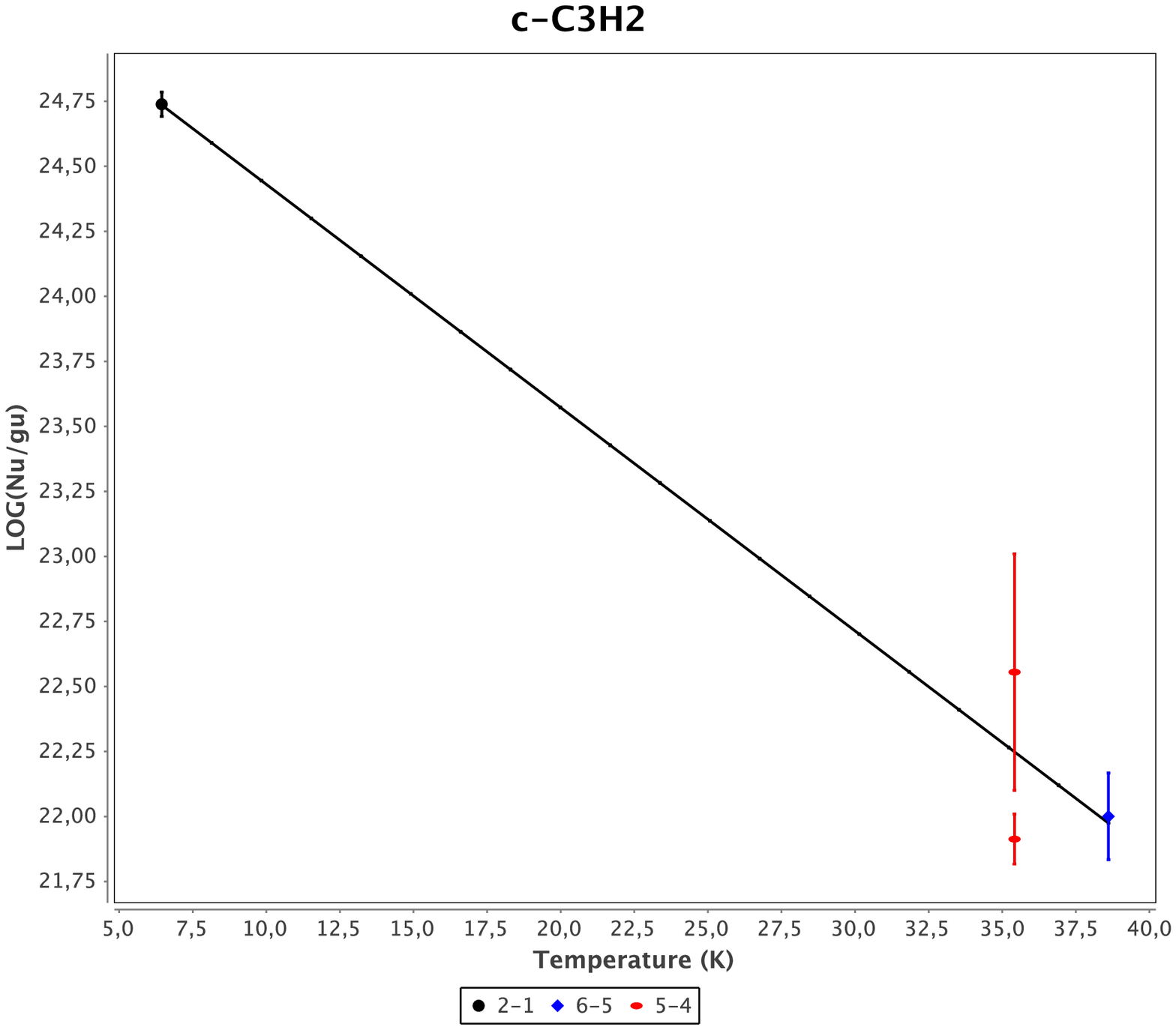}
\includegraphics[width=0.33\textwidth]{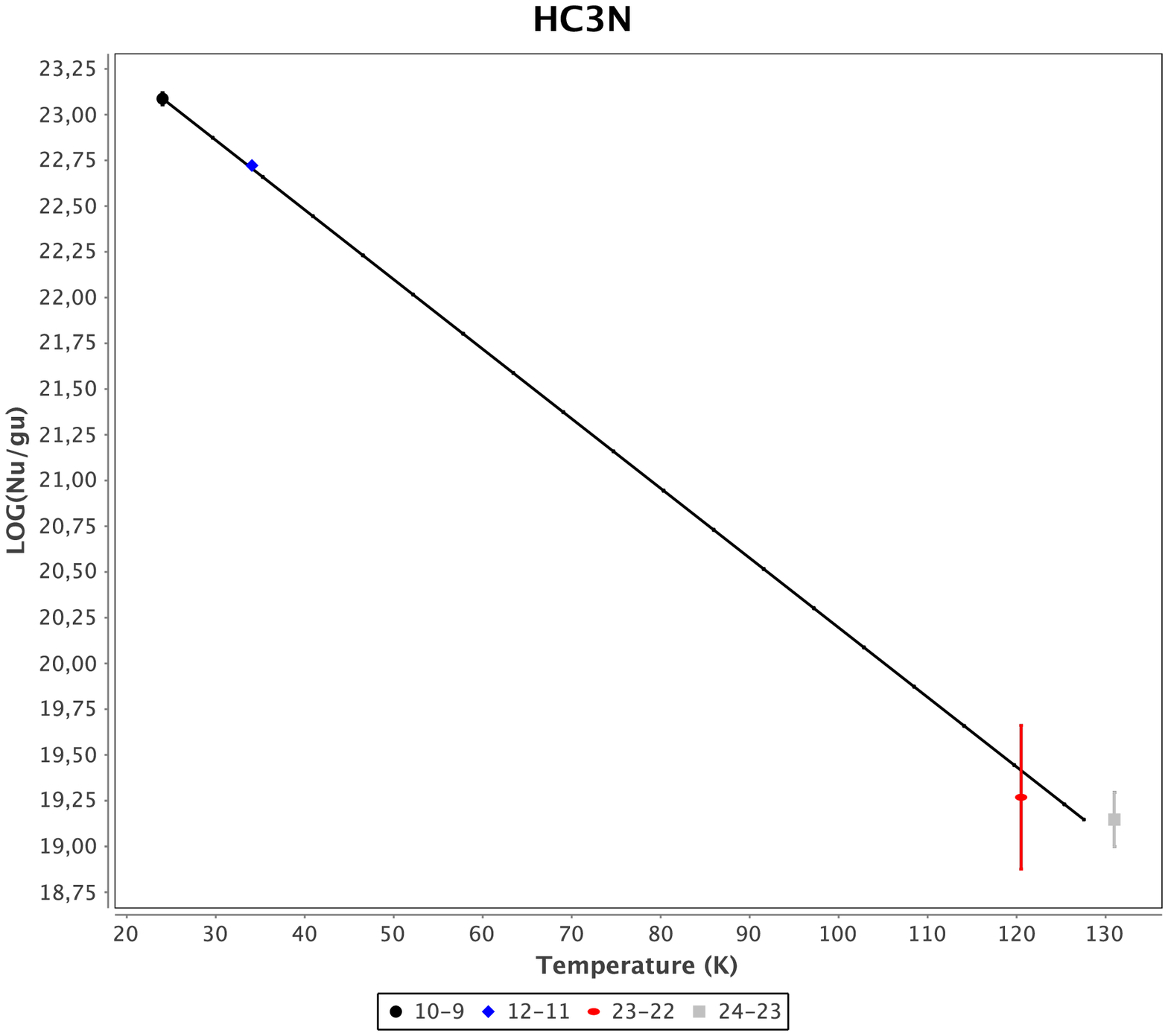}
\includegraphics[width=0.33\textwidth]{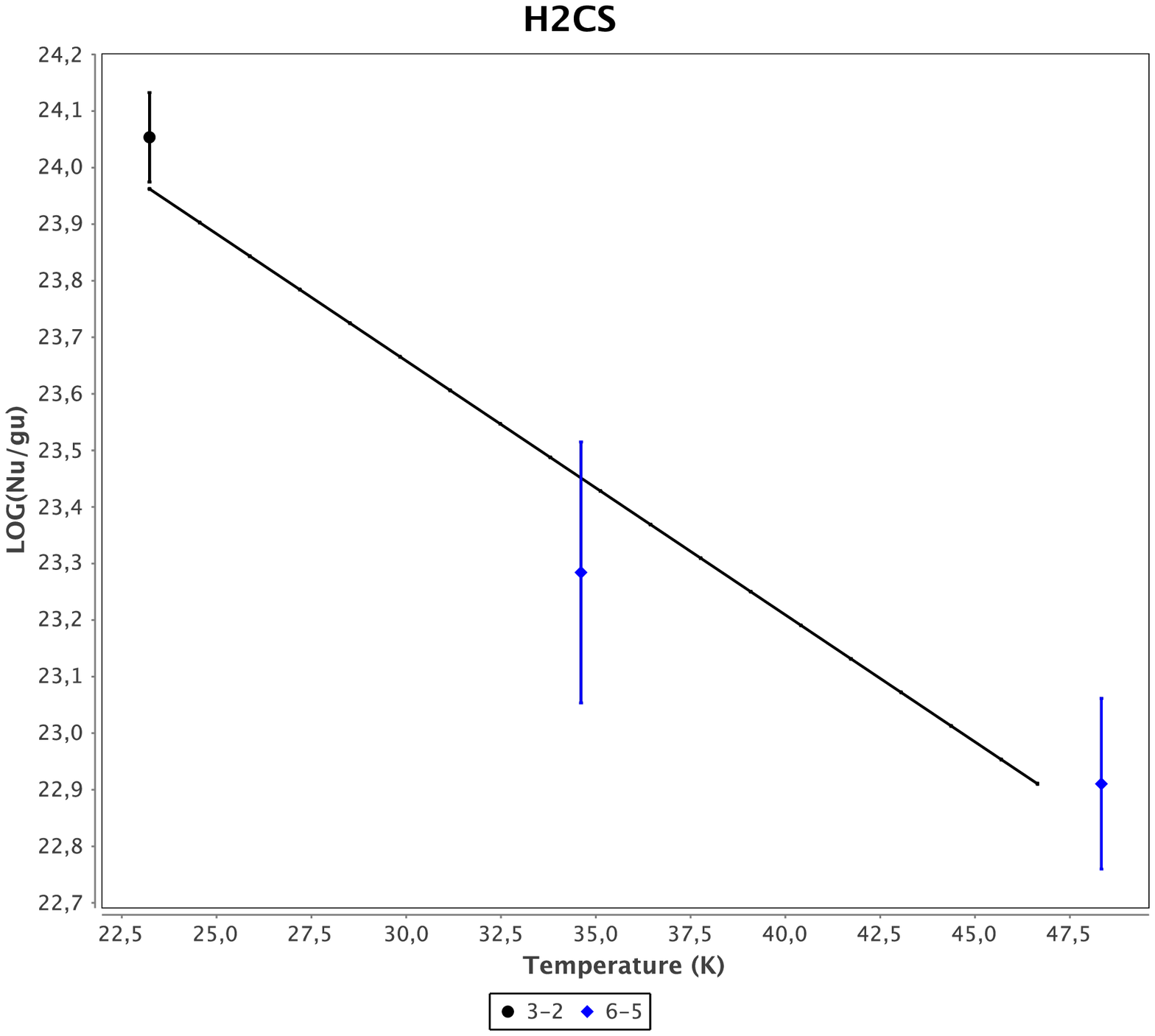}
\includegraphics[width=0.33\textwidth]{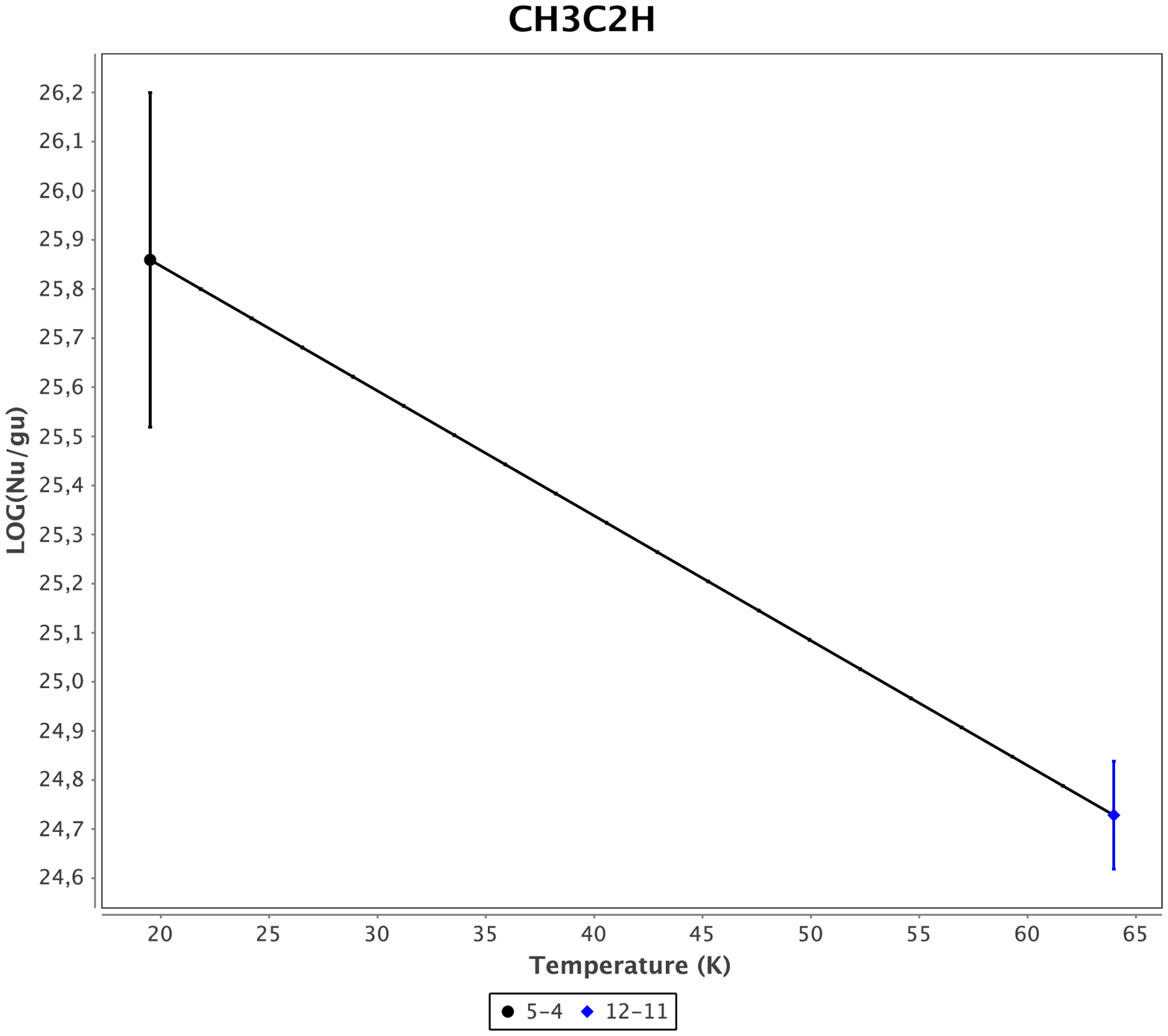}
\includegraphics[width=0.33\textwidth]{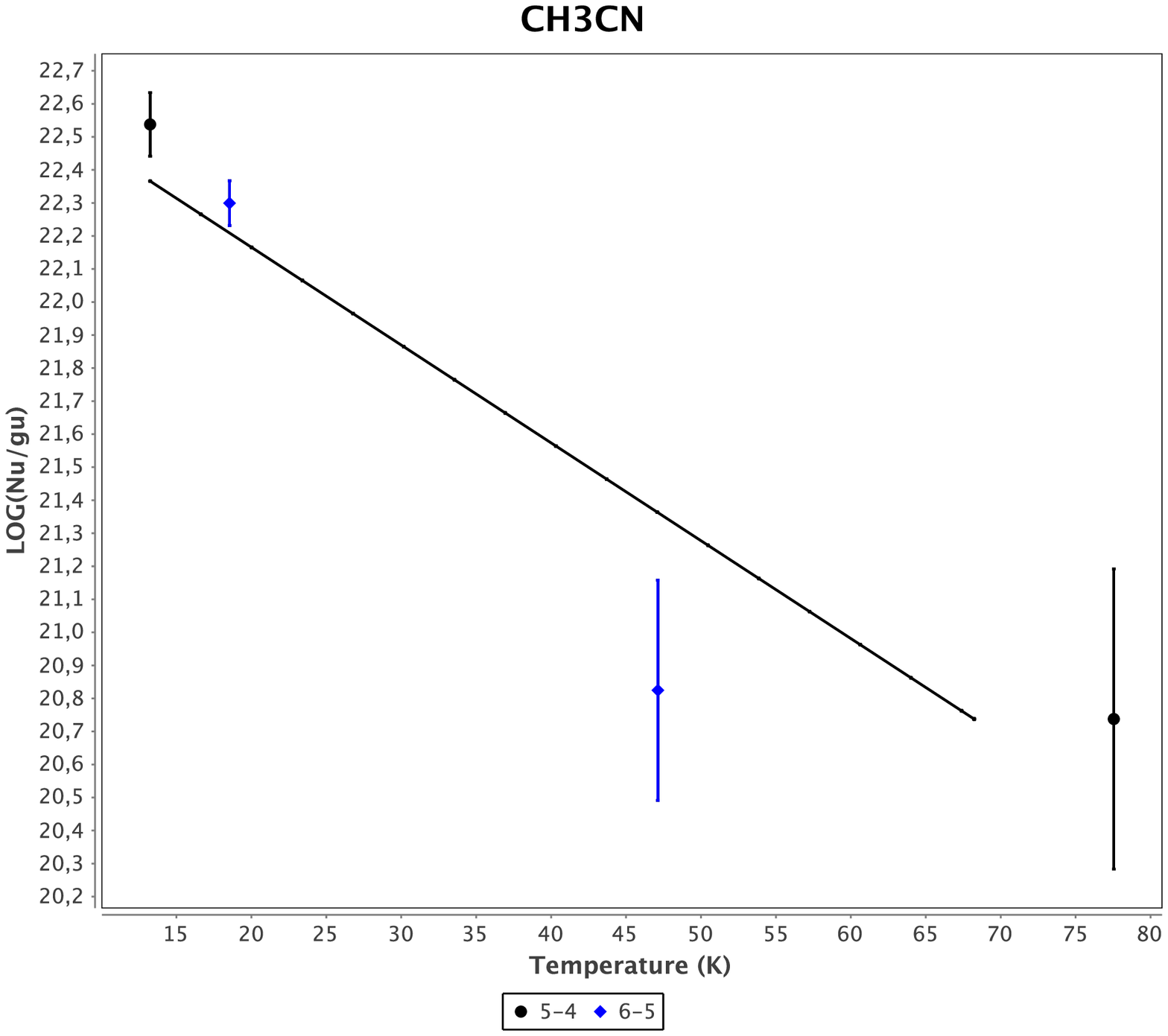}
 \includegraphics[width=0.33\textwidth]{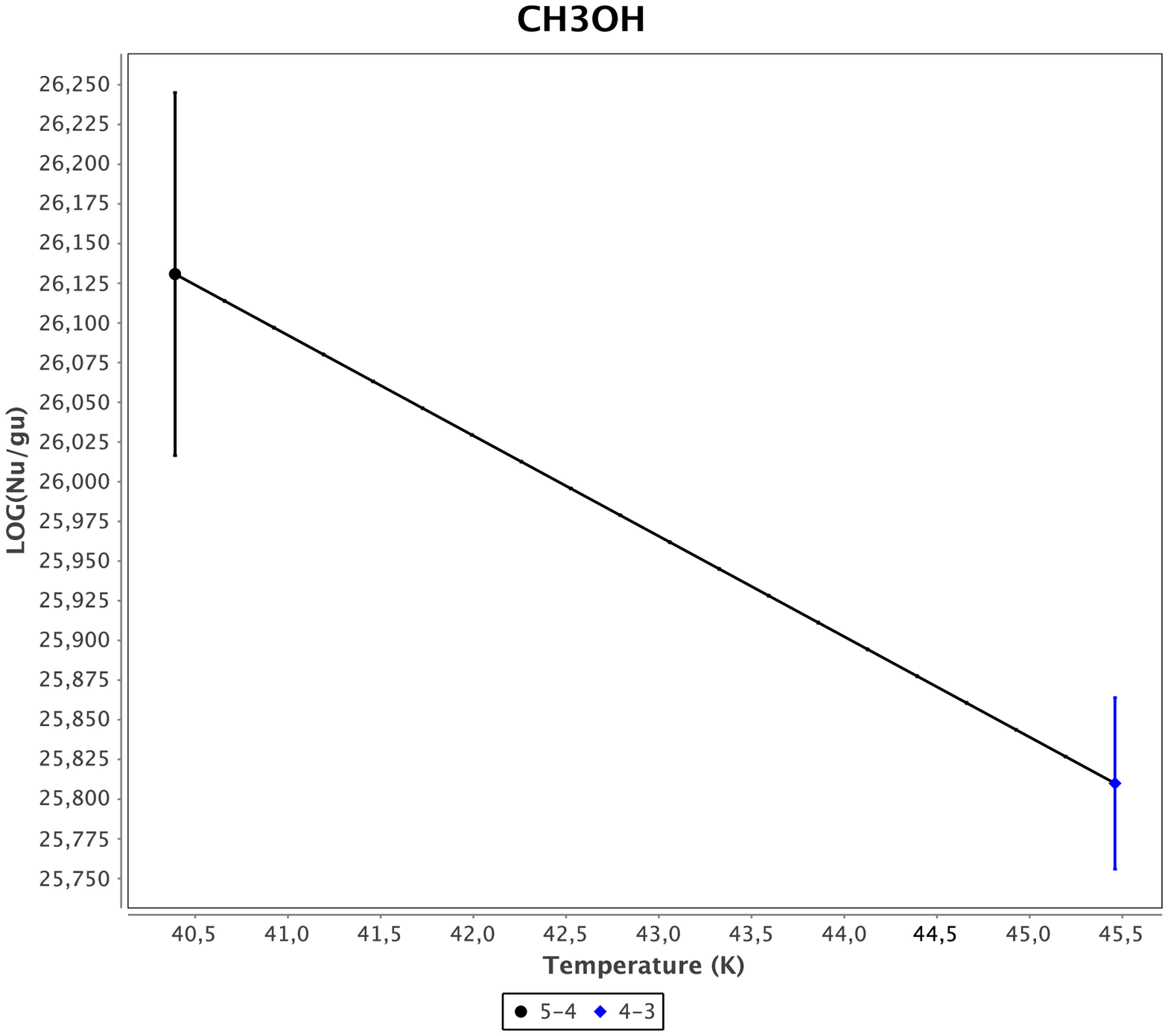}
\includegraphics[width=0.33\textwidth]{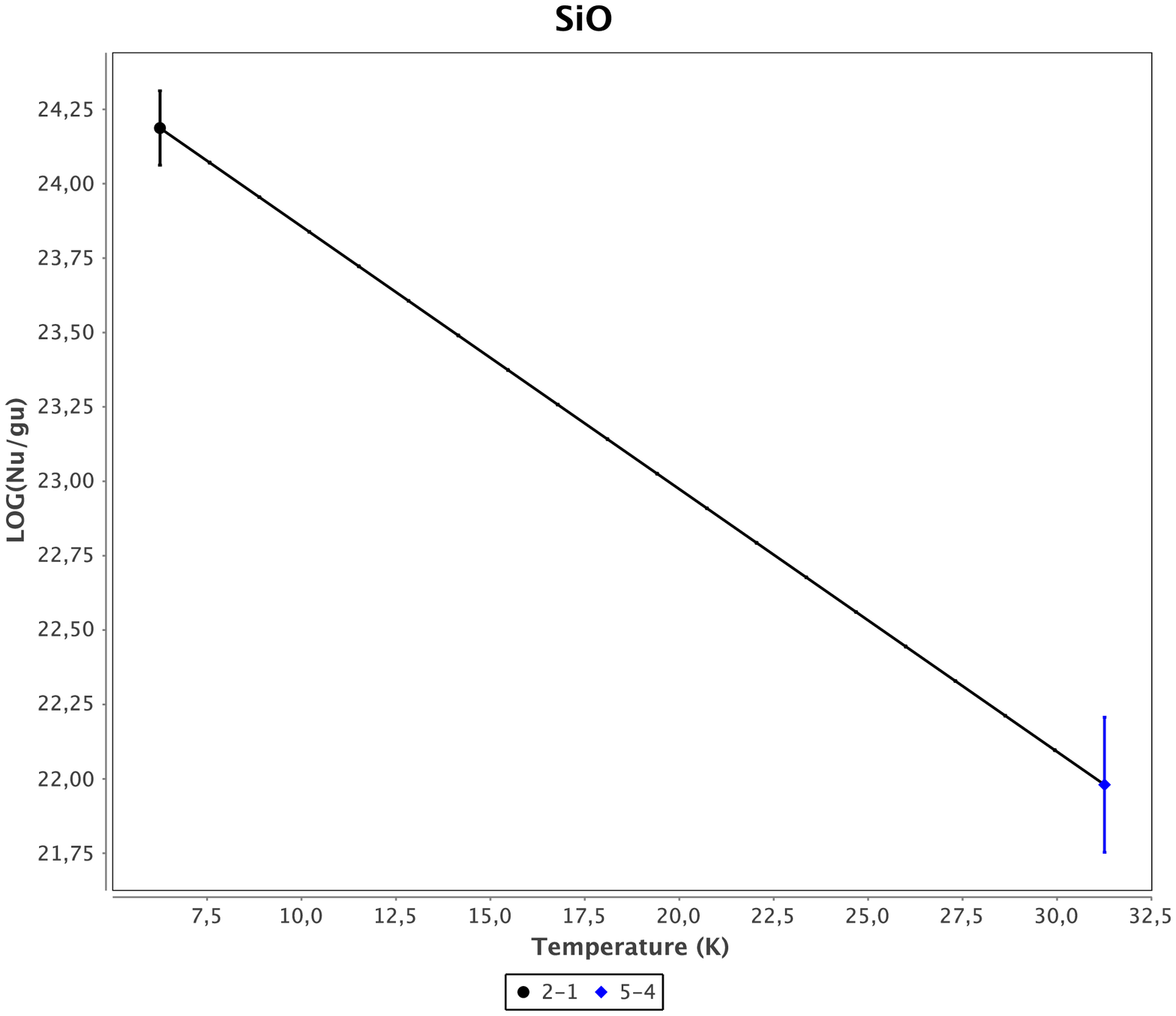}
      \caption{Rotational diagrams of the MP's most representative molecules (Offset ($+$15$\arcsec$, $-$15$\arcsec$)). }
 \end{figure*}
 
\begin{figure*}
\includegraphics[width=0.33\textwidth]{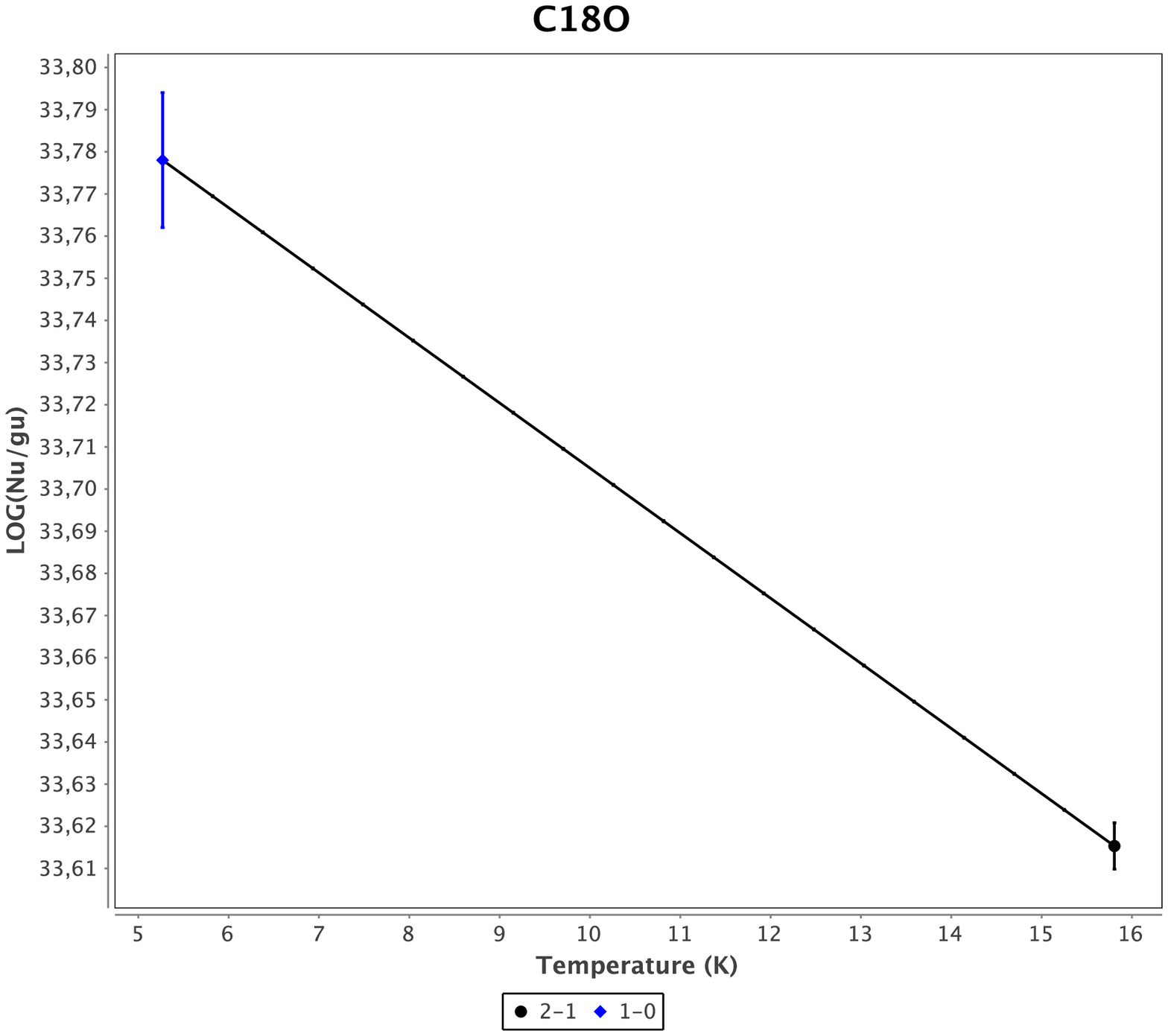}
\includegraphics[width=0.33\textwidth]{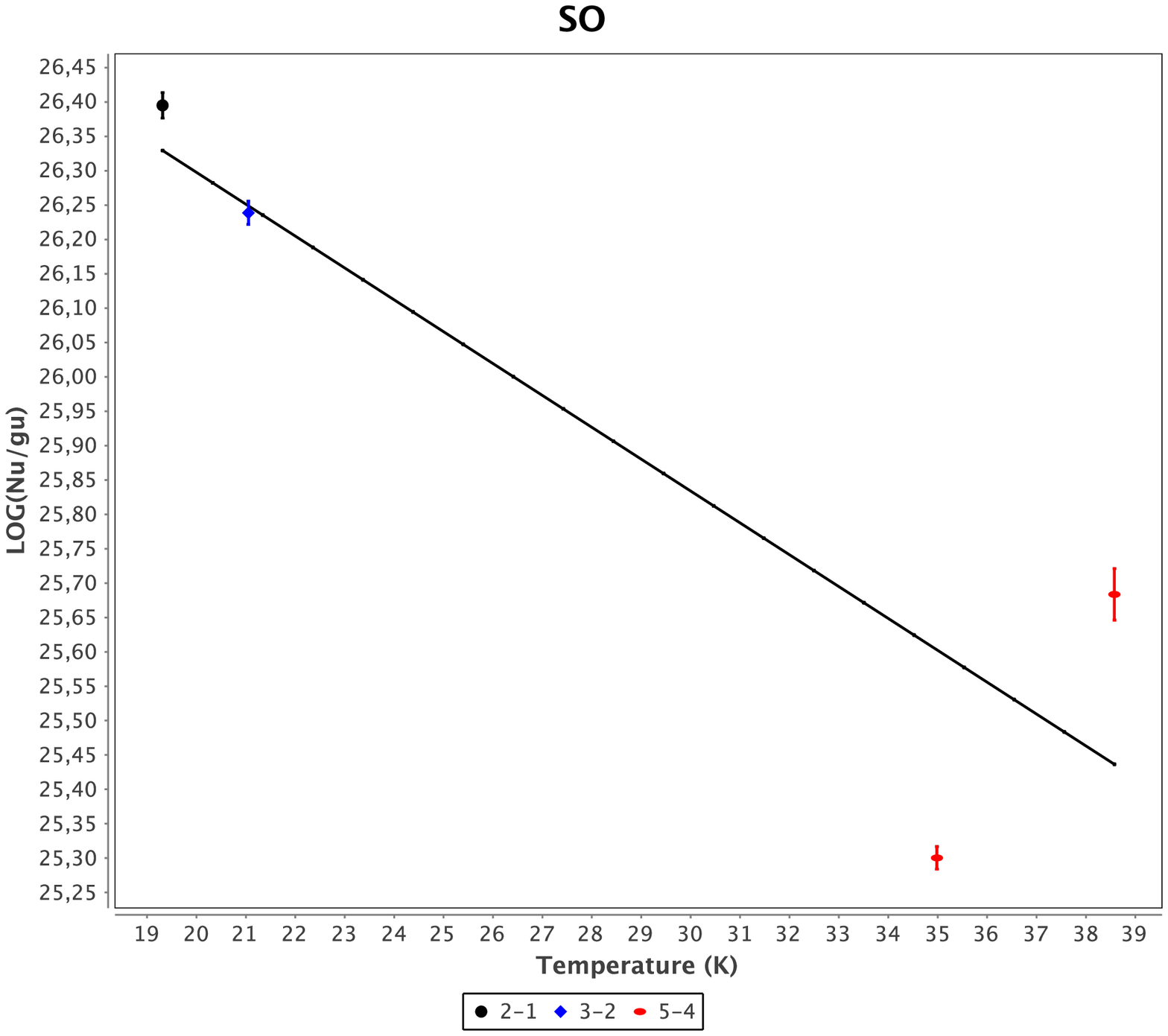}
\includegraphics[width=0.33\textwidth]{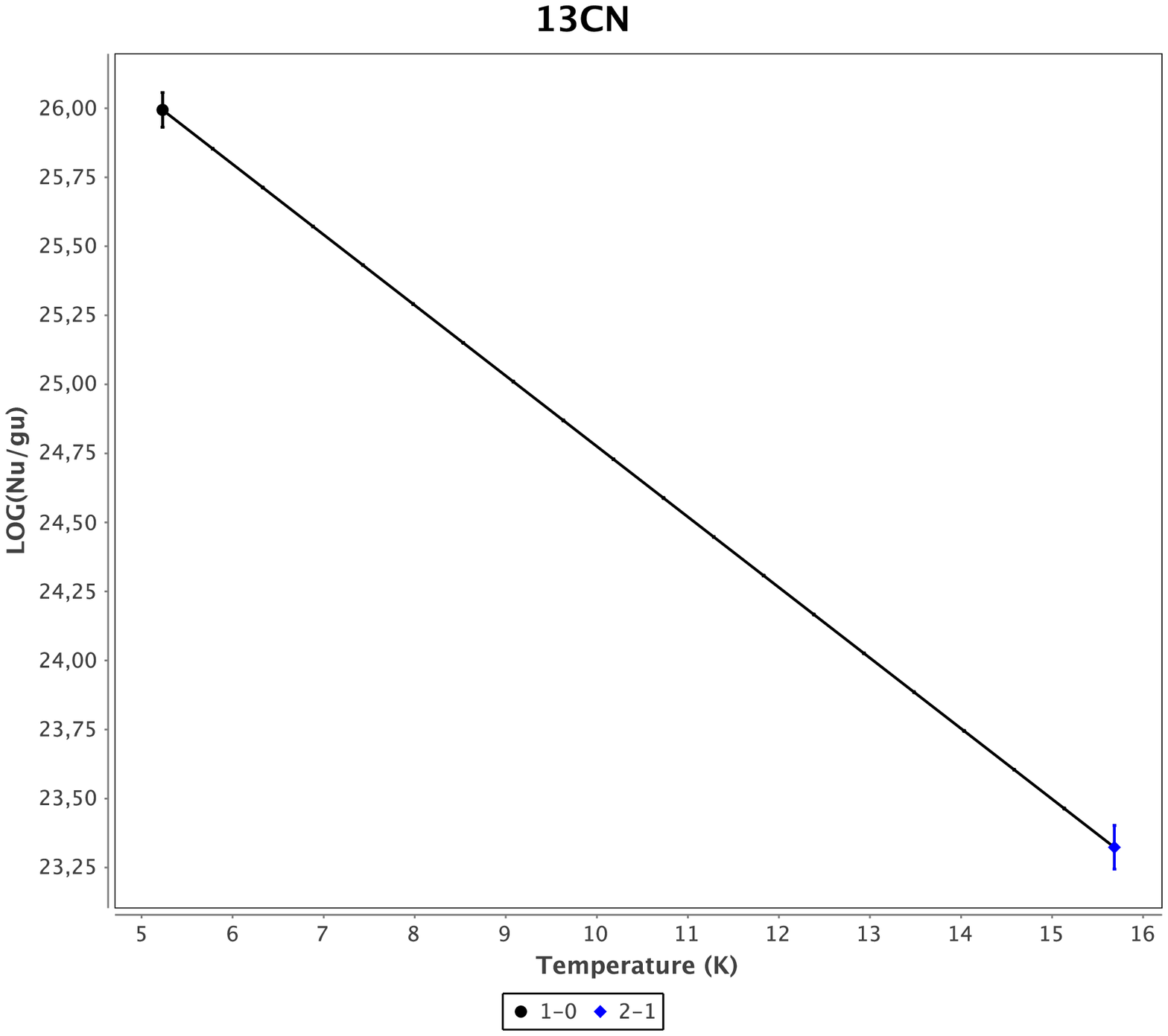}
\includegraphics[width=0.33\textwidth]{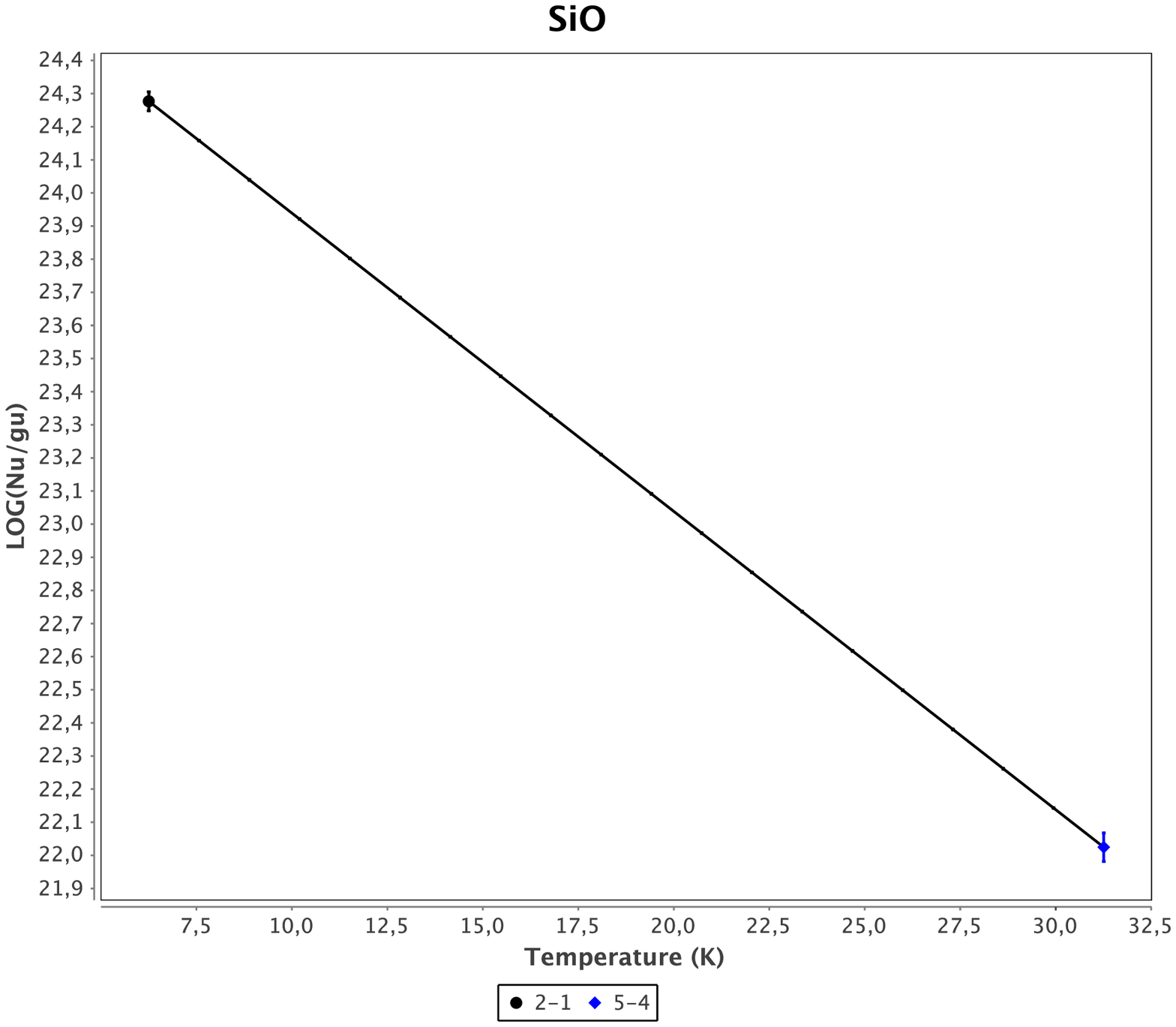}
\includegraphics[width=0.33\textwidth]{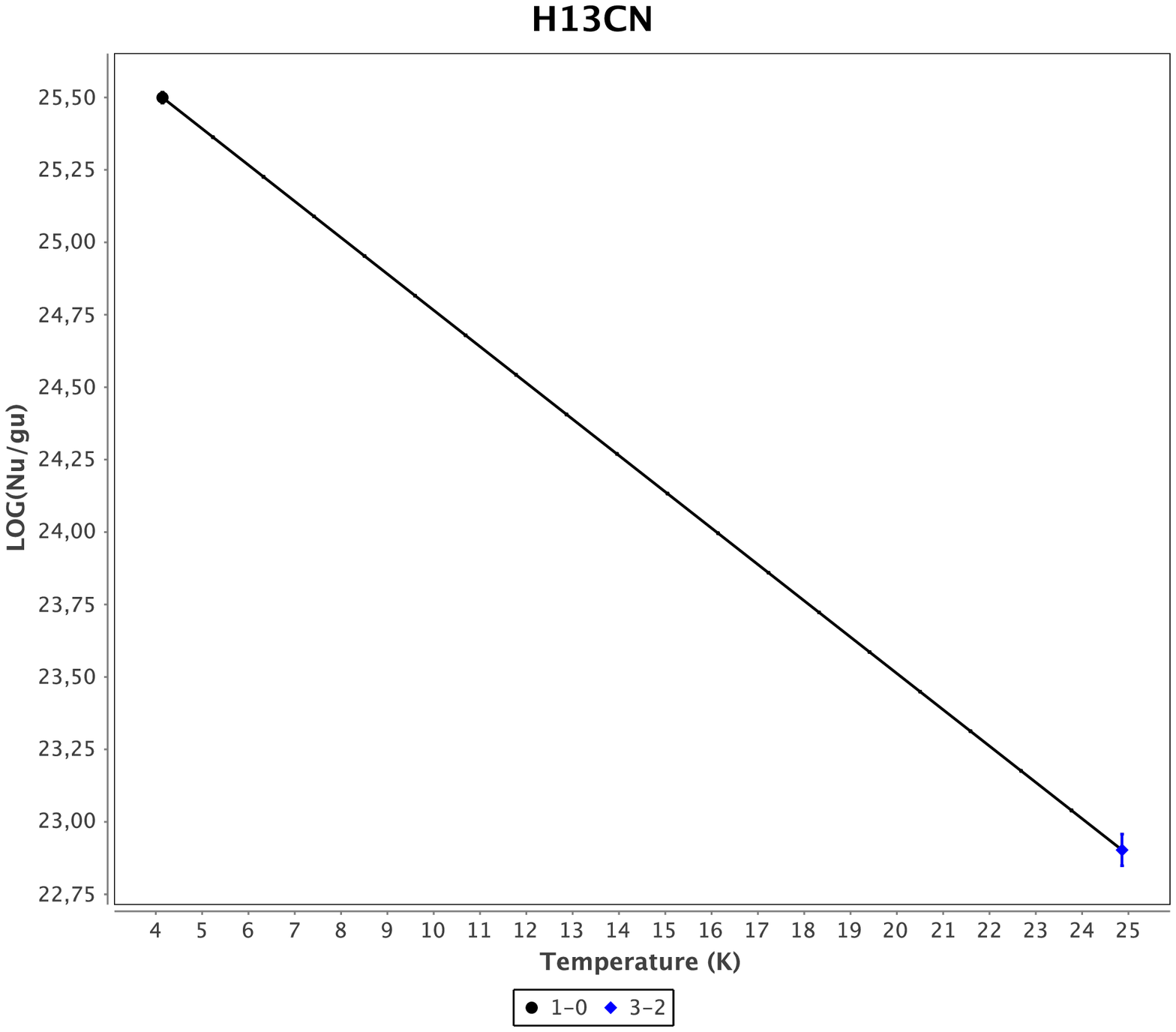}
\includegraphics[width=0.33\textwidth]{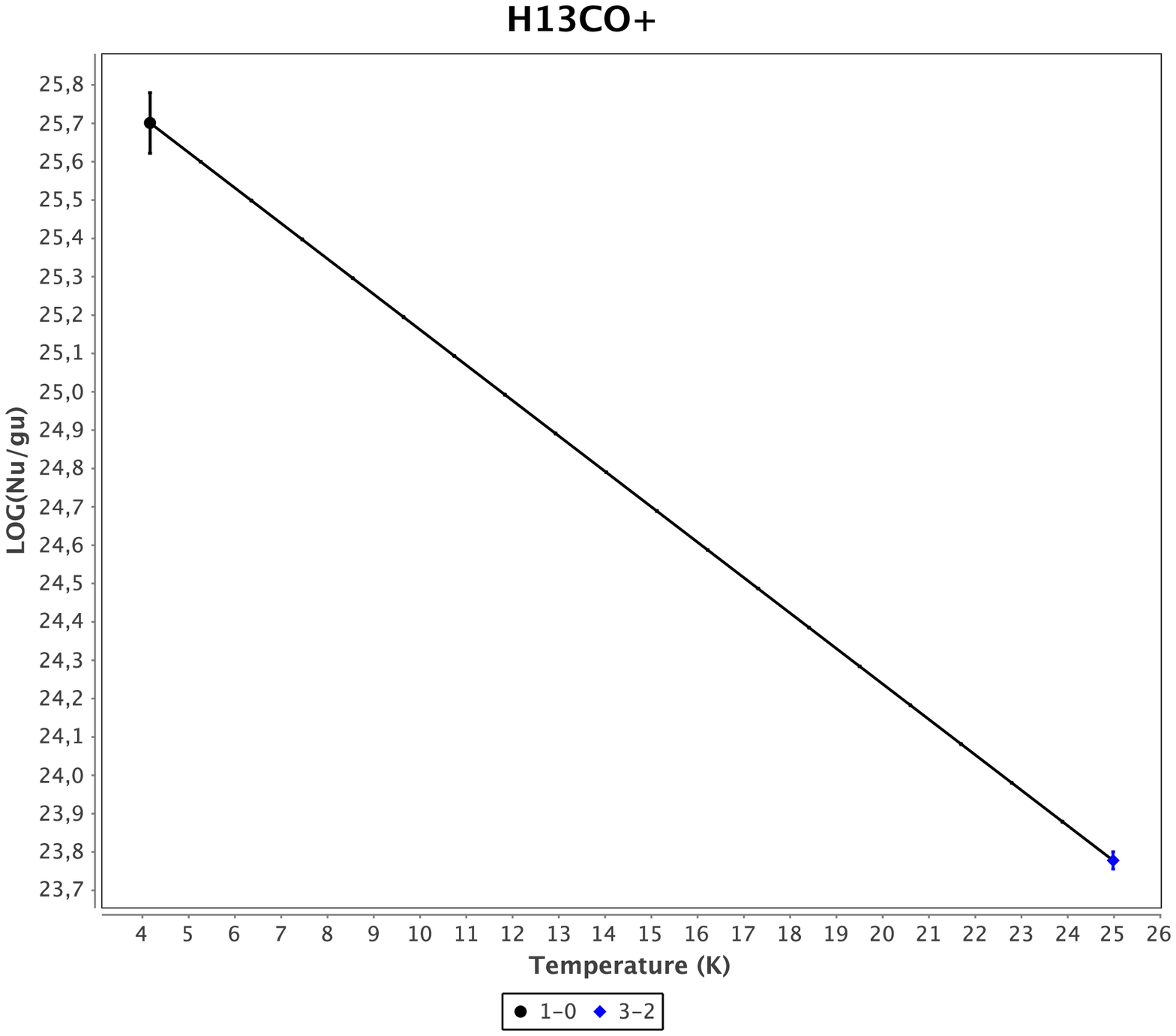}
\includegraphics[width=0.33\textwidth]{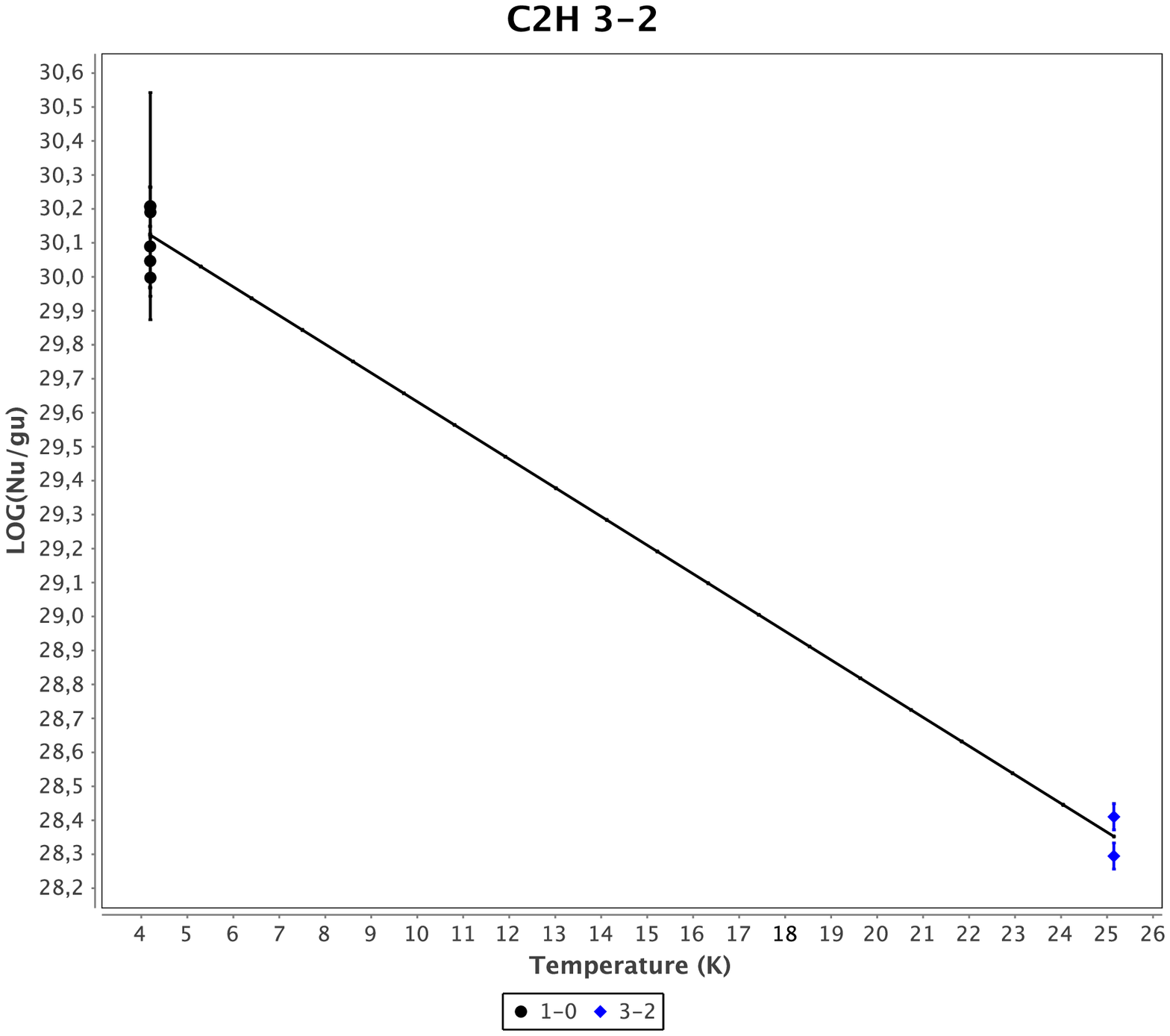}
\includegraphics[width=0.33\textwidth]{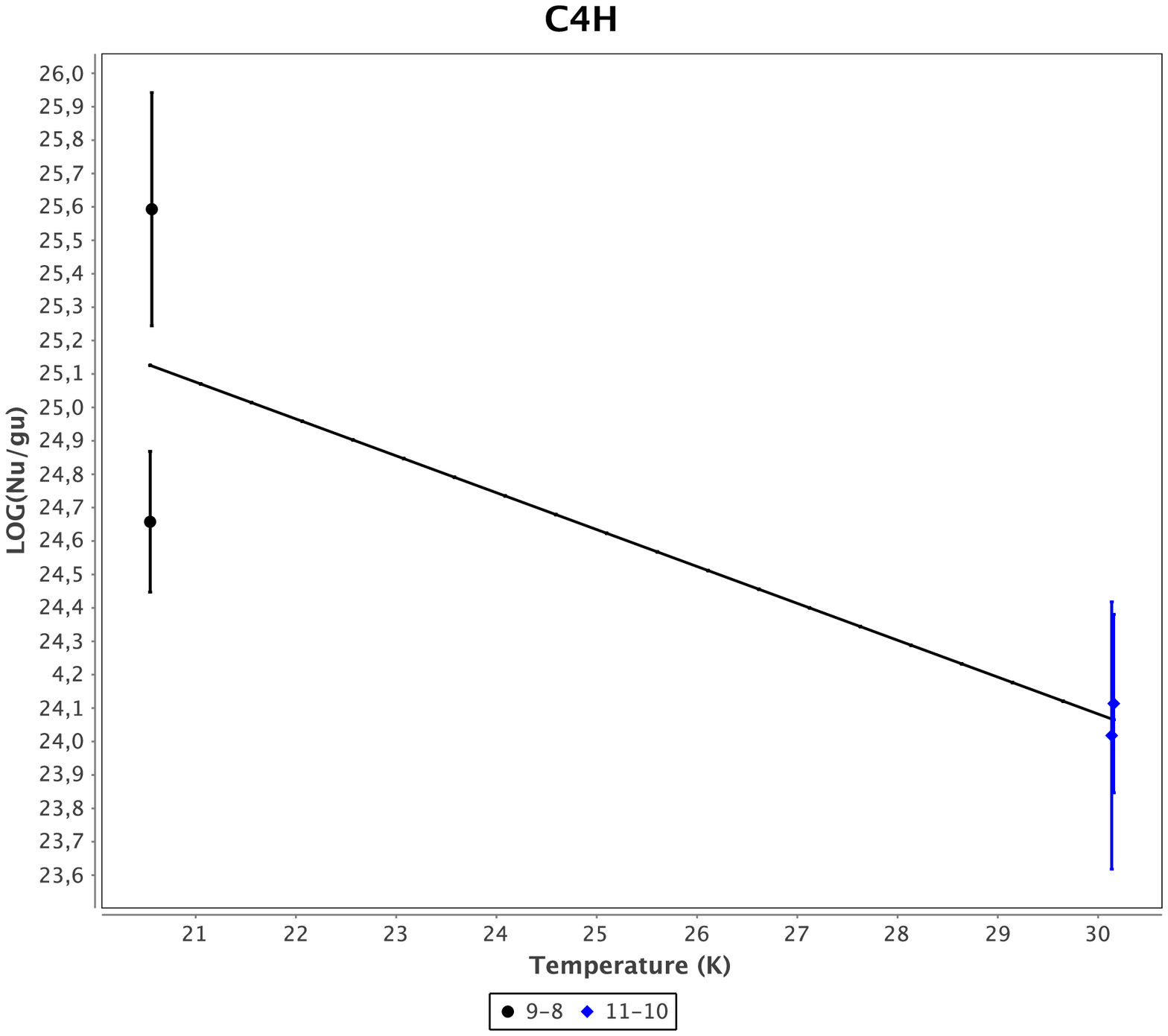}
\includegraphics[width=0.33\textwidth]{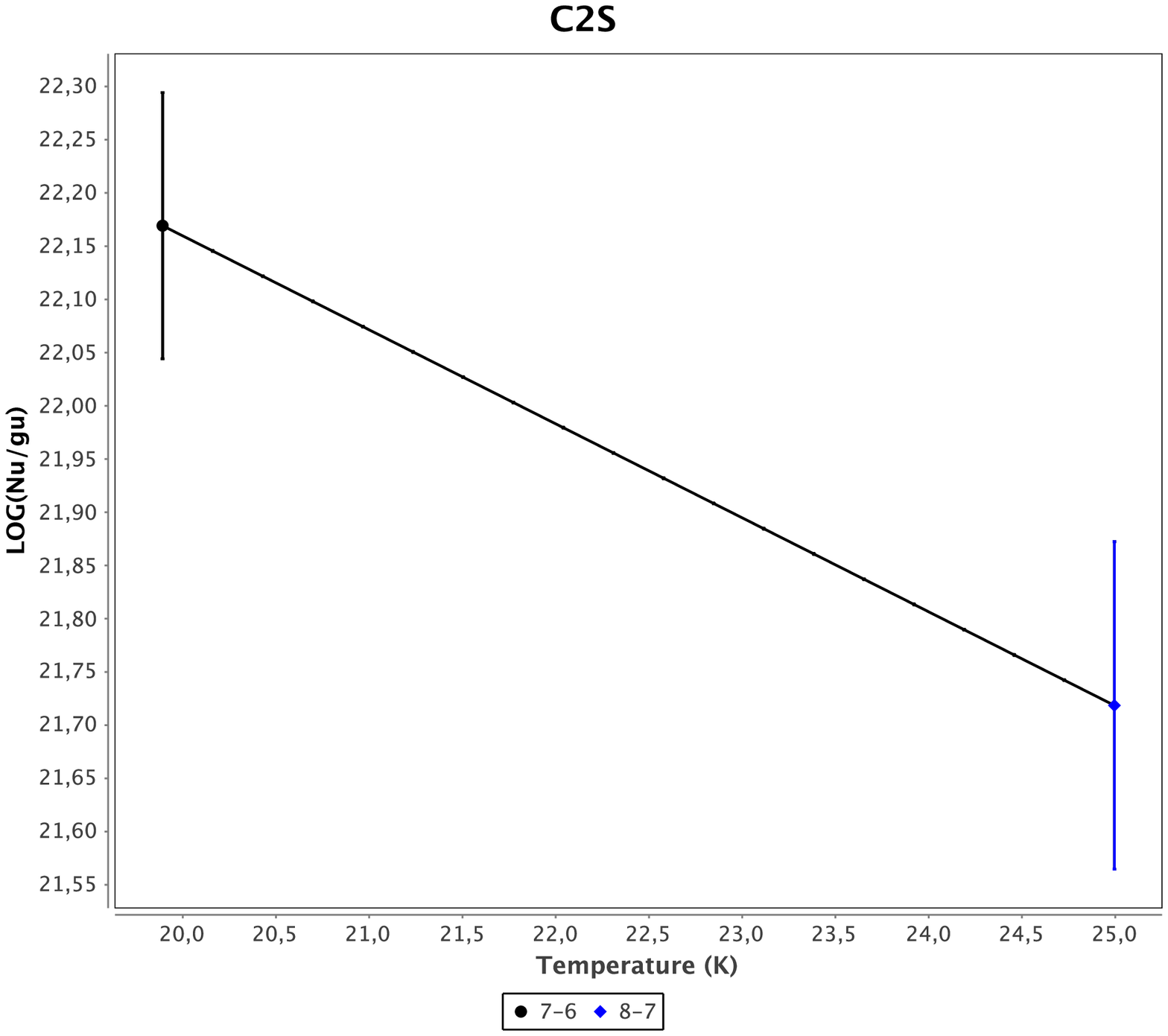}
\includegraphics[width=0.33\textwidth]{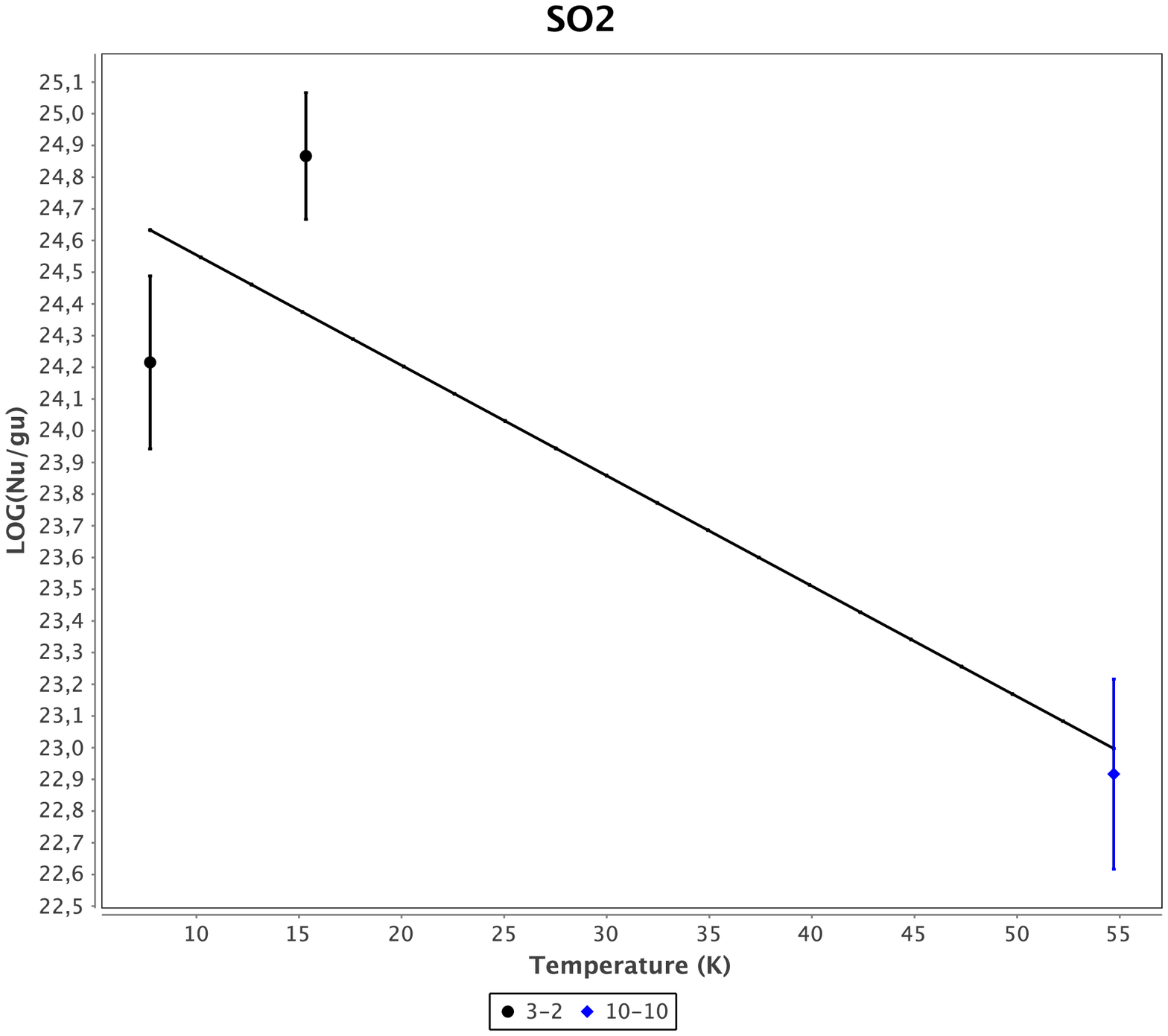}
\includegraphics[width=0.33\textwidth]{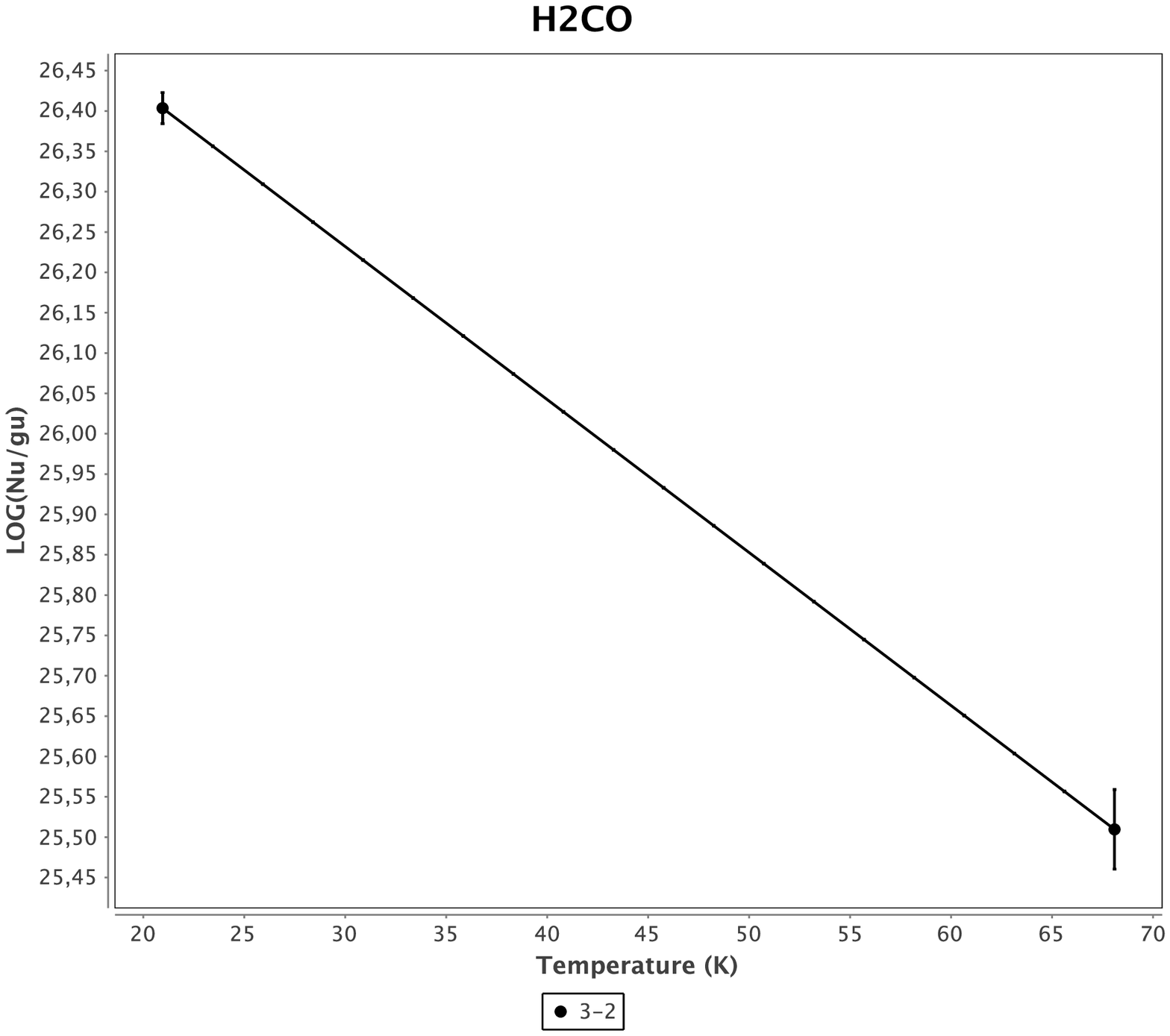}
\includegraphics[width=0.33\textwidth]{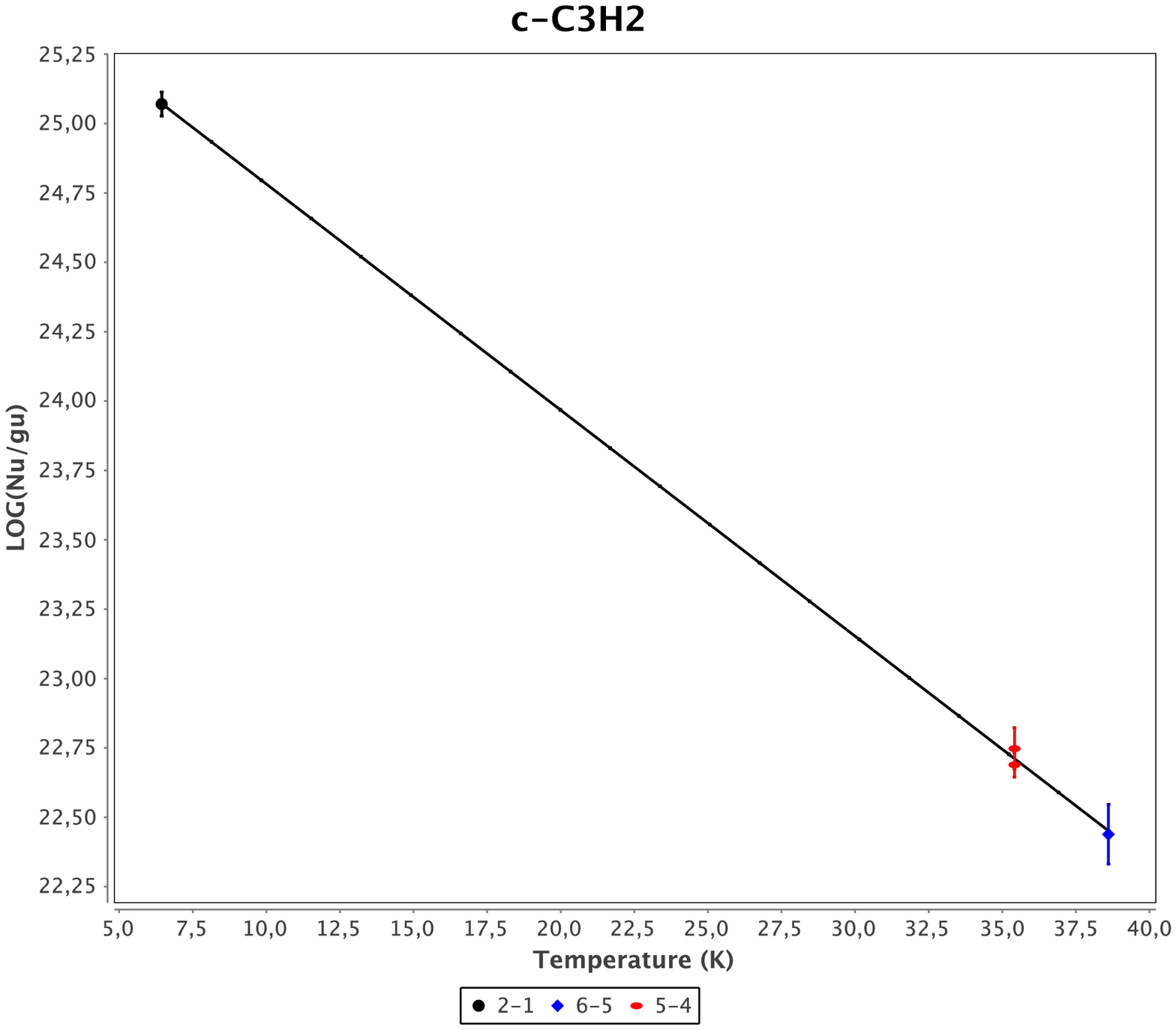}
\includegraphics[width=0.33\textwidth]{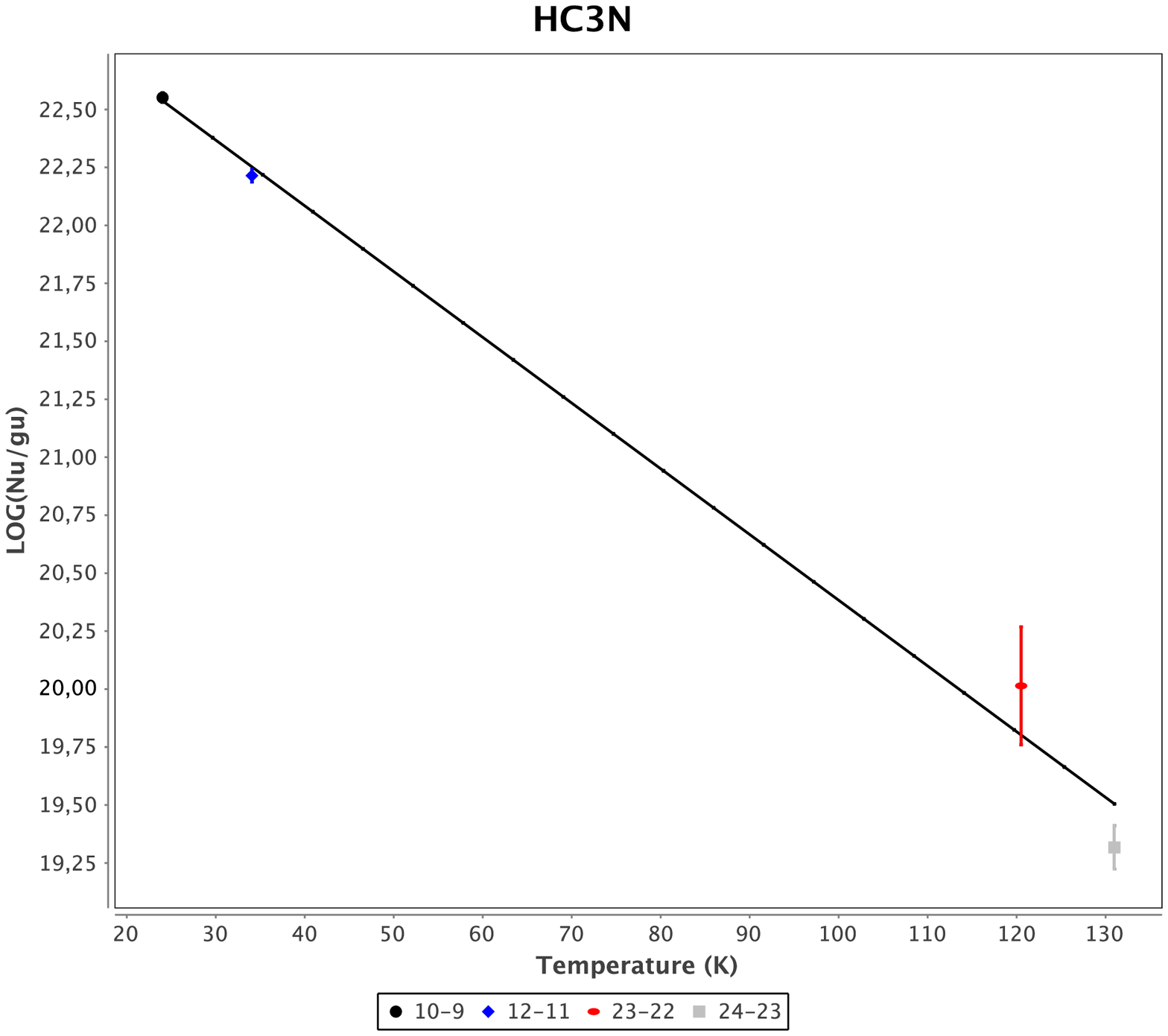}
\includegraphics[width=0.33\textwidth]{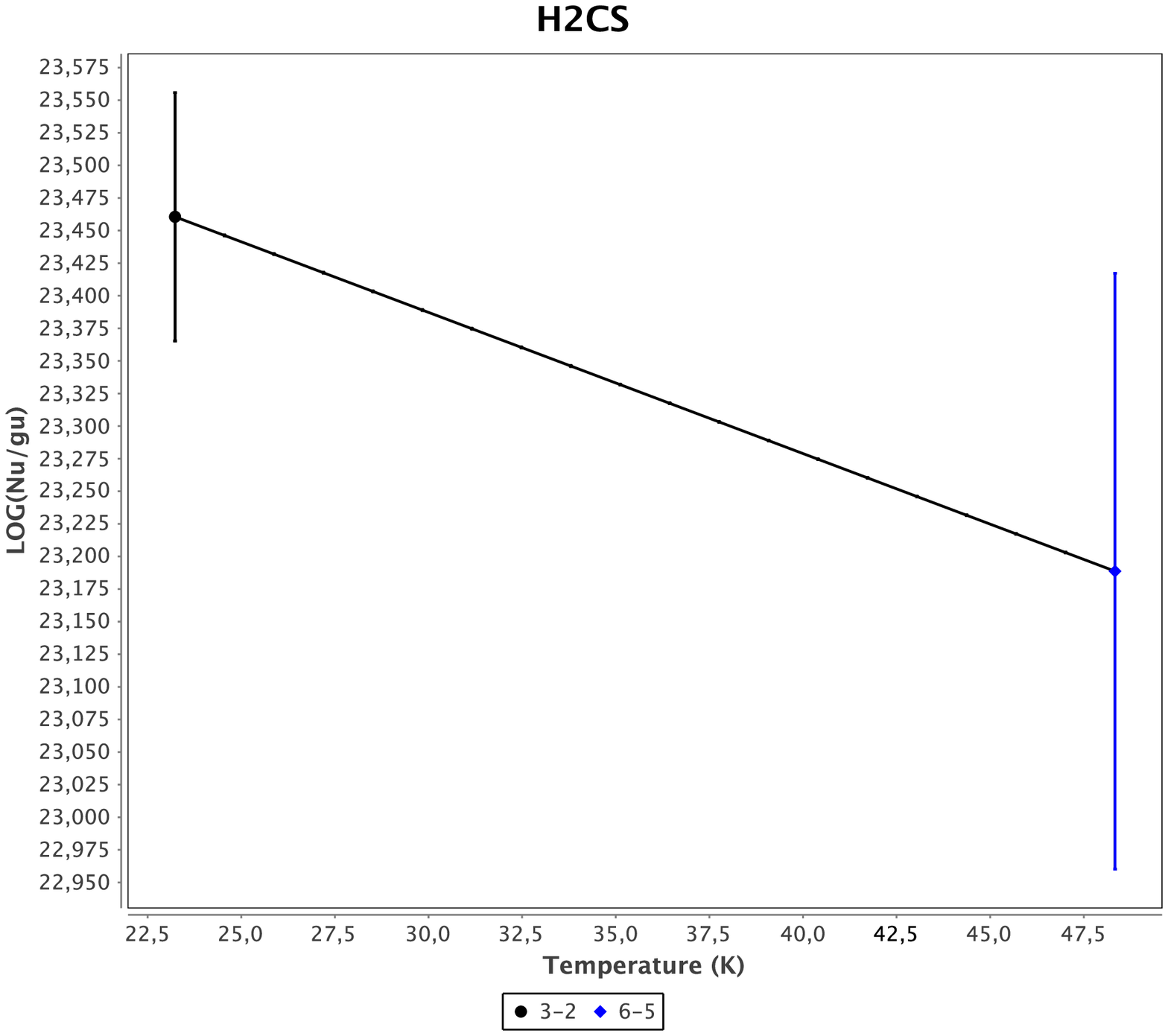}
\includegraphics[width=0.33\textwidth]{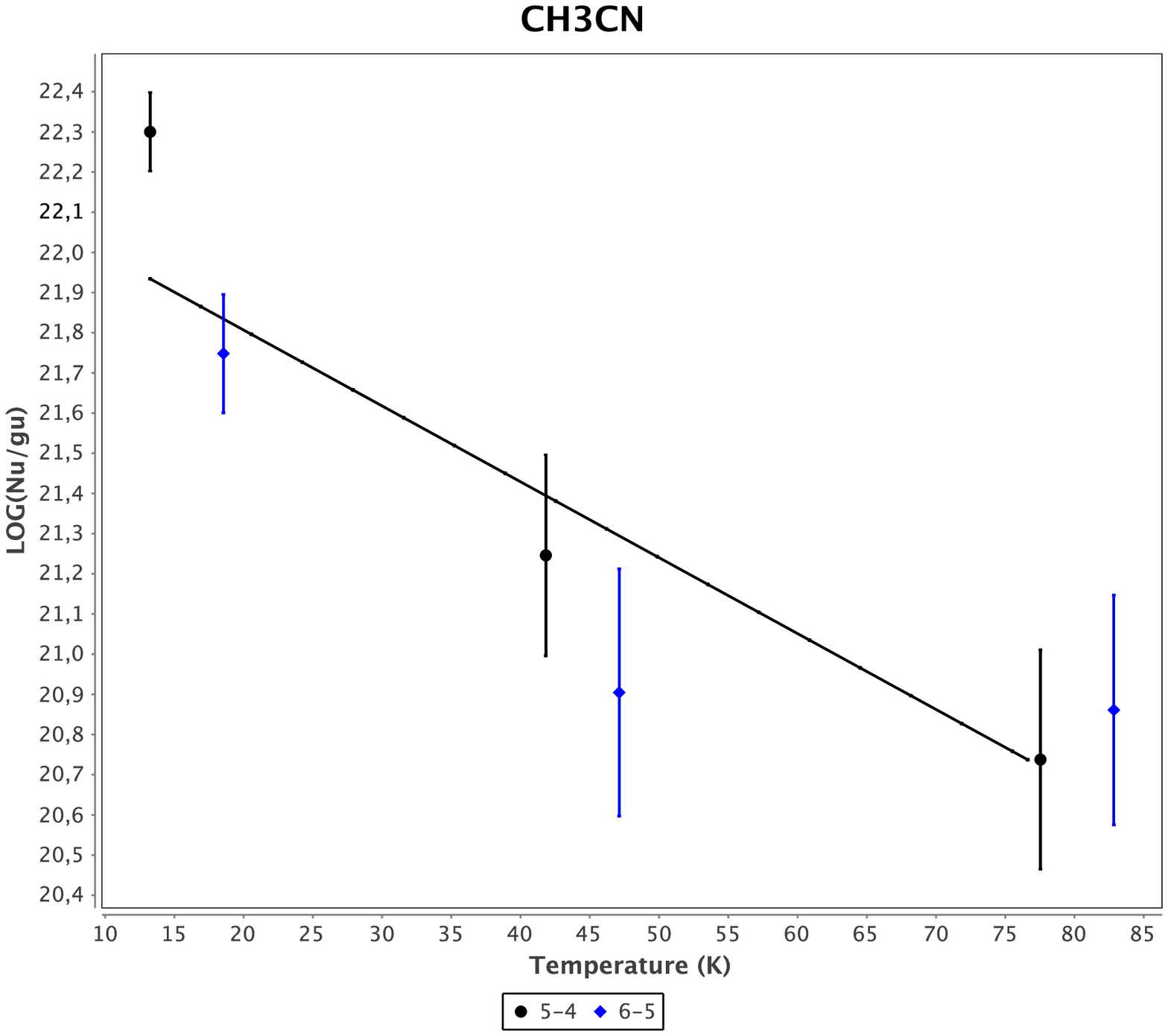}
\end{figure*}
\begin{figure*}
\includegraphics[width=0.33\textwidth]{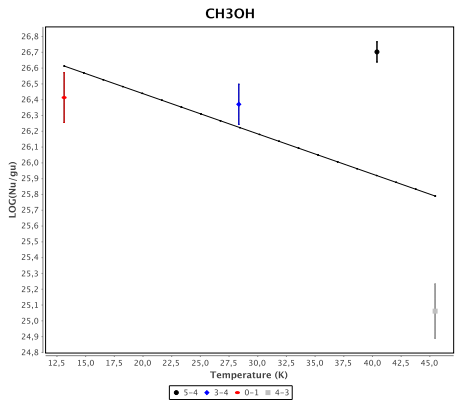}
   \caption{Rotational diagrams of the  most representative molecules towards MP2 (Offset ($+$0$\arcsec$, $+$40$\arcsec$)). }
\end{figure*}
\end{appendix}
\end{document}

%% file: table-paolo.tex
%\caption{Offset (0$\arcsec$,0$\arcsec$)} % title of Table
%\begin{center} % used for centering table
\section{Identified lines}
\begin{table*}[!h]
\caption{Observed transitions and line intensities}
\begin{tabular}{l l ccc } % centered columns (3 columns)

\hline\hline
Frequency   & Transition  & Intensity (IF) & Intensity (MP1) & Intensity (MP2)\\
(MHz)	& & (K$\times$km~s$^{-1}$)& (K$\times$km~s$^{-1}$)& (K$\times$km~s$^{-1}$)\\

 \hline 

\hline % inserts single horizontal line
       83802 &  H(66)$\delta$  &  1.23(0.11)  &  0.24(0.06)  & \\
       84107 &  C(40)$\alpha$+He(40)$\alpha$ in the image band  &  & 
 0.66(0.14)  & \\
       84521 &  CH$_3$OH \textit{5$_{-1,5}$$\rightarrow$4$_{0,4}$}  & 
 0.48(0.08)  &  0.35(0.04)  &  0.62(0.04) \\
       84914 &  H(60)$\gamma$  &  2.26(0.12)  &  0.70(0.04)  & \\
       85197 &  CS \textit{2$\rightarrow$1} in the image band  &  &  0.39(0.19) 
 &  0.39(0.09) \\
       85339 &  c-C$_3$H$_2$ \textit{2$_{1,2}$$\rightarrow$1$_{0,1}$}  & 
 1.18(0.20)$^a$  &  1.50(0.07)$^a$  &  2.09(0.09)$^a$ \\
       85348 &  HCS$^+$ \textit{2$\rightarrow$1}  &  0.76(0.09)  &  0.32(0.02) 
 &  0.33(0.09) \\
       85457 &  CH$_3$C$_2$H \textit{5$_k$$\rightarrow$4$_k$} k=0,1  & 
 0.41(0.07)  &  0.47(0.16)  &  0.53(0.07) \\
       85634 & 
 C$_4$H \textit{9$\rightarrow$8} J=\textit{19/2$\rightarrow$17/2}  &  &  & 
 0.29(0.08) \\
       85673 & 
 C$_4$H \textit{9$\rightarrow$8} J=\textit{17/2$\rightarrow$15/2}  &  &  & 
 0.53(0.15) \\
       85688 &  H(42)$\alpha$  &  15.97(0.13)  &  4.08(0.13)  &  0.89(0.16)\\
       85731 &  C(42)$\alpha$  &  &  0.23(0.06)  & \\
       85925 &  NH$_2$D \textit{1$_{1,1}$-1$_{0,1}$}$^{t}$  &  &  0.19(0.04)  & 
\\
%       86054 &  HC$^{15}$N \textit{1$\rightarrow$0}  &  &  &  \\
       86055 &  HC$^{15}$N \textit{1$\rightarrow$0}  &  0.29(0.07)  & 
 0.52(0.14)  &  0.37(0.06)\\
       86094 &  SO \textit{2$_2$$\rightarrow$1$_1$}  &  0.38(0.05)  & 
 0.31(0.12)  &  0.54(0.01) \\
       86339 &  H$^{13}$CN \textit{1$\rightarrow$0}  &  1.78(0.08)  & 
 2.11(0.03)  & 1.64(0.03) \\
%       86340 &  H$^{13}$CN \textit{1$\rightarrow$0}  &  &  &  1.64(0.03) \\
       86488 &  H(70)$\epsilon$  &  0.95(0.12)  &  & \\
       86671 & 
 HCO \textit{1$_{0,1}$$\rightarrow$0$_{0,0}$} J=\textit{3/2$\rightarrow$1/2} F=\textit{2$\rightarrow$1} 
 &  &  0.22(0.05)  &  0.54(0.05) \\
       86690 &  H(74)$\zeta$  &  0.99(0.17)  &  & \\
       86708 & 
 HCO \textit{1$_{0,1}$$\rightarrow$0$_{0,0}$} J=\textit{3/2$\rightarrow$1/2} F=\textit{1$\rightarrow$0} 
 &  &  0.15(0.03)  &  0.37(0.05) \\
%       86709 & 
% HCO \textit{1$_{0,1}$$\rightarrow$0$_{0,0}$} J=\textit{3/2$\rightarrow$1/2} F=\textit{1$\rightarrow$0} 
% &  &  0.15(0.03)  & \\
       86754 &  H$^{13}$CO$^+$ \textit{1$\rightarrow$0}  &  0.55(0.09)  & 
 1.43(0.02)  &  1.14(0.09) \\
       86777 & 
 HCO \textit{1$_{0,1}$$\rightarrow$0$_{0,0}$} J=\textit{1/2$\rightarrow$1/2} F=\textit{1$\rightarrow$1} 
 &  &  0.12(0.03)  &  0.31(0.04) \\
       86847 &  SiO \textit{2$\rightarrow$1}  &  &  0.32(0.04)  &  0.35(0.01) \\
       87091 &  HN$^{13}$C \textit{1$\rightarrow$0}  &  0.31(0.06)  & 
 0.57(0.04)  &  0.35(0.04) \\
       87284 & 
 C$_2$H N=\textit{1$\rightarrow$0} J=\textit{3/2$\rightarrow$1/2} F=\textit{1$\rightarrow$1} 
 &  0.42(0.08)  &  0.68(0.02)  &  0.69(0.04) \\
       87317 & 
 C$_2$H N=\textit{1$\rightarrow$0} J=\textit{3/2$\rightarrow$1/2} F=\textit{2$\rightarrow$1} 
 &  4.37(0.38)  &  5.47(0.49)  &  5.77(0.46) \\
       87329 & 
 C$_2$H N=\textit{1$\rightarrow$0} J=\textit{3/2$\rightarrow$1/2} F=\textit{1$\rightarrow$0} 
 &  2.12(0.11)  &  2.97(0.06)  &  3.00(0.36) \\
       87402 & 
 C$_2$H N=\textit{1$\rightarrow$0} J=\textit{1/2$\rightarrow$1/2} F=\textit{1$\rightarrow$1} 
 &  2.17(0.14)  &  3.24(0.06)  &  3.38(0.13) \\
       87407 & 
 C$_2$H N=\textit{1$\rightarrow$0} J=\textit{1/2$\rightarrow$1/2} F=\textit{0$\rightarrow$1} 
 &  0.84(0.15)  &  1.15(0.05)  &  1.10(0.06) \\
       87446 & 
 C$_2$H N=\textit{1$\rightarrow$0} J=\textit{1/2$\rightarrow$1/2} F=\textit{1$\rightarrow$0} 
 &  &  0.77(0.06)  &  0.68(0.05) \\
       87615 &  H(65)$\delta$  &  1.86(0.08)  &  0.41(0.09)  & \\
       87706 &  Unidentified (F$_{image}$=110433~MHz)  &  1.47(0.30)  &  & \\
       87938 &  $^{13}$CO \textit{1$\rightarrow$0} in the image band  & 
 0.96(0.10)  &  1.05(0.05)  &  0.84(0.05) \\
       88358 &  C$^{18}$O \textit{1$\rightarrow$0} in the image band  &  & 
 0.36(0.03)  &  0.27(0.04) \\
       88406 &  H(52)$\beta$  &  6.77(0.08)  &  1.01(0.07)  &  0.47(0.06) \\
       88578 &  Unidentified (F$_{image}$=109560~MHz)  &  0.47(0.06)  &  & \\
       88632 &  HCN \textit{1$\rightarrow$0} F=1-1, F=2-1 and F=0-1  & 
 8.47(0.06)  &  20.58(0.05)  &  23.24(0.04) \\
       88866 &  H$^{15}$NC \textit{1$\rightarrow$0} $^t$  &  &  0.14(0.04)  & \\
       88942 &  Unidentified (F$_{image}$=109196~MHz)  &  &  &  0.20(0.04) \\
       89112 &  Unidentified (F$_{image}$=109027~MHz)  &  0.24(0.05)  &  & \\
       89165 &  Unidentified (F$_{image}$=108974.5~MHz)  &  0.66(0.02)  &  & \\
       89189 &  HCO$^+$ \textit{1$\rightarrow$0}  &  11.59(0.02)$^a$ &  13.90(0.22)  & 
 17.67(0.11)$^a$ \\
%       89190 &  HCO$^+$ \textit{1$\rightarrow$0}  &  11.59(0.02)$^a$  &  & \\
       89199 &  H(59)$\gamma$  &  2.50(0.21)  &  & \\
       90174 &  H(69)$\epsilon$ and   H(73)$\zeta$ &  1.63(0.12)  &  & \\
       90664 &  HNC \textit{1$\rightarrow$0}  &  3.19(0.05)  &  10.58(0.14)  & 
 8.33(0.06) \\
       90979 &  HC$_3$N \textit{10$\rightarrow$9}  &  0.90(0.06)  &  2.41(0.08) 
 &  1.41(0.03) \\
       91406 &  H(39)$\alpha$ in the image band  &  0.61(0.13)  &  & \\
       91663 &  H(64)$\delta$  &  1.86(0.11)  &  0.52(0.08)  & \\
       91942 &  CH$_3$$^{13}$CN \textit{5$_0$$\rightarrow$4$_0$}  & 
 0.59(0.16)  &  & \\
       91971 &  CH$_3$CN \textit{5$_3$$\rightarrow$4$_3$}  &  0.64(0.14)  & 
 0.11(0.05)  &  0.11(0.03) \\
       91980 &  CH$_3$CN \textit{5$_2$$\rightarrow$4$_2$}  &  &  &  0.12(0.03) 
\\
       91987 &  CH$_3$CN \textit{5$_k$$\rightarrow$4$_k$} k=0,1  &  & 
 0.52(0.05)  &  0.41(0.04) \\
         \hline %inserts single line
\end{tabular}

$^a$ The profile shows self-absorption; $^t$ Tentative detection; $^b$ obtained by degrading the HERA maps to the angular resolution of the 3mm lines. 
\end{table*}

\setcounter{table}{0}
\begin{table*}
\caption{Observed transitions and line intensities (continuation)}

\begin{tabular}{l l ccc } % centered columns (3 columns)

\hline\hline
Frequency   & Transition  & Intensity (IF) & Intensity (MP1) & Intensity (MP2)\\
(MHz)	& & (K$\times$km~s$^{-1}$)& (K$\times$km~s$^{-1}$)& (K$\times$km~s$^{-1}$)\\

 \hline 

\hline % inserts single horizontal line

       92034 &  H(41)$\alpha$  &  20.45(0.17)  &  4.01(0.18)  &  1.39(0.05) \\
       92080 &  C(41)$\alpha$  &  0.38(0.08)$^t$  &  & \\
       92436 &  Unidentified (F$_{image}$=105702~MHz)  &  0.39(0.05)  &  & \\
       92494 &  $^{13}$CS \textit{2$\rightarrow$1}  &  0.30(0.05)  & 
 0.42(0.06)  &  0.37(0.03) \\
%       93173 & 
% N$_2$H$^+$ \textit{1$\rightarrow$0} F=\textit{1$\rightarrow$1} , \textit{1$\rightarrow$2} and \textit{1$\rightarrow$0} 
% &  1.53(0.06)  &  & \\
       93174 & 
 N$_2$H$^+$ \textit{1$\rightarrow$0} F=\textit{1$\rightarrow$1} , \textit{1$\rightarrow$2} and \textit{1$\rightarrow$0} 
 &  1.53(0.06)   &  8.66(0.09)  &  2.47(0.04) \\
       93607 &  H(51)$\beta$  &  6.04(0.10)  &  1.06(0.08)  &  0.35(0.08) \\
       93776 &  H(58)$\gamma$  &  2.77(0.17)  &  0.56(0.09)  & \\
       93827 &  H(72)$\zeta$  &  0.89(0.09)  &  & \\
       93870 &  C$_2$S \textit{7$_8$$\rightarrow$6$_7$}  &  0.14(0.04)  &  & 
 0.16(0.02) \\
       94073 &  H(68)$\epsilon$  &  1.03(0.07)  &  & \\
      103252 &  Unidentified (F$_{image}$=94889~MHz)  &  0.52(0.04)  &  & \\
      103266 &  C$_4$H v=1 \textit{21/2$\rightarrow$19/2} $^{t}$  &  0.47(0.05) 
 &  & \\
      103915 &  H(56)$\gamma$  &  3.02(0.07)  &  0.46(0.07)  & \\
      104029 &  SO$_2$ \textit{3$_{1,3}$$\rightarrow$2$_{0,2}$}$^t$  &  & 
 0.16(0.06)  &  0.11(0.03) \\
      104062 &  Unidentifed (F$_{image}$=93900~MHz)  &  &  &  0.16(0.03) \\
      104239 &  SO$_2$ \textit{10$_{1,9}$$\rightarrow$10$_{0,10}$}$^t$  &  &  & 
 0.10(0.03) \\
      104533 &  H(51)$\beta$ in the image band  &  0.44(0.03)  &  & \\
      104617 &  H$_2$CS \textit{3$_{1,2}$$\rightarrow$2$_{1,1}$}  &  0.19(0.03) 
 &  0.38(0.03)  &  0.21(0.02) \\
      104667 & 
 C$_4$H \textit{11$\rightarrow$10} J=\textit{23/2$\rightarrow$21/2}  &  & 
 0.19(0.03)  &  0.30(0.06) \\
      104705 & 
 C$_4$H \textit{11$\rightarrow$10} J=\textit{21/2$\rightarrow$19/2}  &  & 
 0.27(0.05)  &  0.30(0.04) \\
      104966 &  N$_2$H$^+$ \textit{1$\rightarrow$0} in the image band  &  & 
 0.52(0.03)  & \\
      105302 &  H(49)$\beta$  &  6.22(0.10)  &  0.66(0.06)  &  0.39(0.05) \\
      105410 &  H(61)$\delta$  &  1.70(0.07)  &  & \\
      105516 &  Unidentified (F$_{image}$=92625~MHz)  &  0.36(0.10)  &  & \\
      106079 &  H(69)$\zeta$  &  0.71(0.09)  &  & \\
      106104 &  H(41)$\alpha$ in the image band  &  0.45(0.10)  &  & \\
      106348 &  C$_2$S \textit{N=8$\rightarrow$7}  \textit{J=9$\rightarrow$8}      &  &  &  0.13(0.02) \\
      106737 &  H(39)$\alpha$  &  18.80(0.11)  &  2.33(0.07)  &  0.57(0.06) \\
      107014 &  CH$_3$OH \textit{3$_{1,3}$$\rightarrow$4$_{0,4}$}  &  & 
 0.14(0.06)  &  0.55(0.07) \\
      107206 &  H(65)$\epsilon$  &  1.23(0.12)  &  & \\
      107254 &  Unidentified (F$_{image}$=90887~MHz)  &  0.59(0.18)  &  & \\
      107424 &  Unidentified (F$_{image}$=90720~MHz)  &  0.51(0.03)  &  & \\
      107476 &  HNC \textit{1$\rightarrow$0} in the image band  &  0.25(0.07) 
 &  0.46(0.03)  &  0.48(0.04) \\
      108427 & 
 $^{13}$CN N=\textit{1$\rightarrow$0} J=\textit{1/2$\rightarrow$1/2} 
 &  &  0.27(0.01)  &  0.16(0.01) \\
      108780 & 
 $^{13}$CN N=\textit{1$\rightarrow$0} J=\textit{3/2$\rightarrow$1/2}  
 &  &  0.24(0.01)  &  0.39(0.03) \\
      108894 &  CH$_3$OH \textit{0$_{0,0}$$\rightarrow$1$_{-1,1}$}  &  & 
 0.20(0.09)  &  0.19(0.03) \\
      108950 &  HCO$^+$ \textit{1$\rightarrow$0} in the image band  & 
 0.21(0.21)  &  0.47(0.13)  &  0.72(0.05) \\
      109174 &  HC$_3$N \textit{12$\rightarrow$11}  &  0.70(0.04)  & 
 2.41(0.04)  &  1.45(0.04) \\
      109206 &  Unidentified (F$_{image}$=88935~MHz)  &  &  &  0.13(0.03) \\
      109252 &  SO \textit{2$_3$$\rightarrow$1$_2$}  &  0.21(0.18)  & 
 0.34(0.01)  &  0.59(0.01) \\
      109536 &  H(55)$\gamma$  &  3.25(0.10)  &  & \\
      109782 &  C$^{18}$O \textit{1$\rightarrow$0}  &  3.47(0.06)  & 
 4.63(0.05)  &  3.75(0.06) \\
      109906 &  HNCO \textit{5$\rightarrow$4}  &  &  0.13(0.04)  & \\
      109998 &  Unidentified (F$_{image}$=88142~MHz)  &  0.43(0.05)  &  & \\
%      110024 & 
% C$^{15}$N N=\textit{1$\rightarrow$0} J=\textit{3/2$\rightarrow$1/2}  &  & 
% 0.13(0.03)  & \\
      110025 & 
 C$^{15}$N N=\textit{1$\rightarrow$0} J=\textit{3/2$\rightarrow$1/2} F=\textit{2$\rightarrow$1} and \textit{1$\rightarrow$0} 
 &  & 0.13(0.03)  &  0.25(0.04) \\
      110119 &  Unidentified (F$_{image}$=88022~MHz)  &  &  &  0.18(0.04) \\
      110201 &  $^{13}$CO \textit{1$\rightarrow$0}  &  42.31(0.03)  & 
 49.18(1.07)  &  41.61(0.15) \\
      110364 &  CH$_3$CN \textit{6$_3$$\rightarrow$5$_3$}  &  &  &  0.21(0.06) 
\\
      110375 &  CH$_3$CN \textit{6$_2$$\rightarrow$5$_2$}  &  &  0.12(0.04)  & 
 0.13(0.04) \\
      110383 &  CH$_3$CN \textit{6$_k$$\rightarrow$5$_k$} k=0,1  &  1.13(0.13) 
 &  0.59(0.04)  &  0.34(0.05) \\
      110601 &  H(60)$\delta$  &  1.66(0.13)  &  & \\
      205081 &  CH$_3$C$_2$H \textit{12$_0$$\rightarrow$11$_0$}  &  0.49(0.08) 
 &  0.71(0.10)  & \\
      205292 &  C$^{18}$O \textit{2$\rightarrow$1} in the image band  &  &  & 
 1.01(0.08) \\
      205986 &  H$_2$CS \textit{6$_{0,6}$$\rightarrow$5$_{0,5}$}  &  & 
 0.26(0.06)  & \\

         \hline %inserts single line
\end{tabular}

$^a$ The profile shows self-absorption; $^t$ Tentative detection; $^b$ obtained by degrading the HERA maps to the angular resolution of the 3mm lines. 
\end{table*}

\setcounter{table}{0}

\begin{table*}
\caption{Observed transitions and line intensities (continuation)}
\begin{tabular}{l l ccc } % centered columns (3 columns)

\hline\hline
Frequency   & Transition  & Intensity (IF) & Intensity (MP1) & Intensity (MP2)\\
(MHz)	& & (K$\times$km~s$^{-1}$)& (K$\times$km~s$^{-1}$)& (K$\times$km~s$^{-1}$)\\

 \hline 

\hline % inserts single horizontal line

       206176 &  SO \textit{4$_5$$\rightarrow$3$_4$}  &  1.41(0.09)  & 
 1.30(0.07)  &  1.60(0.09) \\
      206676 &  Unidentified (F$_{image}$=218175~MHz)  &  1.31(0.12)  & 
 0.79(0.06)  &  0.65(0.05) \\
      206907 &  Unidentified (F$_{image}$=217944~MHz)  &  0.49(0.11)  &  & \\
      207836 &  Unidentified (F$_{image}$=217015~MHz)  &  &  &  0.40(0.08) \\
      208700 &  SO$_2$ \textit{3$_{2,2}$$\rightarrow$2$_{1,1}$}  &  &  & 
 0.35(0.07) \\
      208767 &  Unidentified (F$_{image}$=221286~MHz)  &  &  0.29(0.07)  & \\
      209097 &  Unidentified F$_{image}$=220956~MHz)  &  0.64(0.06)  & 
 0.76(0.05)  &  0.86(0.12) \\
      209201 &  H$_2$CS \textit{6$_{1,5}$$\rightarrow$5$_{1,4}$}  &  0.39(0.04) 
 &  0.53(0.08)  &  0.70(0.16) \\
      209230 &  HC$_3$N \textit{23$\rightarrow$22}  &  0.22(0.10)  & 
 0.28(0.11)  &  0.59(0.15) \\
      209277 &  Unidentified (F$_{image}$=220777~MHz)  &  0.37(0.08)  & 
 0.32(0.08)  & \\
      209591 &  Unidentified (F$_{image}$=220462~MHz)  &  1.84(0.54)  &  & \\
      209655 &  $^{13}$CO \textit{2$\rightarrow$1} in the image band  & 
 1.97(0.10)  &  2.13(0.07)  &  2.18(0.11) \\
      209894 &  H(44)$\gamma$  &  2.37(0.16)  &  & \\
      209999 &  C(44)$\gamma$  &  &  &  0.51(0.11) \\
      210495 &  C$^{18}$O \textit{2$\rightarrow$1} in the image band  & 1.18(0.07)  & 
 0.32(0.06)  & 1.28(0.09) \\
%      210496 &  C$^{18}$O \textit{2$\rightarrow$1} in the image band  & 
% 1.18(0.07)  &  & \\
      210502 &  H(31)$\alpha$  &  15.18(0.20)  &  & \\
      216373 &  CCD N=\textit{3$\rightarrow$2} J=\textit{7/2$\rightarrow$5/2} 
 &  &  0.31(0.05)  &  0.49(0.10) \\
      216428 &  CCD N=\textit{3$\rightarrow$2} J=\textit{5/2$\rightarrow$3/2} 
 &  &  0.31(0.05)  & 0.40(0.07) \\
%      216431 &  CCD N=\textit{3$\rightarrow$2} J=\textit{5/2$\rightarrow$3/2} 
% &  &  &  0.40(0.07) \\
      216711 & 
 H$_2$S \textit{2$_{2,0}$$\rightarrow$2$_{1,1}$}
 &  &  &  0.21(0.05) \\
      217105 &  SiO \textit{5$\rightarrow$4}  &  &  0.22(0.05)  &  0.23(0.01) \\
%      217238 &  DCN \textit{3$\rightarrow$2}  &  &  &  1.32(0.06) \\
      217239 &  DCN J=\textit{3$\rightarrow$2}  &  1.23(0.11)  &  2.94(0.06)  & 1.32(0.06) 
\\
      217467 & 
 $^{13}$CN N=\textit{2$\rightarrow$1} J=\textit{5/2$\rightarrow$3/2} F=\textit{4$\rightarrow$3}, \textit{3$\rightarrow$2} and \textit{2$\rightarrow$1} 
 &  &  0.12(0.01)  &  0.38(0.03) \\
      217822 & 
 c-C$_3$H$_2$ \textit{6$_{1,6}$$\rightarrow$5$_{0,5}$} and \textit{6$_{0,6}$$\rightarrow$5$_{1,5}$} 
 &  0.80(0.20)  &  1.20(0.20)  &  1.86(0.20) \\
      217835 &  CO \textit{2$\rightarrow$1} in the image band  &  1.80(0.16)  &  1.95(0.45)
 &  3.04(0.40) \\
%      217837 &  CO \textit{2$\rightarrow$1} in the image band  &  &  & 
% 3.04(0.40) \\
%     217840 &  CO \textit{2$\rightarrow$1} in the image band  &  &  1.95(0.45) 
% & \\
      217940 &  c-C$_3$H$_2$ \textit{5$_{1,4}$$\rightarrow$4$_{2,3}$}  &  & 
 0.52(0.05)  &  1.13(0.05) \\
      218160 &  c-C$_3$H$_2$ \textit{5$_{2,4}$$\rightarrow$4$_{1,3}$}  &  & 
 0.33(0.15)  &  0.40(0.03) \\
      218182 &  Unidentified (F$_{image}$=230192~MHz)  &  &  0.50(0.09) &  0.58(0.11) \\
      %      218183 &  Unidentified (F$_{image}$=230191~MHz)  &  &  0.50(0.09)  & \\
      218222 &  H$_2$CO \textit{3$_{0,3}$$\rightarrow$2$_{0,2}$}  &  3.78(0.13) 
 &  7.71(0.10)  &  6.24(0.12) \\
      218325 &  HC$_3$N \textit{24$\rightarrow$23}  &  0.17(0.06)  & 
 0.27(0.04)  &  0.32(0.03) \\
      218440 &  CH$_3$OH \textit{4$_{2,2}$$\rightarrow$3$_{1,2}$}  & 
 0.63(0.13)  &  0.74(0.04)  &  0.35(0.06) \\
      218476 &  H$_2$CO \textit{3$_{2,2}$$\rightarrow$2$_{2,1}$}  &  1.02(0.12) 
 &  1.77(0.09)  &  1.42(0.07) \\
      218667 &  Unidentified (F$_{image}$=229707~MHz)  &  0.90(0.11)  &  & \\
      218760 &  H$_2$CO \textit{3$_{2,1}$$\rightarrow$2$_{2,0}$}  &  0.87(0.08) 
 &  1.75(0.05)  &  1.41(0.06) \\
      219560 &  C$^{18}$O \textit{2$\rightarrow$1}  &  10.68(0.17) & 
 14.0(0.07)  &  12.74(0.07) \\
      219597 &  Unidentified (F$_{image}$=228776~MHz)  &  1.90(0.13)  &  & \\
      219852 &  H$_2$CN \textit{3$_{0,3}$$\rightarrow$2$_{0,2}$} $^{t}$  &  1.03(0.29) & 
 0.57(0.05) &  0.48(0.06) \\
%      219853 &  H$_2$CN \textit{3$_{0,2}$$\rightarrow$2$_{0,2}$} $^{t}$  & 
% 1.03(0.29)  &  0.57(0.05)  & \\
      219911 &  H$_2$$^{13}$CO \textit{3$_{1,2}$-2$_{1,1}$} $^{t}$  & 
 1.06(0.15)  &  & \\
      219949 &  SO \textit{6$_5$$\rightarrow$5$_4$}  &  1.45(0.07)  & 
 2.07(0.01)  &  1.83(0-03) \\
      259012 &  H$^{13}$CN \textit{3$\rightarrow$2} $^b$  &  1.65(0.21)  & 
 2.50(0.25)  &  1.10(0.06) \\
      260255 &  H$^{13}$CO$^+$ \textit{3$\rightarrow$2} $^b$  &  2.00(0.04)  & 
 2.12(0.04)  &  1.52(0.03) \\
      262004 & 
 C$_2$H N=\textit{3$\rightarrow$2} J=\textit{7/2$\rightarrow$5/2} $^b$  & 
 14.59(0.60)  &  15.60(0.60)  &  12.89(0.47) \\
 \hline %inserts single line
 
\end{tabular}

$^a$ The profile shows self-absorption; $^t$ Tentative detection; $^b$ obtained by degrading the HERA maps to the angular resolution of the 3mm lines. \end{table*}